\documentclass[british,english]{aa}
\usepackage[T1]{fontenc}
\usepackage{verbatim}
\usepackage{amstext}
\usepackage{graphicx}

\makeatletter

\usepackage{graphicx}
\usepackage[varg]{txfonts}

\@ifundefined{showcaptionsetup}{}{%
 \PassOptionsToPackage{caption=false}{subfig}}
\usepackage{subfig}
\makeatother

\usepackage{babel}
\begin{document}
\title{Rotational mixing in carbon-enhanced metal-poor stars with \emph{s}-process enrichment}\titlerunning{Rotational mixing in CEMP-\emph{s} stars}

\author{E. Matrozis\thanks{Member of the International Max Planck Research School (IMPRS) for Astronomy and Astrophysics at the Universities of Bonn and Cologne}\and 
R. J. Stancliffe}

\institute{Argelander-Institut für Astronomie (AIfA), University of Bonn, Auf
dem Hügel 71, DE-53121, Bonn, Germany\\
\email{elvijs@astro.uni-bonn.de}}

\date{Received ; accepted}

\abstract{Carbon-enhanced metal-poor (CEMP) stars with s-process enrichment
(CEMP-\emph{s}) are believed to be the products of mass transfer from
an asymptotic giant branch (AGB) companion, which has long since become
a white dwarf. The surface abundances of CEMP-\emph{s} stars are thus
commonly assumed to reflect the nucleosynthesis output of the first
AGB stars. We have previously shown that, for this to be the case,
some physical mechanism must counter atomic diffusion (gravitational
settling and radiative levitation) in these nearly fully radiative
stars, which otherwise leads to surface abundance anomalies clearly
inconsistent with observations. Here we take into account angular
momentum accretion by these stars. We compute in detail the evolution
of typical CEMP-\emph{s }stars from the zero-age main sequence, through
the mass accretion, and up the red giant branch for a wide range of
specific angular momentum $j_{\textrm{a}}$ of the accreted material,
corresponding to surface rotation velocities, $v_{\textrm{rot}}$,
between about $0.3$ and $300\ \textrm{km}\thinspace\textrm{s}^{-1}$.
We find that only for $j_{\textrm{a}}\gtrsim10^{17}\ \text{cm}^{2}\thinspace\text{s}^{-1}$
($v_{\text{rot}}>20\ \text{km}\thinspace\text{s}^{-1}$, depending
on mass accreted) angular momentum accretion directly causes chemical
dilution of the accreted material. This could nevertheless be relevant
to CEMP-\emph{s} stars, which are observed to rotate more slowly,
if they undergo continuous angular momentum loss akin to solar-like
stars. In models with rotation velocities characteristic of CEMP-\emph{s}
stars, rotational mixing primarily serves to inhibit atomic diffusion,
such that the maximal surface abundance variations (with respect to
the composition of the accreted material) prior to first dredge-up
remain within about $0.4\ \text{dex}$ without thermohaline mixing
or about $0.5\text{--}1.5$\ dex with thermohaline mixing. Even in
models with the lowest rotation velocities ($v_{\text{rot}}\lesssim1\ \text{km}\thinspace\text{s}^{-1}$),
rotational mixing is able to severely inhibit atomic diffusion, compared
to non-rotating models. We thus conclude that it offers a natural
solution to the problem posed by atomic diffusion and cannot be neglected
in models of CEMP-\emph{s} stars.}

\maketitle

\section{Introduction}

The tireless hunt for the most metal-poor and thus oldest stars in
the Galaxy \citep[e.g.][]{1985AJ.....90.2089B,2001A&A...375..366C,2009AJ....137.4377Y,2014ApJS..211...17A}
has revealed that a significant fraction of metal-poor stars are highly
enriched in carbon compared to the Sun. While these carbon-enhanced
metal-poor (CEMP) stars \citep[$\text{[C/Fe]}\gtrsim1$;][]{2005ARA&A..43..531B,2010A&A...509A..93M}
make up only about 10\% of all stars at $\text{[Fe/H]}\simeq-2$\footnotemark{}\footnotetext{The relative abundance of element A with respect to element B is $\text{[A/B]}=\log\left(C_\text{A}/C_\text{B}\right)-\log\left(C_\text{A}/C_\text{B}\right)_{\odot}$ where $C$ is the number or mass fraction.},
their prevalence rapidly increases towards lower metallicities to
near 100\% by $\text{[Fe/H]}\simeq-4$ \citep{2006ApJ...652L..37L,2012ApJ...744..195C,2013ApJ...762...27Y,2013AJ....146..132L,2014ApJ...797...21P}.
Spectroscopic studies have subsequently shown that CEMP stars are
also commonly marked by large abundances of elements produced by the
slow (\emph{s}) and/or rapid (\emph{r}) neutron-capture process, such
as barium and europium. Accordingly, CEMP stars are further classified
into CEMP-\emph{s}, CEMP-\emph{r}, and CEMP-\emph{r}/\emph{s} stars
\citep{2005ARA&A..43..531B,2010A&A...509A..93M}.

Many of these CEMP stars are relatively unevolved, being located on
the main sequence or on the red giant branch. The prodigious amounts
of carbon and heavy elements observed in these stars are thus expected
to have external origins. Moreover, given the very different conditions
required for the \emph{s-} and \emph{r}-process to operate – neutron
densities of $n\lesssim10^{7}\ \text{cm}^{-3}$ \citep{1999ARA&A..37..239B}
and $n\gtrsim10^{20}\ \text{cm}^{-3}$ \citep{2015MNRAS.452.1970W},
respectively – the different sub-classes most likely have distinct
formation sites. A key insight into the origin of CEMP-\emph{s} stars
\citep[$\text{[Ba/Fe]}>1$ and $\text{[Ba/Eu]}>0$;][]{2010A&A...509A..93M}
comes from their radial motion. Many studies over the years have shown
that the radial velocity of these objects periodically varies, indicating
the presence of an unseen companion \citep{2005ApJ...625..825L,2014MNRAS.441.1217S,2016A&A...588A...3H}.
The current view on the origin of these stars is thus accretion of
carbon- and \emph{s}-process-rich material from an asymptotic giant
branch (AGB) companion that has since become a white dwarf and faded
from view. This makes CEMP-\emph{s} stars the low-metallicity analogs
of Ba stars and CH stars \citep{1990ApJ...352..709M,2016A&A...586A.158J}.

Carbon-enhanced metal-poor stars with \emph{s}-process enrichment
thus provide a window onto the nucleosynthesis of the earliest generations
of low-mass AGB stars, an important contributor to the chemical evolution
of the Universe \citep[e.g.][]{1999ApJ...521..691T,2001ApJ...549..346T,2011MNRAS.414.3231K,2014ApJ...787...10B}.
To reliably link the surface abundances of CEMP-\emph{s} stars with
the nucleosynthesis output of these long-extinct AGB stars, however,
we must understand what happens to the material after it is accreted
by the less evolved companion. In particular, will it simply remain
on the surface, or mix with the material below? And if it does mix,
are all elements affected similarly, or is the nucleosynthesis signature
of the accreted material altered such that the surface abundances
no longer reflect the accreted composition?

Most studies that aim at linking the abundances of CEMP-\emph{s} stars
to low-metallicity AGB nucleosynthesis models allow for some overall
dilution of the accreted matter in the original material of the star
\citep[e.g.][]{2011MNRAS.418..284B,2012MNRAS.422..849B,2015A&A...581A..22A,2015A&A...576A.118A}.
While this approach works rather well in many cases, in a previous
paper we showed that the competition between gravitational settling
and radiative levitation should considerably modify the surface abundances
of CEMP-\emph{s} stars, in particular distorting the abundance ratios
between different elements, and concluded that either relatively high
mass loss or some additional mixing is required to bring the models
in accord with observations \citep{2016A&A...592A..29M}.

We remained agnostic as to the cause of this extra mixing. But rotational
mixing is a promising candidate, since the accreted material should
carry with it some angular momentum. In fact, the angular momentum
content may be so high that an interesting question is how the accreting
star can deal with it and accrete more than a few hundredths of a
solar mass of material \citep{2017arXiv170708224M}. Here we sidestep
that issue and simply assume that the newly accreted layers of CEMP-\emph{s}
stars, which can have masses up to some tenths of a solar mass, are
spinning more rapidly than their interiors. We then follow the post-mass-transfer
evolution of these stars, in particular noting the evolution of their
surface abundances as a result of rotational mixing, combined with
atomic diffusion and thermohaline mixing.

\section{Methods\label{sec:Methods}}

We use a version of the \textsc{stars} code \citep{1971MNRAS.151..351E,1972MNRAS.156..361E,1995MNRAS.274..964P,2009MNRAS.396.1699S}
to produce all the models presented here. The modifications introduced
for modelling rotating stars are described in \citet{2012MNRAS.419..748P,2012MNRAS.423.1221P}.
In particular, the internal transport of specific angular momentum
$j\propto\Omega r^{2}$ is modelled by a diffusion equation following
\citet{2000ApJ...528..368H}:
\begin{equation}
\frac{\mathrm{d}{\left(\Omega r^{2}\right)}}{\mathrm{d}{t}}=\frac{\partial}{\partial m}\left[\left(4\pi r^{2}\rho\right)^{2}r^{2}\left(D_{\text{conv}}+D_{\text{rot}}\right)\frac{\partial\Omega}{\partial m}\right].\label{eq:djdt}
\end{equation}
Here $\Omega$ is the angular velocity; $t$ is time; $r$ and $m$
are the radial and mass coordinate, respectively; $\rho$ is the density;
$D_{\text{conv}}$ is the convective mixing coefficient \citep[given by mixing length theory;][we set $\alpha_\text{MLT}=2$]{1958ZA.....46..108B};
and $D_{\textrm{rot}}$ is the turbulent viscosity given by:
\begin{equation}
D_{\text{rot}}=D_{\text{ES}}+D_{\text{DSI}}+D_{\text{SSI}}+D_{\text{SHI}}+D_{\text{GSF}}.\label{eq:Drot}
\end{equation}
In this equation each individual term corresponds to turbulent transport
arising from, respectively, Eddington-Sweet (ES) circulation \citep{1974IAUS...66...20K},
dynamical and secular shear instabilities \citep{1974IAUS...59..185Z,1978ApJ...220..279E},
the Solberg-H\o iland instability \citep{1946ApNr....4....1W}, and
the GSF instability \citep{1967ApJ...150..571G,1968ZA.....68..317F}.
We refer the reader to \citet{2000ApJ...528..368H} for a discussion
of the origin and evaluation of each of these terms.

The mass fraction $X_{i}$ of each species $i$ evolves according
to
\begin{equation}
\frac{\mathrm{d}{X_{i}}}{\mathrm{d}{t}}=\frac{\partial}{\partial m}\left[\left(4\pi r^{2}\rho\right)^{2}D_{\text{mix},i}\frac{\partial X_{i}}{\partial m}\right]-\frac{\partial}{\partial m}\left(4\pi r^{2}\rho X_{i}w_{i}\right)+R_{i}.\label{eq:dXdt}
\end{equation}
Here $R_{i}$ accounts for nuclear reactions, $w_{i}$ is the atomic
diffusion velocity as described in \citet{2016A&A...592A..29M}, and
$D_{\text{mix},i}$ in general is given by
\begin{equation}
D_{\text{mix},i}=D_{\text{conv}}+D_{\mu}+D_{i}+f_{c}D_{\text{rot}},\label{eq:Dmix}
\end{equation}
where $D_{\mu}$ and $D_{i}$ are the thermohaline mixing \citep{2010ApJ...723..563D}
and concentration diffusion coefficients, respectively.

There are two adjustable parameters in our adopted prescription for
rotational mixing. First, the parameter $f_{c}$ in Eq.\ \eqref{eq:Dmix}
determines the contribution of the rotationally induced instabilities
to chemical transport. Second, many of the terms in Eq.\ \eqref{eq:Drot}
depend on the molecular weight gradient $\nabla_{\mu}$. The sensitivity
of rotational mixing to $\mu$-gradients is assumed to be reduced
by a factor $f_{\mu}$, i.e. $\nabla_{\mu}$ is replaced by $f_{\mu}\nabla_{\mu}$.
Following \citet{2000ApJ...528..368H} we adopt $f_{c}=1/30$ \citep{1992A&A...253..173C}
and $f_{\mu}=0.05$. The influence of these parameters is examined
in Sect.\ \ref{subsec:fcfmu}.

To save a considerable amount of computational time, we do not use
the OP opacities and radiative accelerations \citep{2005MNRAS.360..458B,2007MNRAS.382..245S}
introduced in the code by \citet{2016A&A...592A..29M}. Instead we
use the OPAL-based \citep{1996ApJ...464..943I} opacity tables of
\citet{2004MNRAS.348..201E} and ignore radiative levitation for now.
This is perfectly sufficient to get a handle on the importance of
atomic diffusion in a given model.

\subsection{Accretion and grid selection}

As in related earlier work \citep{2007A&A...464L..57S,2008MNRAS.389.1828S,2016A&A...592A..29M},
accretion of material is simulated by increasing the mass of the models
at a rate of $10^{-6}\ M_{\odot}\text{yr}^{-1}$. The composition
of the added mass is set to the average composition of the AGB models
of \citet{2012ApJ...747....2L}. In particular, we use the yields
from their models with initial masses $M_{1}=0.9$, 1.0, 1.25 and
$1.5\ M_{\odot}$. The age, at which mass accretion starts is $t_{\text{mt}}=9.1$,
6.3, 3.06, and 1.8\ Gyr, respectively \citep[see table~1 in][]{2016A&A...592A..29M}.
All models have a zero-age main sequence (ZAMS) metallicity of $Z=10^{-4}$
\citep[scaled to $\text{[Fe/H]}=-2.14$]{2009ARA&A..47..481A}.

We adopt the same grid as in \citet{2016A&A...592A..29M}: the initial
secondary masses are $M_{2,\text{i}}=0.6\text{--}0.8\ M_{\odot}$
in steps of $0.05\ M_{\odot}$, and the accreted masses span $\Delta M=0.05\text{--}0.3\ M_{\odot}$
(with some $M_{2,\text{i}}=0.8\ M_{\odot}$ models with $\Delta M=10^{-3},10^{-2}\ M_{\odot}$),
resulting in final CEMP star masses $M_{2,\text{f}}=0.8\text{--}0.95\ M_{\odot}$
\citep{2015A&A...581A..62A}. For each combination of $M_{1}$, $M_{2,\text{i}}$
and $\Delta M$ we have one more dimension: the specific angular momentum
$j_{\text{a}}$ of the added material. We investigate ten values of
$j_{\text{a}}$ in the range $\left(0.001\text{--}1\right)\times10^{18}\ \text{cm}^{2}\thinspace\text{s}^{-1}$.
While the specific angular momentum of the accreted material in real
systems is likely closer to the higher end of these values (Sect.\ \ref{subsec:real-cemps}),
this range can be interpreted as representing different degrees of
angular momentum loss during accretion and is suitable to produce
CEMP-\emph{s} models with surface rotation velocities between $v_{\text{rot}}\lesssim0.5\ \text{km}\thinspace\text{s}^{-1}$
(i.e. nearly stationary) and $v_{\text{rot}}\gtrsim300\ \text{km}\thinspace\text{s}^{-1}$
(close to critical rotation), once they have settled on the post-mass-transfer
main sequence. On the ZAMS the models are uniformly rotating with
a surface rotation velocity $v_{\text{rot}}\simeq0.3\ \text{km}\thinspace\text{s}^{-1}$.

\section{Results\label{sec:Results}}

Prior to summarizing the general features of our calculations (Sect.\ \ref{subsec:results-summary}),
we consider in depth the evolution of a system characterized by $M_{1}=1.25\ M_{\odot}$,
$M_{2,\text{i}}=0.75\ M_{\odot}$, $\Delta M=0.05\ M_{\odot}$, first
under the influence of rotational mixing alone (Sect.~\ref{subsec:illust-mod-seq-r}),
and then together with diffusion and thermohaline mixing (Sect.~\ref{subsec:illust-mod-seq-rdt}).

\subsection{Models with rotational mixing only\label{subsec:illust-mod-seq-r}}

\begin{figure}
\subfloat[Hertzsprung-Russell diagram]{\includegraphics[width=1\columnwidth]{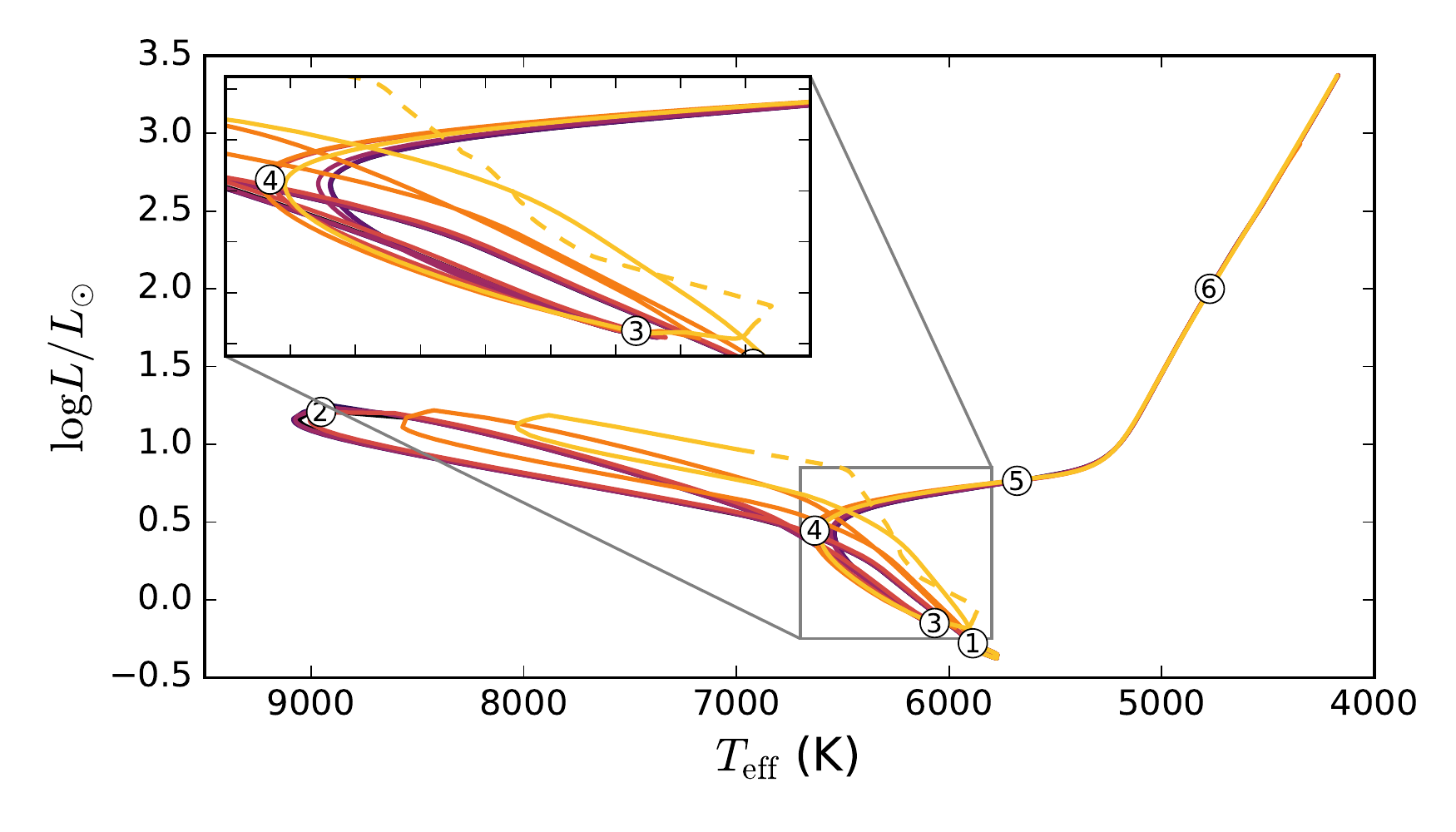}

\label{fig:mp125ms0750dm0050r_hrd}}

\subfloat[Evolution of surface angular rotation velocity]{\includegraphics[width=1\columnwidth]{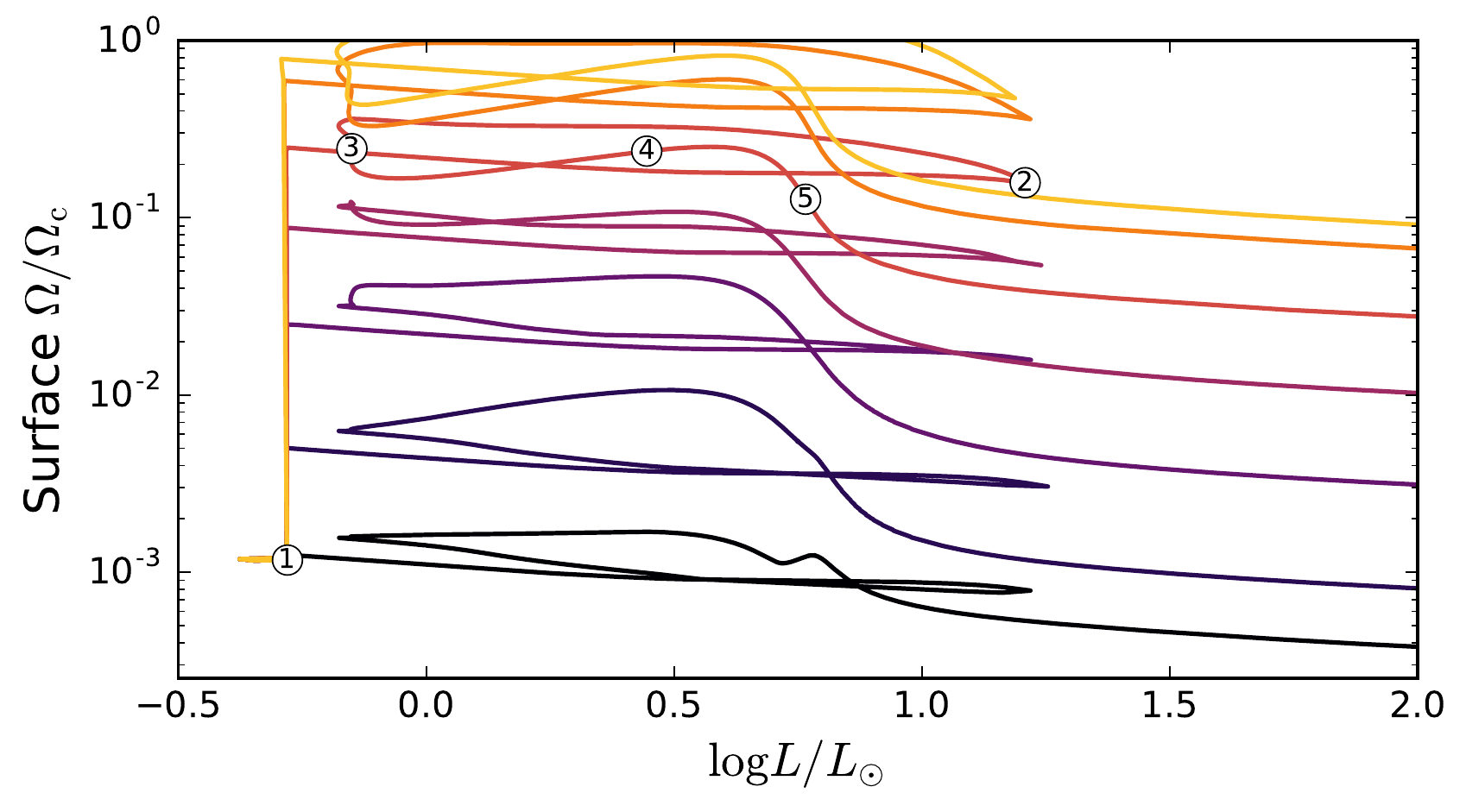}

\label{fig:mp125ms0750dm0050r_5-vs-89}}

\subfloat[Evolution of surface rotation velocity]{\includegraphics[width=1\columnwidth]{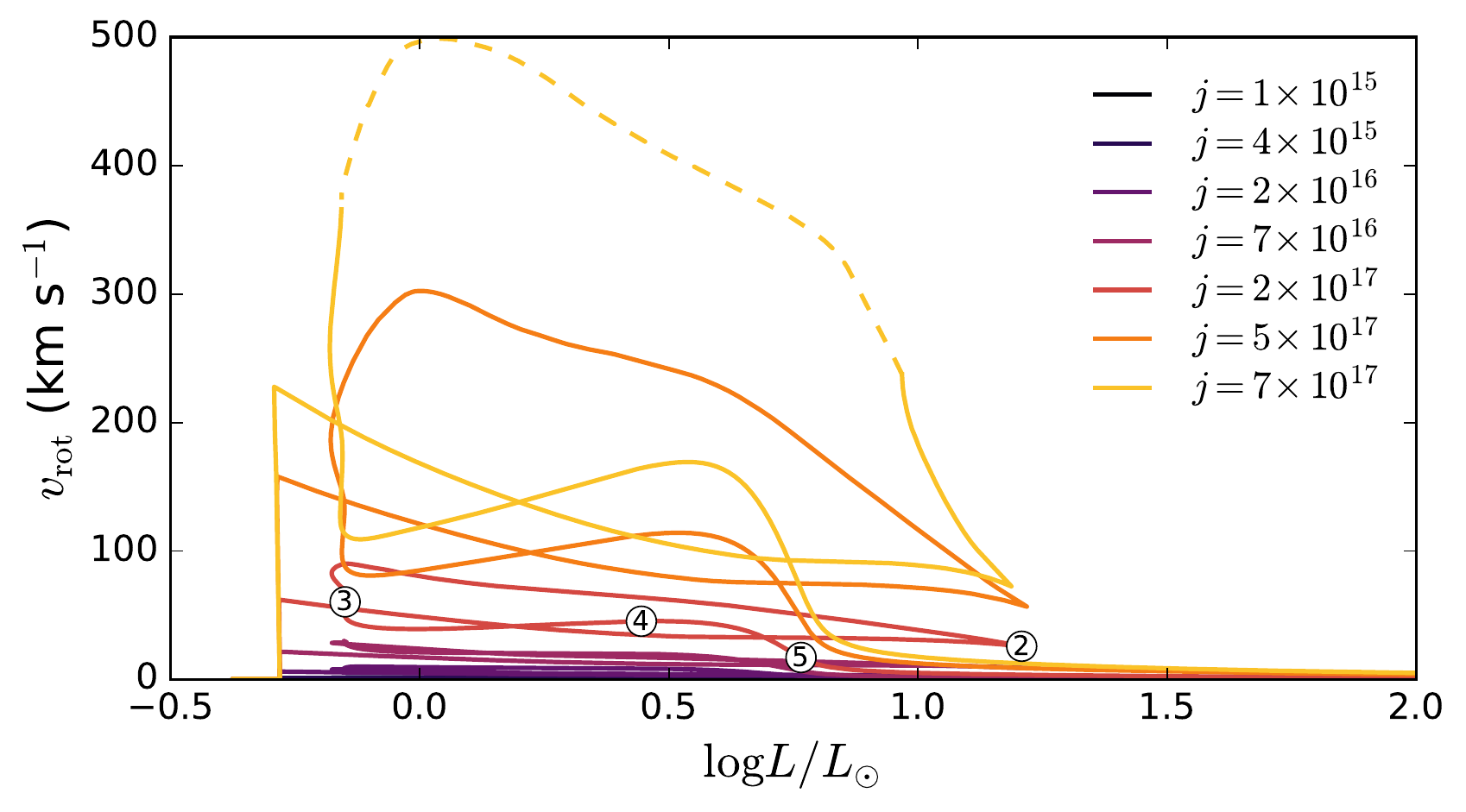}

\label{fig:mp125ms0750dm0050r_5-vs-84}}

\caption{Evolution of a $M_{2,\text{i}}=0.75\ M_{\odot}$ secondary accreting
$\Delta M=0.05\ M_{\odot}$ of material from a primary with initial
mass $M_{1}=1.25\ M_{\odot}$. Tracks distinguish different specific
angular momentum of the accreted material. The numbers on top of the
$j_{\text{a}}=2\times10^{17}\ \text{cm}^{2}\thinspace\text{s}^{-1}$
track mark the beginning of mass transfer (`1'), end of mass transfer
(`2'), return to the main sequence (`3'), main sequence turn-off (`4'),
beginning of first dredge-up (`5'), end of first dredge-up (`6').
The dashed part of the $j_{\text{a}}=7\times10^{17}\ \text{cm}^{2}\thinspace\text{s}^{-1}$
track marks the phase of the evolution where the star is formally
above critical rotation (see text).\label{fig:mp125ms0750dm0050r}}
\end{figure}

Figure\ \ref{fig:mp125ms0750dm0050r} shows the Hertzsprung-Russel
diagram (HRD) and the evolution of the rotation velocity of the secondary
of this system when the accreted material is assigned different values
of specific angular momentum. In all systems with $M_{1}=1.25\ M_{\odot}$
the mass transfer is assumed to start at $t=3.06$\ Gyr. This point
is identified by the circle numbered `1' in the figure. The $M_{2,\text{i}}=0.75\ M_{\odot}$
secondary at this age is still early on in its main sequence with
a central hydrogen mass fraction of $X_{\text{H},\text{c}}=0.59$
(down from the ZAMS value of 0.758). Once mass transfer starts, the
tracks corresponding to different values of $j_{\text{a}}$ separate.

The accretion rate of $\dot{M}=10^{-6}\ M_{\odot}\text{yr}^{-1}$
is high enough that the accretion timescale $\tau\simeq M_{2}/\dot{M}$
is always much shorter (by a factor of ten or more) than the thermal
timescale of any CEMP star progenitor. Accreting material (part of
the track between `1' and `2' in Fig.\ \ref{fig:mp125ms0750dm0050r_hrd})
at this rate therefore drives the star out of thermal equilibrium
to higher luminosity and effective temperature. Once accretion ends
(`2'), the star attempts to return to equilibrium, becoming fainter
and cooler in the process.

While returning to thermal equilibrium, the stars spin up for a time,
both in absolute terms and as a fraction of the critical velocity
(Figs.\ \ref{fig:mp125ms0750dm0050r_5-vs-89},\subref*{fig:mp125ms0750dm0050r_5-vs-84}).
For the highest values of specific angular momentum the stars may
reach critical rotation at some point, as in the $j_{\text{a}}=7\times10^{17}\ \text{cm}^{2}\thinspace\text{s}^{-1}$
case in this system (yellow line). Since our primary interest is the
long-term evolution after relaxation, we do not attempt to model this
brief ($\delta t\simeq\tau_{\text{KH}}$) phase accurately, assuming
that it is not very important for the subsequent evolution.\footnote{Presumably the star must shed the super-critical layers as it contracts.
Unless the material is re-accreted later, the star then ends up with
a slightly lower mass. In this particular case, losing about $0.004\ M_{\odot}$
of material during the contraction suffices to keep the star below
critical rotation. This amount of mass loss has only a small effect
on the subsequent evolution.} Instead, we simply limit the centrifugal acceleration, and the resulting
structural deformation, while the star is formally rotating at super-critical
velocities (dashed portion of the line). After relaxation, this particular
model never exceeds $\Omega/\Omega_{\text{c}}\simeq0.8$.

\begin{figure*}
\subfloat[Evolution of carbon]{\includegraphics[width=1\columnwidth]{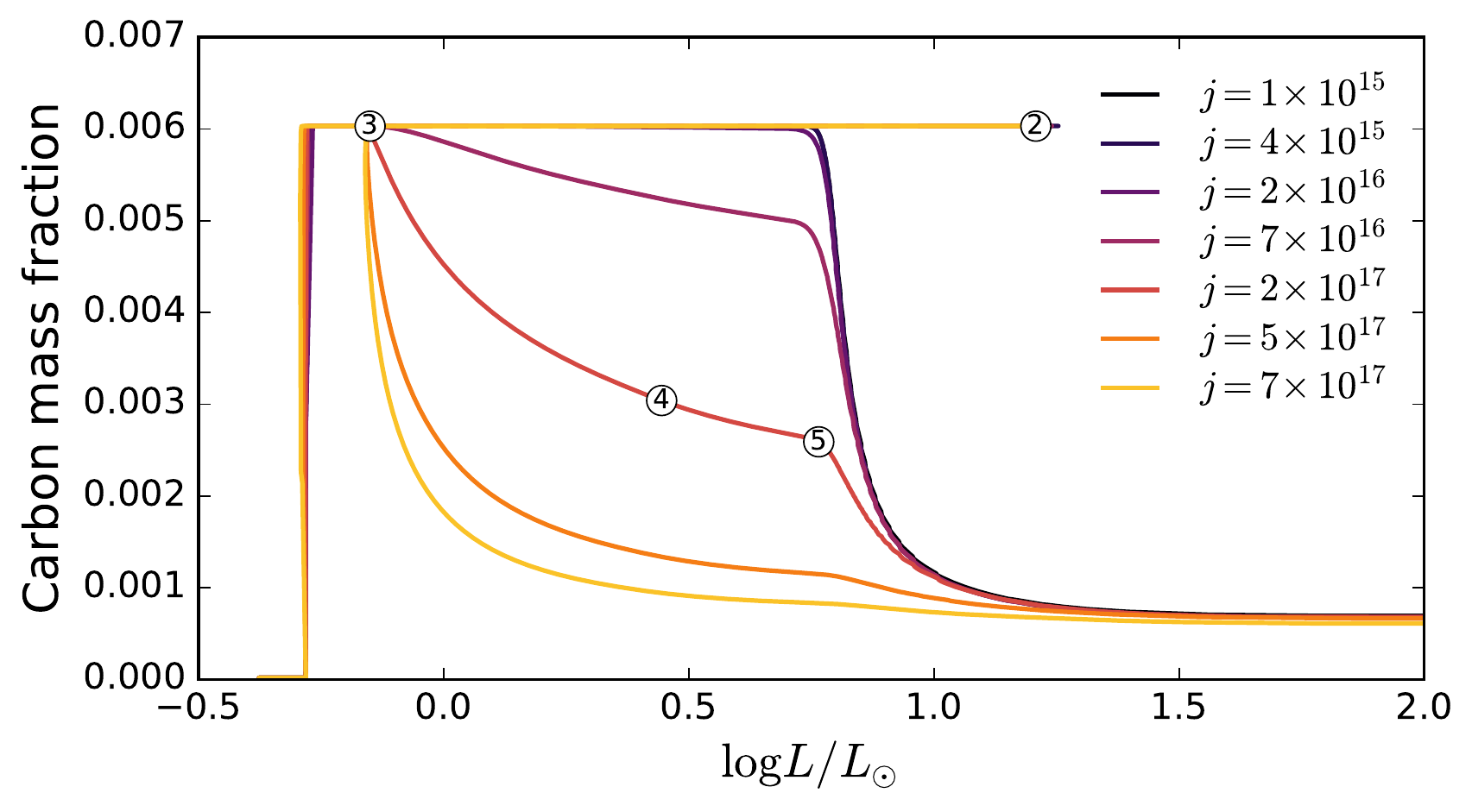}

\label{fig:mp125ms0750dm0050r_5-vs-30}}\hspace{\columnsep}\subfloat[Evolution of nitrogen]{\includegraphics[width=1\columnwidth]{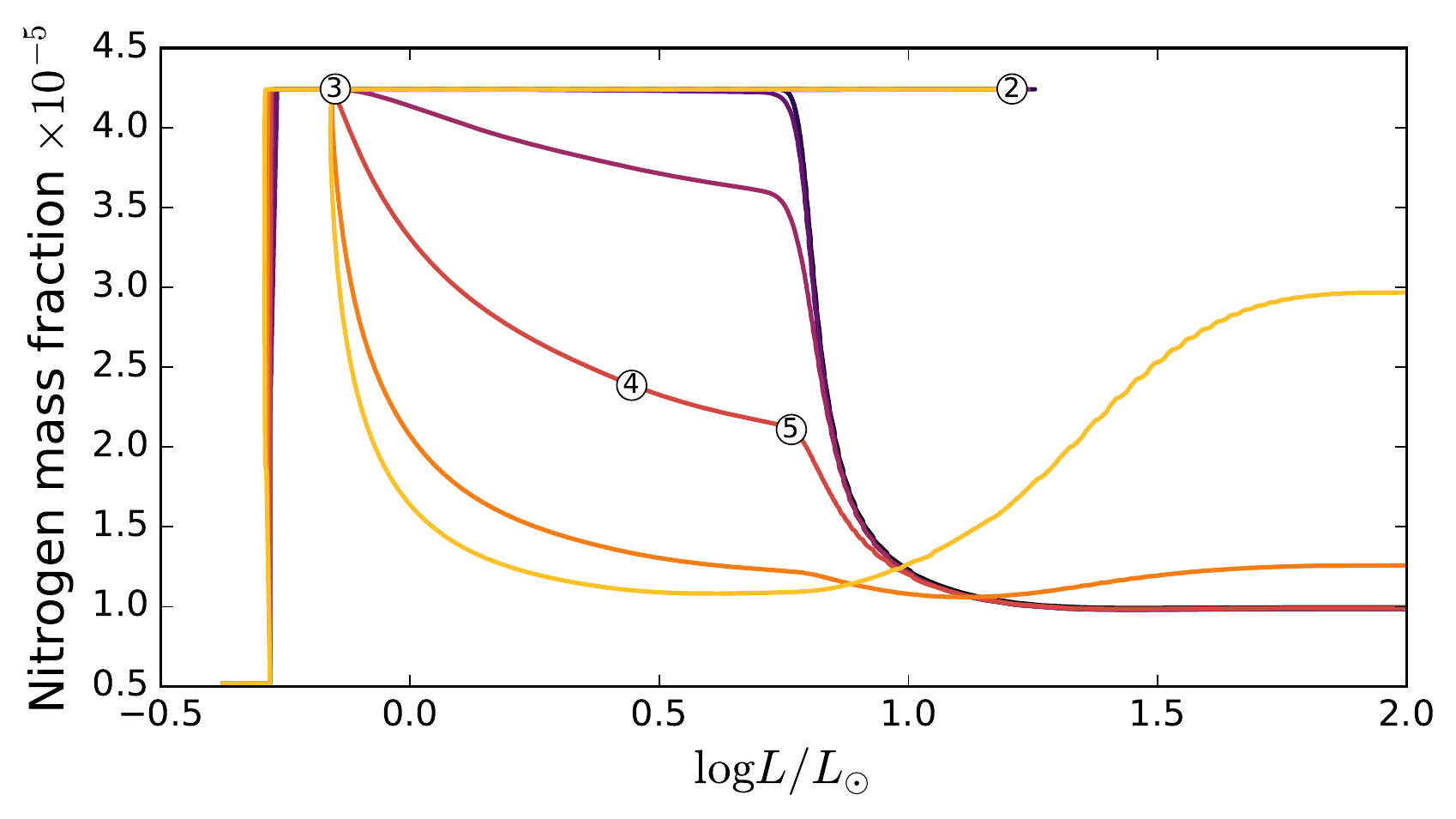}

\label{fig:mp125ms0750dm0050r_5-vs-31}}

\caption{As Fig.\ \ref{fig:mp125ms0750dm0050r} but showing the evolution
of the carbon and nitrogen surface mass fractions.\label{fig:mp125ms0750dm0050r_X-vs-logl}}
\end{figure*}

Various aspects of the evolution of the star after it settles back
on the main sequence (`3') depend on the angular momentum accreted.
First, owing to the extra support against gravity from the centrifugal
force, rotating stars are normally cooler and less luminous than non-rotating
stars – they resemble non-rotating stars of lower mass \citep{1970A&A.....8...76S}.
Here this effect is largely compensated for by the unusual chemical
structure. The non-rotating model is fairly cool ($T_{\text{eff}}\simeq6550$\ K
at turn-off, same as the $j_{\text{a}}=10^{15}\ \text{cm}^{2}\thinspace\text{s}^{-1}$
track and about 200\ K less than a regular $0.8\ M_{\odot}$ star
at $Z=10^{-4}$) because of the high metallicity of the accreted material.
Making the star rotate leads to dilution of this material by rotational
mixing (Fig.\ \ref{fig:mp125ms0750dm0050r_X-vs-logl}), and the corresponding
change in the opacity actually makes the star hotter (although not
more luminous) at turn-off (`4') than in the non-rotating case.\footnote{To elaborate, models with rotation are indeed slightly cooler and
less luminous early on in the post-mass-transfer main sequence. But
by the time they reach turn-off, most of them are hotter as a result
of the mixing.} Only for very rapid rotation rates ($\Omega/\Omega_{\text{c}}\gtrsim0.5$
or $v_{\text{rot}}>100\ \text{km}\thinspace\text{s}^{-1}$), when
rotational mixing does not lead to significant further changes in
the structure of the star, do the mechanical effects shift the track
back to cooler temperatures. The resulting spread in turn-off temperatures
between all models of this system is only about 100\ K.

Second, the internal transport of angular momentum leads to different
initial rotational evolution for rapid rotators. Normally, the surface
value of $\Omega/\Omega_{\text{c}}$ somewhat increases during the
main sequence evolution of low-mass stars. This is a consequence of
their slight expansion, which reduces the critical rotation velocity
($\Omega_{\text{c}}\propto R^{-3/2}$). Here, in the rapid rotators,
$\Omega/\Omega_{\text{c}}$ and $v_{\text{rot}}$ first decrease because
the accreted momentum is transported inwards. After reaching a minimum,
$\Omega/\Omega_{\text{c}}$ then increases for the rest of the main
sequence. All models reach a maximum in the rotation rate somewhere
between the turn-off (`4') and the beginning of first dredge-up (FDU;
`5').

\begin{figure}
\includegraphics[width=1\columnwidth]{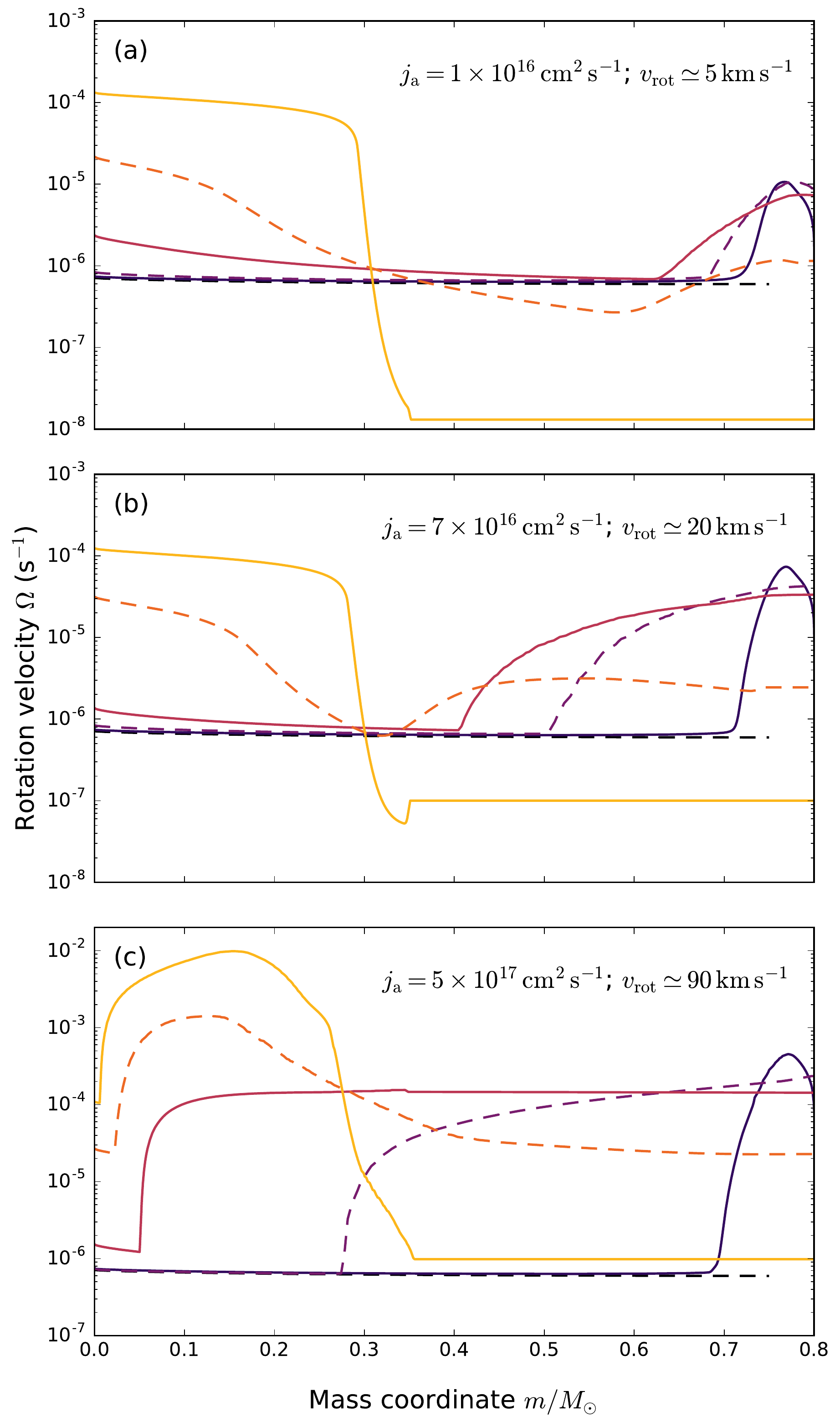}

\caption{Evolution of angular rotation velocity profiles for three different
values of $j_{\text{a}}$. The dashed black profile corresponds to
just before mass transfer (`1') and subsequent profiles, plotted in
alternating line types of progressively lighter colours, correspond
to end of mass transfer (`2'), return to thermal equilibrium (`3'),
main sequence (between `3' and `4'), before first dredge-up (`5')
and after first dredge-up (`6'). The last two profiles demonstrate
the spin-up of the core after the main sequence. Note the different
y scales.\label{fig:mp125ms0750dm0050r_omega-vs-m_evo}}
\end{figure}

\begin{figure*}
\subfloat[Specific angular momentum]{\includegraphics[width=1\columnwidth]{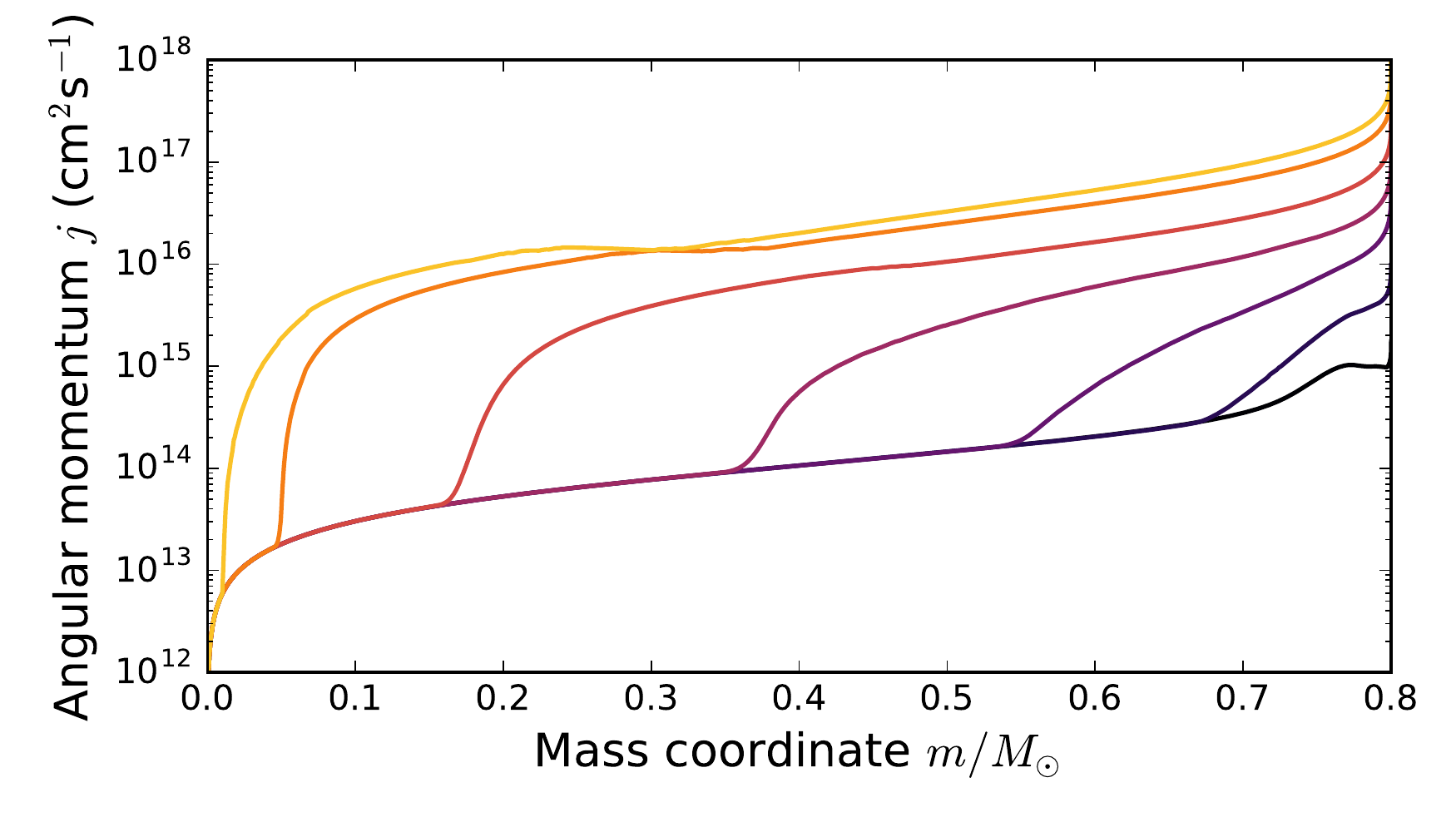}

\label{fig:mp125ms0750dm0050r_j-vs-m_msto}}\hspace{\columnsep}\subfloat[Angular rotation velocity]{\includegraphics[width=1\columnwidth]{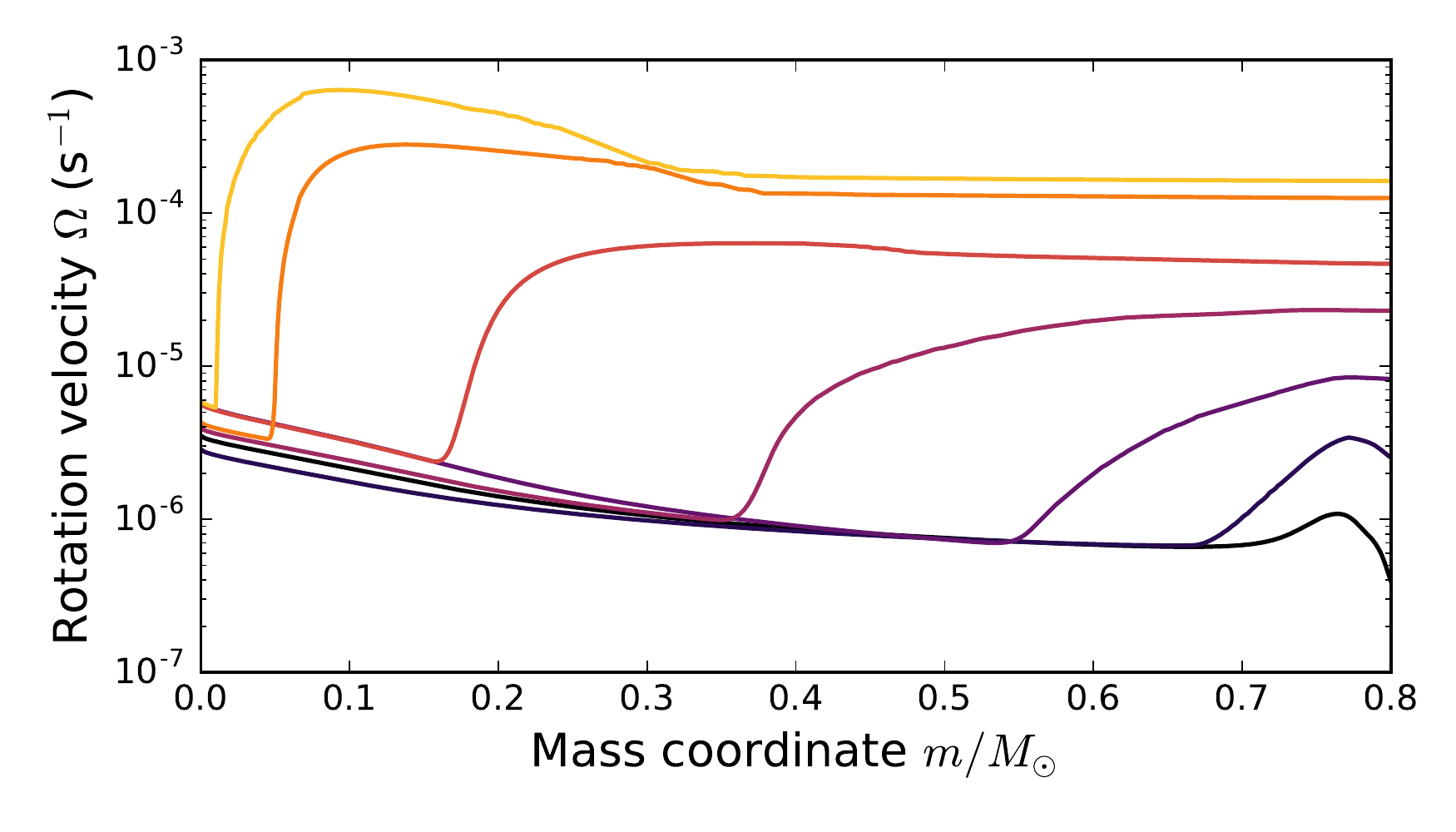}

\label{fig:mp125ms0750dm0050r_omega-vs-m_msto}}

\subfloat[Carbon abundances]{\includegraphics[width=1\columnwidth]{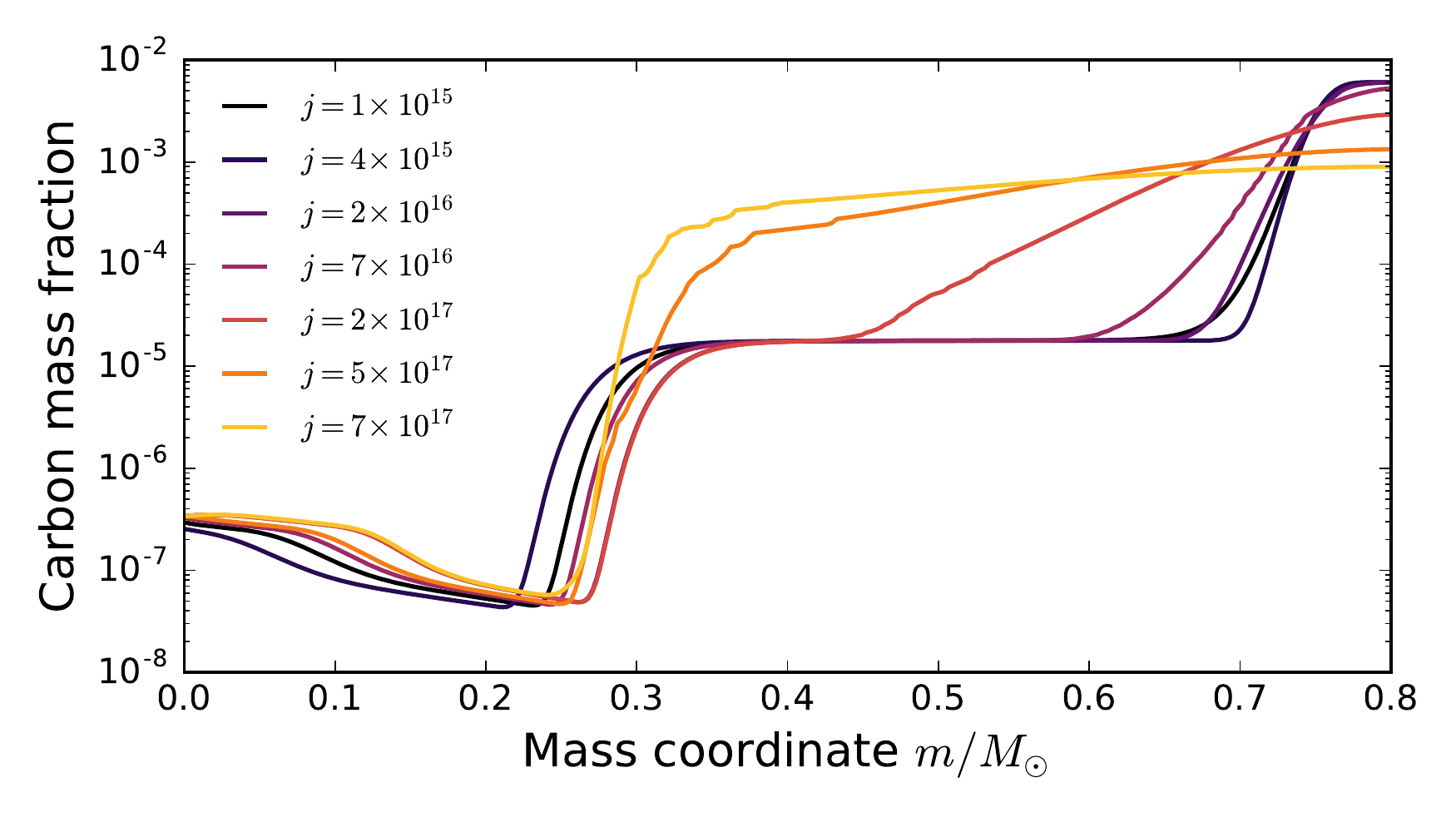}

\label{fig:mp125ms0750dm0050r_XC-vs-m_msto}}\hspace{\columnsep}\subfloat[Nitrogen abundances]{\includegraphics[width=1\columnwidth]{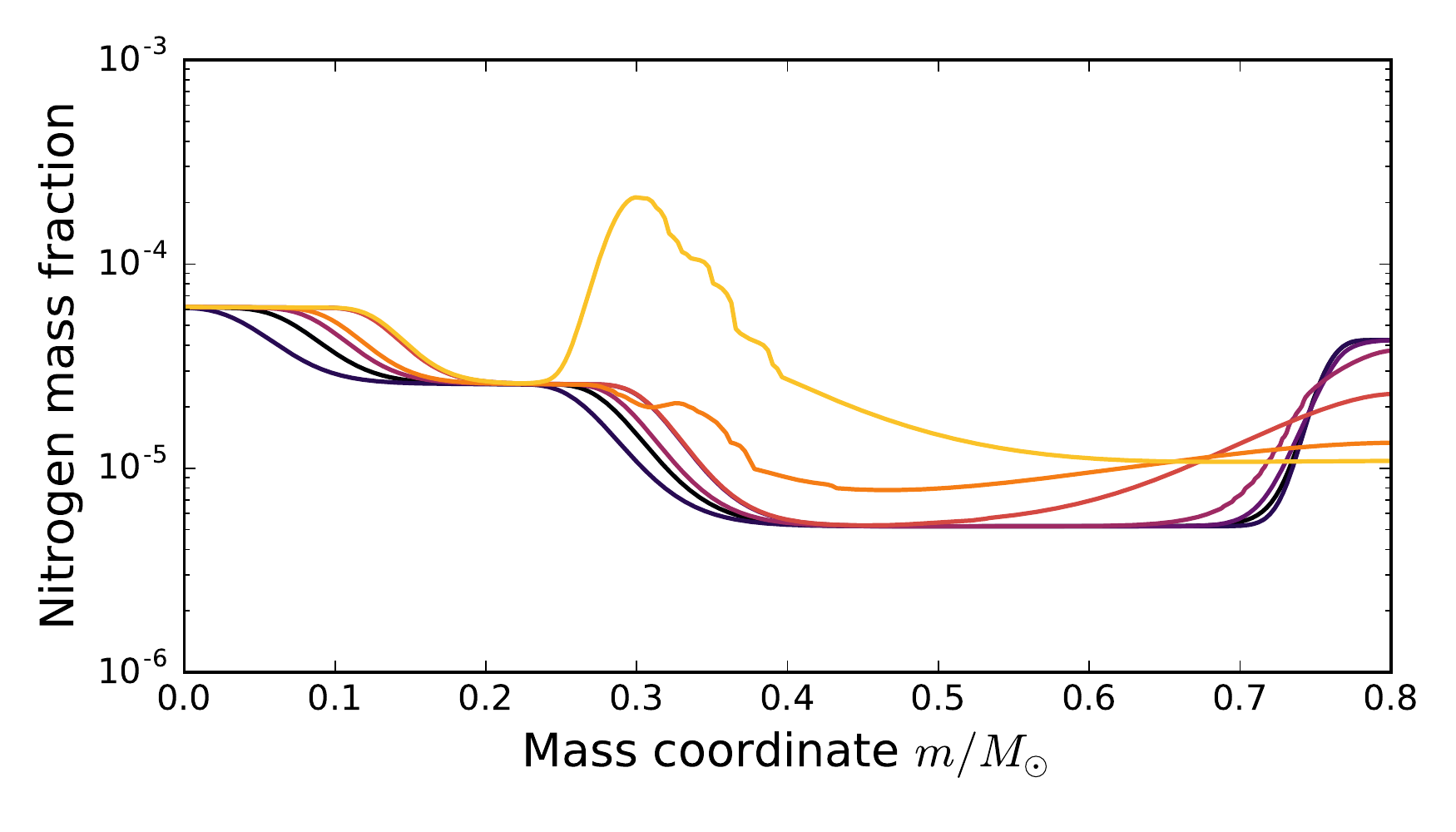}

\label{fig:mp125ms0750dm0050r_XN-vs-m_msto}}

\caption{Distribution of various quantities near the end of the main sequence
in the rotating models shown in Fig.\ \ref{fig:mp125ms0750dm0050r}.\label{fig:mp125ms0750dm0050r_profiles-vs-m_msto}}
\end{figure*}

Figure\ \ref{fig:mp125ms0750dm0050r_omega-vs-m_evo} shows the internal
evolution of the rotational velocity profile in the star between the
beginning of mass transfer and prior to FDU, and Fig.\ \ref{fig:mp125ms0750dm0050r_profiles-vs-m_msto}
shows the internal profiles of the specific angular momentum, angular
rotation velocity, and carbon and nitrogen abundances near the main
sequence turn-off for the different values of $j_{\text{a}}$. Naturally,
accreting more angular momentum leads to deeper and faster transport
of angular momentum. By the time the star reaches the turn-off in
the $j_{\text{a}}=7\times10^{16}\ \text{cm}^{2}\thinspace\text{s}^{-1}$
($v_{\text{rot}}\simeq20\ \text{km}\thinspace\text{s}^{-1}$) case,
the angular momentum is mixed to about the half-way point in the star
by mass. For the highest values of $j_{\text{a}}$ ($v_{\text{rot}}\gtrsim100\ \text{km}\thinspace\text{s}^{-1}$)
nearly all of the star is mixed by then.

The mixing of angular momentum and of chemical elements does not occur
to the same depth (cf. Figs.\ \ref{fig:mp125ms0750dm0050r_j-vs-m_msto}
and \ref{fig:mp125ms0750dm0050r_XC-vs-m_msto}). For example, in the
lowest angular momentum case that shows a change in the surface abundances
on the main sequence, $j_{\text{a}}=7\times10^{16}\ \text{cm}^{2}\thinspace\text{s}^{-1}$
(Fig.\ \ref{fig:mp125ms0750dm0050r_X-vs-logl}), the chemical elements
have been partially mixed down to a mass coordinate of $m\simeq0.6\ M_{\odot}$
by the time the star has reached the turn-off. Meanwhile, the angular
momentum has been transported about twice as deep. This is a direct
consequence of the choice of $f_{c}$, the fraction of total angular
momentum diffusion coefficient applied to chemical transport, in Eq.\ \eqref{eq:Dmix}.
Had we chosen $f_{c}=1$, the depth of chemical and angular momentum
transport would coincide. Instead, the timescale for chemical transport
is much longer than that for angular momentum transport, and thus
angular momentum has been transported to a greater depth at a given
time. We return to the influence of $f_{c}$ and $f_{\mu}$ in Sect.\ \ref{subsec:fcfmu}.

\begin{figure*}
\includegraphics[width=1\textwidth]{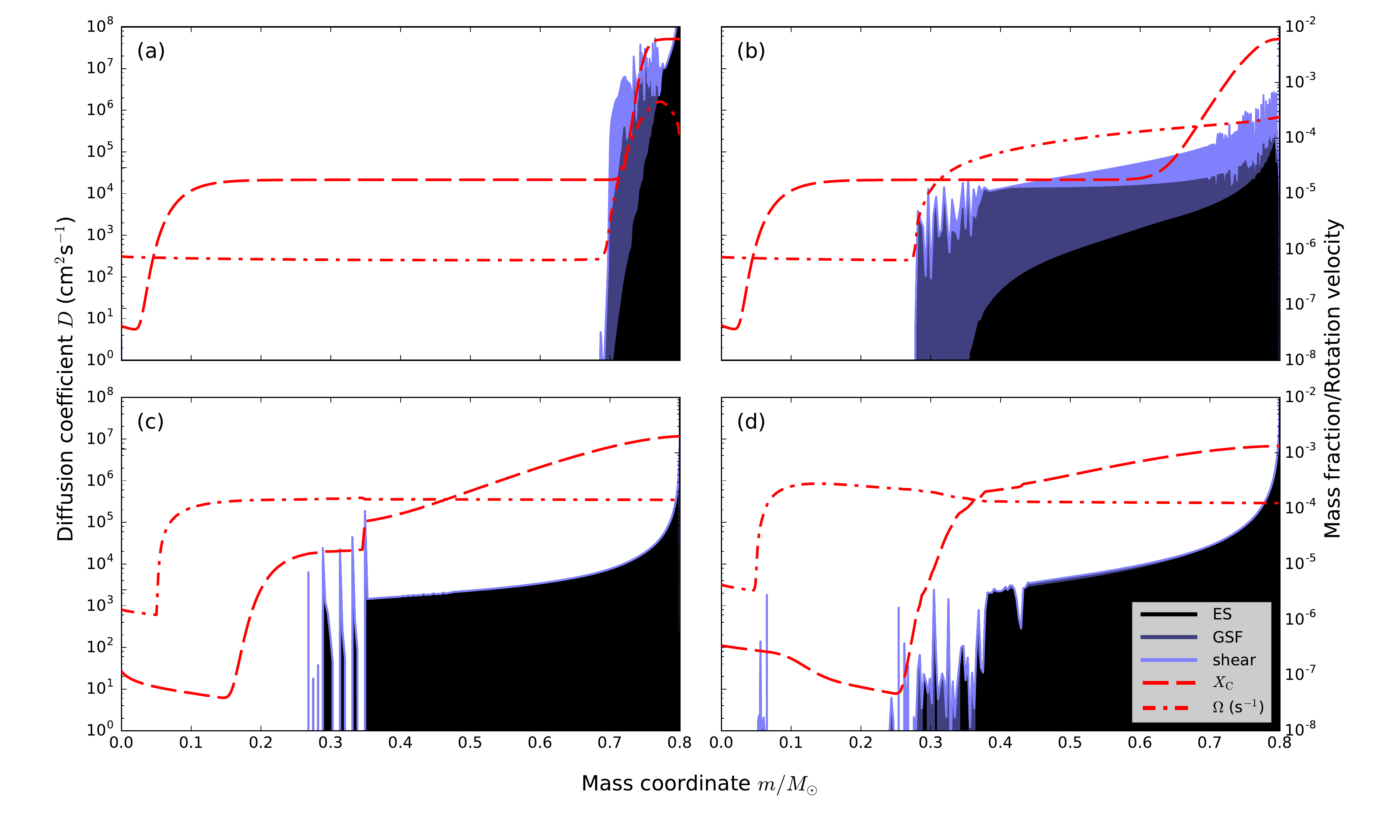}

\caption{Individual contributions of the most important rotational instabilities
to the total angular momentum transport coefficient in the $j_{\text{a}}=5\times10^{17}\ \text{cm}^{2}\thinspace\text{s}^{-1}$
($v_{\text{rot}}\simeq100\ \text{km}\thinspace\text{s}^{-1}$) case
at four instances: a) shortly after mass transfer ends (`2'); b) near
return to equilibrium (`3'); c) main sequence (between `3' and `4');
d) main sequence turn-off (`4'). Initially, large $\Omega$-gradients
favour the shear and GSF instabilities. Later on, when the $\Omega$-gradients
are erased, the overall transport is dominated by the Eddington-Sweet
circulation. The lines show the angular velocity (dash-dotted) and
the carbon abundance (dashed) profiles at the corresponding times.\label{fig:mp125ms0750dm0050j5e17r_Dmix-vs-m_evo_stack}}
\end{figure*}

Which of the different processes dominates the transport of angular
momentum? The answer changes over time (Fig.\ \ref{fig:mp125ms0750dm0050j5e17r_Dmix-vs-m_evo_stack}).
At first, a steep $\Omega$-gradient is present at the interface between
the accreted layer and the original surface of the star. This induces
shear mixing, which dominates the initial transport with some contribution
from the GSF instability. But the transport also smears out the $\Omega$-gradient,
quenching the shear instability. For a while, the GSF instability
is responsible for the continuing inward transport of angular momentum,
until eventually much of the $\Omega$-gradient is removed, and mixing
proceeds over longer timescales by the Eddington-Sweet circulation
(which is the only term in Eq.\ \eqref{eq:Drot} that depends on
$\Omega$ and not its gradient). Some $\Omega$-gradients always remain
(e.g. Fig.\ \ref{fig:mp125ms0750dm0050r_omega-vs-m_evo}), but these
are either too small to contribute to further mixing and/or the mixing
is inhibited by molecular weight gradients. Given the rapid removal
of the $\Omega$-gradients, Eddington-Sweet circulation is responsible
for most of the chemical mixing.

Once first dredge-up starts (`5'), the tracks again converge, and
the evolution up the red giant branch (RGB) is similar in all cases.
This is largely the result of FDU erasing most of the differences
in the chemical structure between the model sequences. Furthermore,
the accreted angular momentum is insufficient to result in large rotation
rates (with respect to the critical rate) of any part of a giant because
of its much larger moment of inertia.

Overall, below $j_{\text{a}}\lesssim2\times10^{16}\ \text{cm}^{2}\thinspace\text{s}^{-1}$
(turn-off velocity of about $10\ \text{km}\thinspace\text{s}^{-1}$)
the evolution of the CEMP star in this system is essentially unaffected
by the rotation. This is because both the centrifugal acceleration
is too small to substantially lower the effective gravity, and the
contrast between the accreted layer and the region below is too small
to trigger significant chemical mixing of the two (although there
is some angular momentum mixing; Fig.\ \ref{fig:mp125ms0750dm0050r_j-vs-m_msto}).
For higher values of $j_{\text{a}}$ the timescale for chemical transport
is finally short enough, compared to the nuclear timescale, that chemical
dilution of the accreted material can occur before FDU (Fig.\ \ref{fig:mp125ms0750dm0050r_X-vs-logl}),
which, as explained above, also shifts the track in the HRD. Naturally,
the greater $j_{\text{a}}$, the more extensive and rapid the mixing.
For the highest values of $j_{\text{a}}$ mixing is deep enough that
FDU no longer plays a significant role in diluting the accreted material.
However, if rotational mixing on the main sequence is extensive enough
to bring the accreted carbon down to regions where it can be burnt,
FDU can bring the produced nitrogen, visible in Fig.\ \ref{fig:mp125ms0750dm0050r_XN-vs-m_msto}
around $m\simeq0.3\ M_{\odot}$, to the surface. In the two most rapidly
rotating models, the nitrogen abundance after FDU (`6') is thus higher
than in all others (Fig.\ \ref{fig:mp125ms0750dm0050r_5-vs-31}).

The evolution further up the giant branch is not very eventful in
these models. No further abundance changes occur once FDU is over.
The contracting core spins up and the expanding envelope slows down
– the surface velocities decrease to $v_{\text{rot}}\lesssim10\ \text{km}\thinspace\text{s}^{-1}$.
There thus develops a large and ever-increasing contrast between the
rotation rate of the core and the envelope (Fig.\ \ref{fig:mp125ms0750dm0050r_omega-vs-m_evo}).
This is not consistent with asteroseismic measurements of red giant
core rotation rates \citep{2012A&A...548A..10M,2014A&A...564A..27D},
which find much greater coupling between the core and the envelope.
This coupling is thought to come about as a result of magnetic fields
\citep{2002A&A...381..923S,2008A&A...481L..87S} and/or gravity waves
\citep{2003A&A...405.1025T,2008A&A...482..597T,2014ApJ...796...17F},
neither of which we have modelled at this time. In terms of surface
abundances, these processes seem more likely to manifest by altering
the importance of rotational mixing than leading to mixing directly
\citep{2005A&A...440..981T,2005A&A...440.1041M,2010A&A...519A.116E},
an effect that to some extent we probe by considering different rotation
rates.

\subsection{Models with atomic diffusion and thermohaline mixing\label{subsec:illust-mod-seq-rdt}}

Atomic diffusion will tend to modify the surface abundances following
mass transfer. In \citet{2016A&A...592A..29M} we showed how, in absence
of other mixing processes, in most CEMP-\emph{s} stars the carbon
should settle out of the surface convection zone, while the surface
abundance of iron should increase as a result of radiative levitation.
Near the main sequence turn-off, before the convective envelope begins
to move inwards in mass, the resulting abundances (e.g. $\text{[C/H]}<-1$
and $\text{[Fe/H]}>-1$ so that $\text{[C/Fe]}<0$) can be very different
from those of the accreted material ($\text{[C/H]}\simeq0$ and $\text{[Fe/H]}\simeq-2$
so that $\text{[C/Fe]}\gtrsim2$).

\begin{figure}
\subfloat[Evolution of carbon]{\includegraphics[width=1\columnwidth]{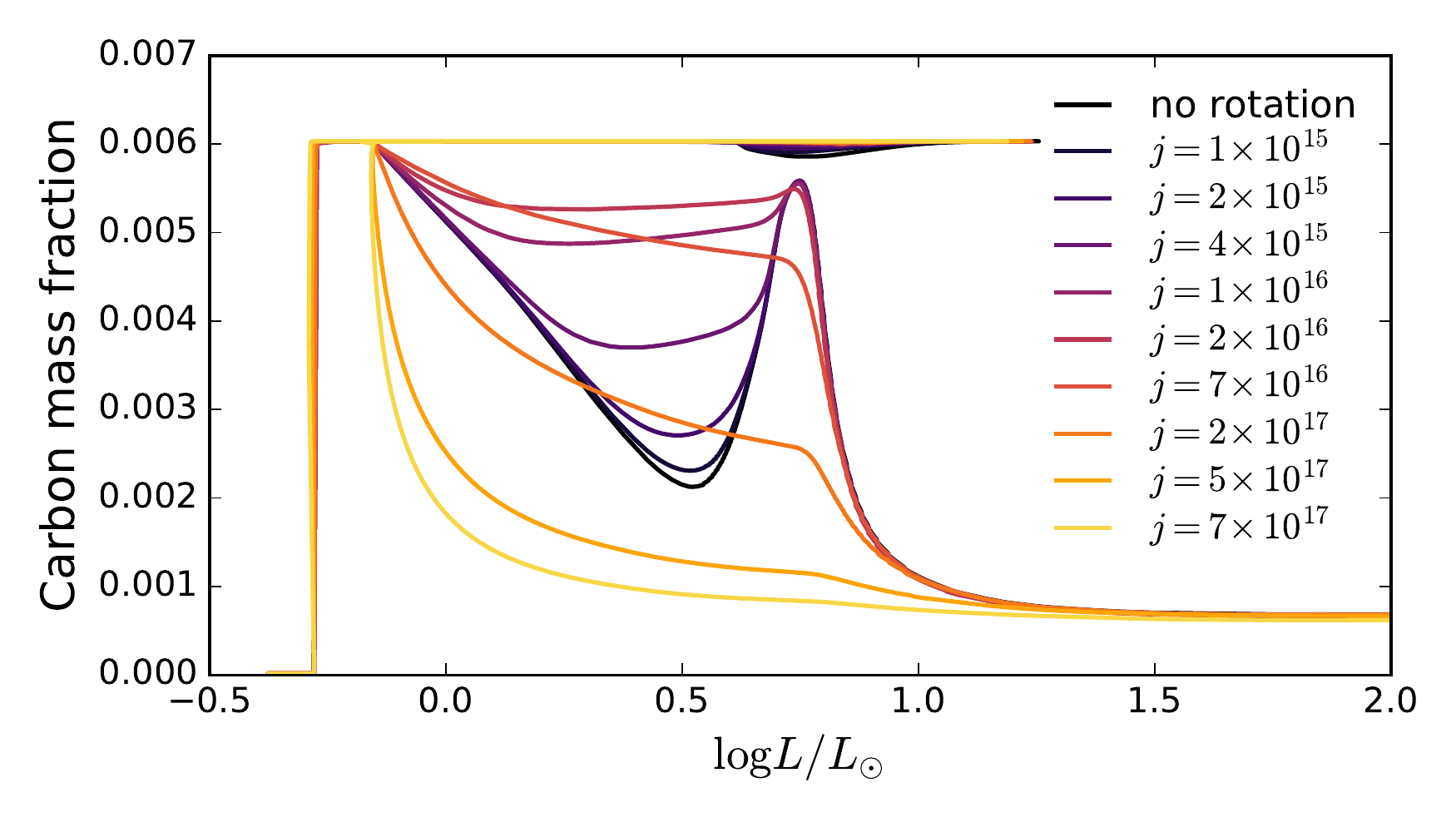}

\label{fig:mp125ms0750dm0050rd_5-vs-30}}

\subfloat[Hertzsprung-Russell diagram]{\includegraphics[width=1\columnwidth]{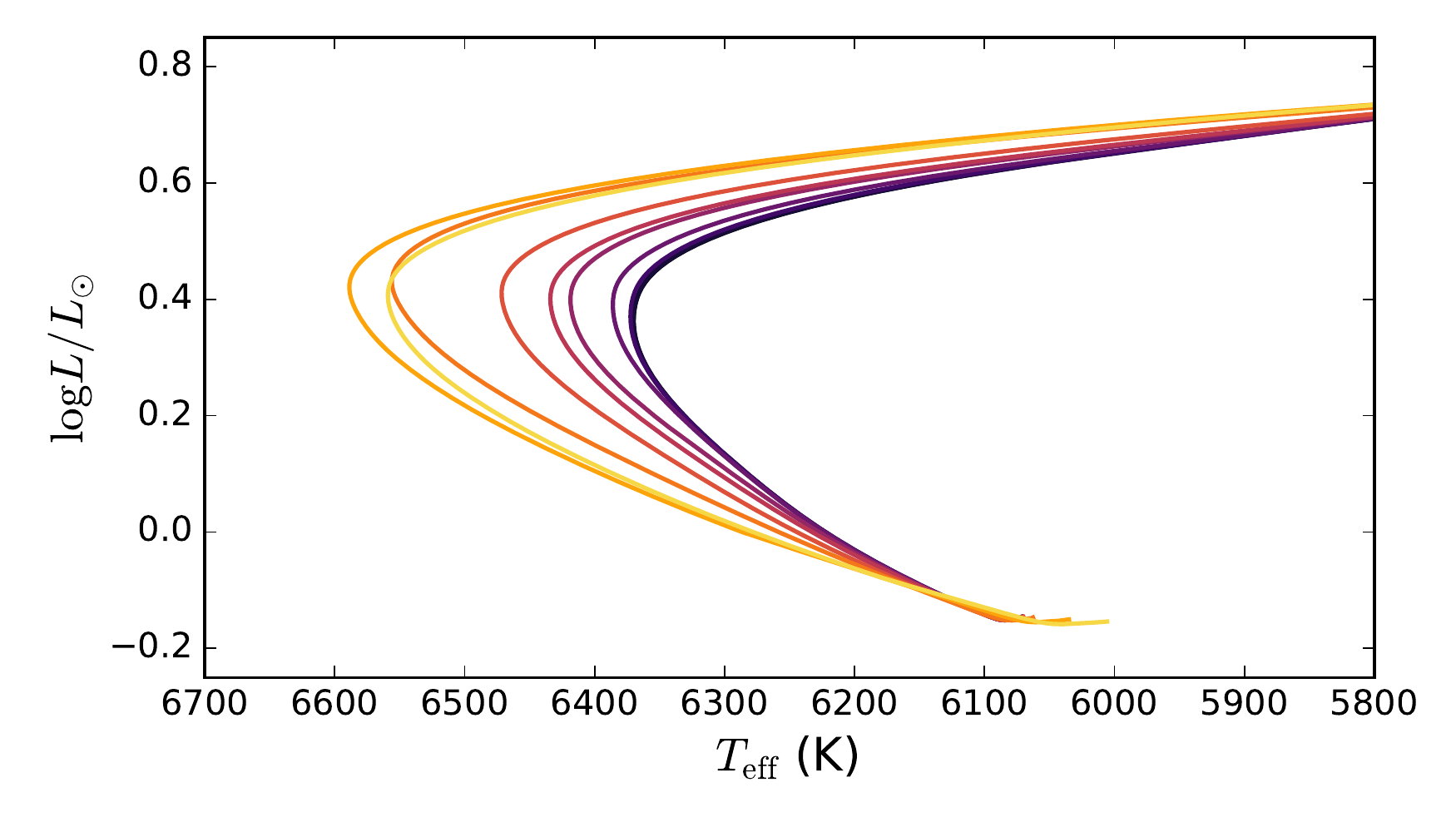}

\label{fig:mp125ms0750dm0050rd_hrd}}

\caption{Evolution and abundances of a $M_{2,\text{i}}=0.75\ M_{\odot}$ secondary
accreting $\Delta M=0.05\ M_{\odot}$ of
material from a $M_{1}=1.25\ M_{\odot}$ primary for different values
of specific angular momentum of accreted material (with atomic diffusion).\label{fig:mp125ms0750dm0050rd}}
\end{figure}

However, atomic diffusion will be counteracted by rotational mixing.
As a result of this competition, abundance variations on the main
sequence are expected in models of all rotation rates (Fig.\ \ref{fig:mp125ms0750dm0050rd_5-vs-30}).
At the lowest rotation velocities atomic diffusion dominates, modifying
the surface abundances of metals until the convective envelope mass
begins to increase shortly after the turn-off. But even a model rotating
at less than a kilometer per second shows a slightly reduced effect
compared to the non-rotating case. As one considers higher rotation
rates, atomic diffusion near the surface is more and more inhibited
up to the $j_{\text{a}}=2\times10^{16}\ \text{cm}^{2}\thinspace\text{s}^{-1}$
case (in this system corresponding to a turn-off velocity of $v_{\text{rot}}\simeq9\ \text{km}\thinspace\text{s}^{-1}$),
where the abundance variations are smallest, and the surface abundances
remain within 15\% of the accreted composition. In non-diffusing models
this is the highest $j_{\text{a}}$ case in which there is practically
no change in the abundances prior to first dredge-up (Fig.\ \ref{fig:mp125ms0750dm0050r_X-vs-logl}).
As $j_{\text{a}}$ is increased still further, rotational mixing takes
over, and the models look more and more like in the purely rotating
case in terms of surface abundances.

The competition between diffusion and rotational mixing also has an
effect on the global properties of the star. It has been shown \citep[e.g.][]{2002ApJ...571..487V,2012MNRAS.427..127B}
that non-rotating models without atomic diffusion are longer-lived
and hotter throughout the main sequence evolution than models with
diffusion. This holds also for models of CEMP-\emph{s} stars. But,
because of partial inhibition of atomic diffusion, accreting even
a small amount of angular momentum makes a model more like a non-diffusing
one and thus hotter throughout the main sequence than a model with
diffusion only (Fig.\ \ref{fig:mp125ms0750dm0050rd_hrd}). When the
angular momentum accreted is sufficient to cause rotational mixing
directly ($j_{\text{a}}\gtrsim7\times10^{16}\ \text{cm}^{2}\thinspace\text{s}^{-1}$),
the models become still hotter. Since the surface abundance anomalies
are actually smaller than in the non-rotating case, this must be due
to inhibition of diffusion deeper in the star. Eventually, the mechanical
effects from rotation take over such that the models with highest
rotation rates are again cooler.

\begin{figure}
\subfloat[Evolution of carbon]{\includegraphics[width=1\columnwidth]{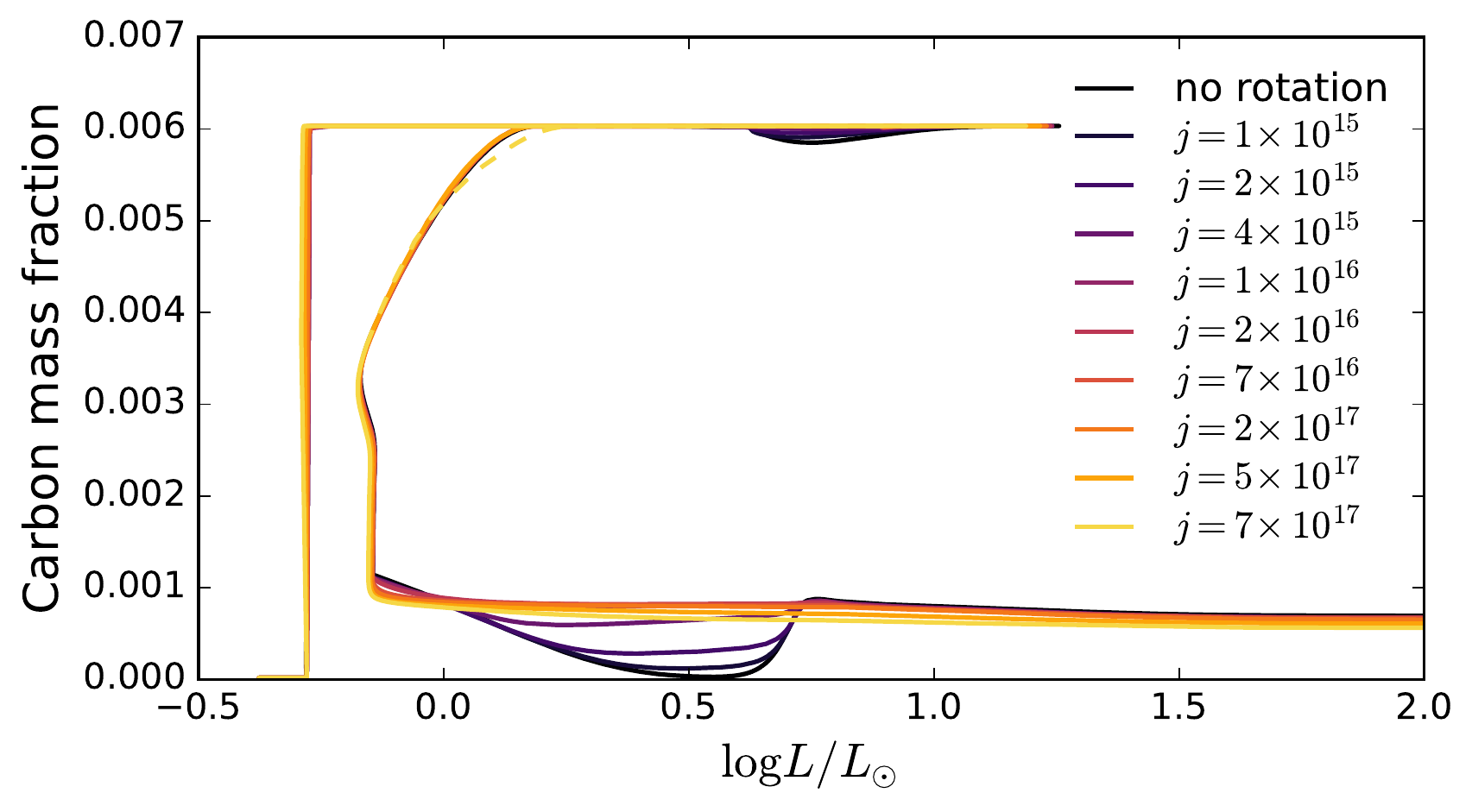}

\label{fig:mp125ms0750dm0050rdt_5-vs-30}}

\subfloat[Evolution of nitrogen]{\includegraphics[width=1\columnwidth]{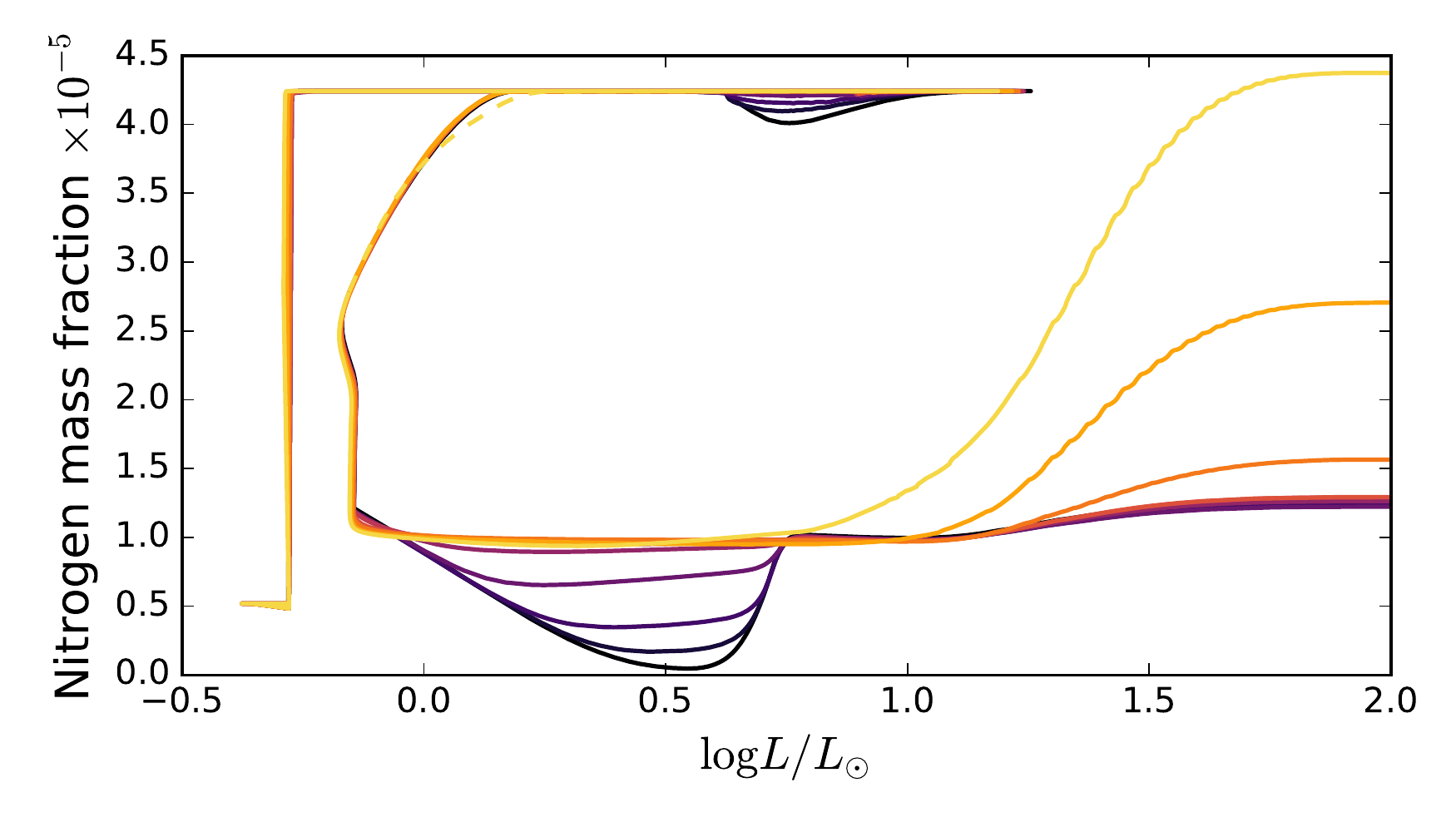}

\label{fig:mp125ms0750dm0050rdt_5-vs-31}}

\subfloat[Hertzsprung-Russell diagram]{\includegraphics[width=1\columnwidth]{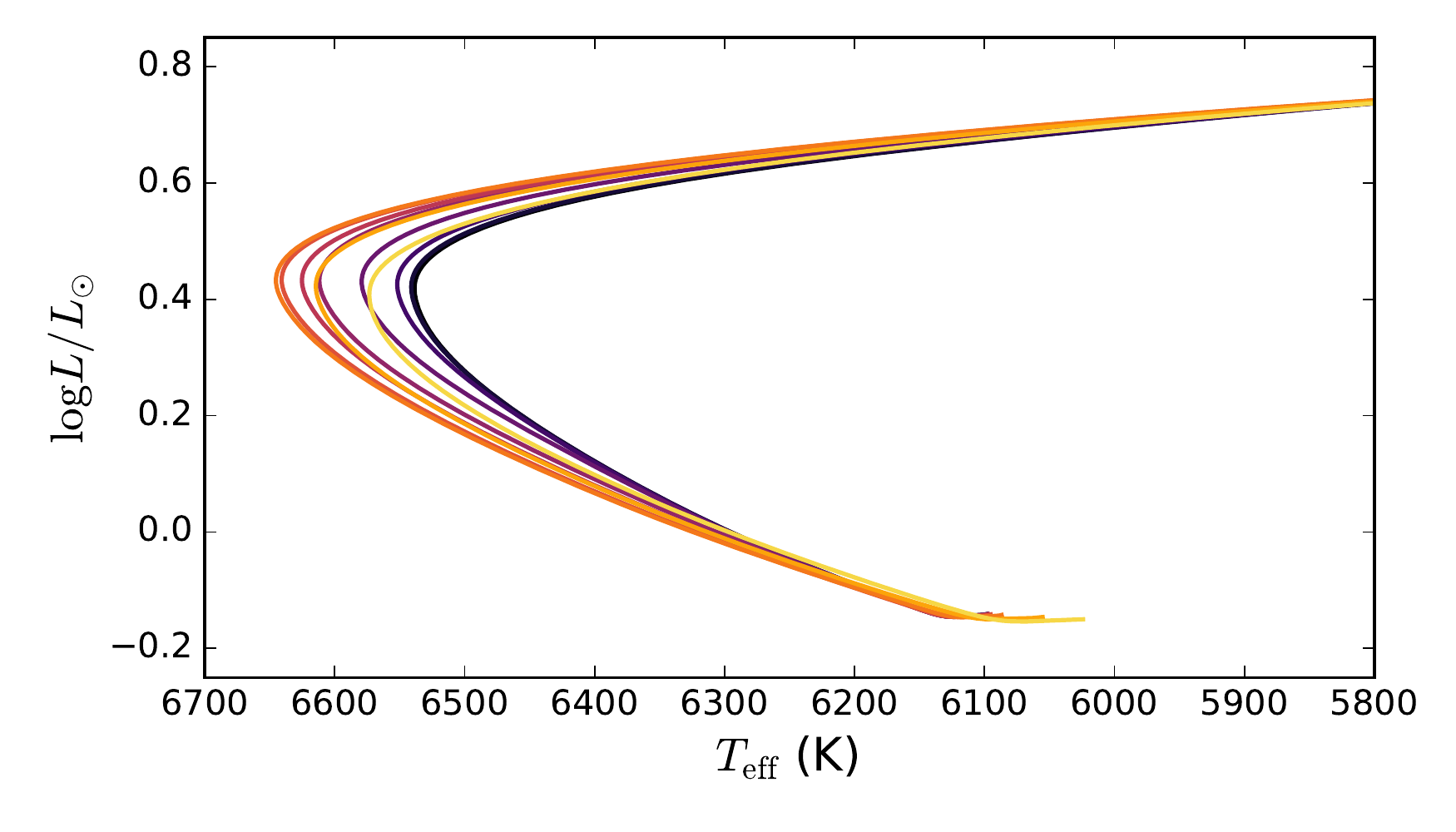}

\label{fig:mp125ms0750dm0050rdt_hrd}}

\caption{As Fig.\ \ref{fig:mp125ms0750dm0050rd} but with thermohaline mixing.\label{fig:mp125ms0750dm0050rdt}}
\end{figure}

Thermohaline mixing, when modelled as an independent process (see
Sect.\ \ref{subsec:observations}), always dominates over diffusion
and rotational mixing. In this system thermohaline mixing rapidly
reduces the carbon abundance by a factor of about six (between $\log L\simeq0.2$
and $-0.1$; Figs.\ \ref{fig:mp125ms0750dm0050rdt_5-vs-30},\subref*{fig:mp125ms0750dm0050rdt_5-vs-31}).
Once thermohaline mixing has leveled the inverse $\mu$-gradient,
diffusion modifies the surface abundances still further in the slowly
rotating models. In the rapidly rotating models the subsequent abundance
evolution depends on the depth of thermohaline mixing. When this depth
is at least comparable to that of rotational mixing (such as in this
system, where thermohaline mixing proceeds down to a mass coordinate
of $m\simeq0.39\ M_{\odot}$), the chief role of rotational mixing
is to inhibit atomic diffusion – it does not lead to significant further
abundance changes before FDU. In systems where thermohaline mixing
is not as deep (because of a smaller $\mu$-gradient), rotational
mixing can lead to further dilution of the accreted material.

A slight increase in the surface nitrogen abundance following FDU
is found in this system even without rotation (Fig.\ \ref{fig:mp125ms0750dm0050rdt_5-vs-31}).
This nitrogen has been produced from the accreted carbon transported
deep into the star by thermohaline mixing \citep{2007A&A...464L..57S}.
Rotational mixing replenishes the carbon at these depths after thermohaline
mixing has shut off so that more nitrogen can be produced, boosting
the amount of nitrogen that is brought to the surface during FDU.
Thermohaline mixing can once again activate following the RGB luminosity
bump because of the $^{3}\text{He}(^{3}\text{He},2\text{p})^{4}\text{He}$
reaction, which reduces the mean molecular weight just above the hydrogen
burning shell \citep{2006Sci...314.1580E,2009MNRAS.396.2313S}. A
further increase in surface nitrogen abundance follows, but normally
by no more than a factor of two because of the low value of the thermohaline
mixing coefficient \citep{2010ApJ...723..563D}.

The mechanical effects are visible earlier and more clearly in models
with thermohaline mixing, because most of the chemical mixing is caused
by the thermohaline instability. Since this makes rotational mixing
of chemical composition largely superfluous, only the mechanical effects
from rotation remain. The spread in temperature around the main sequence
turn-off is smaller than in models without thermohaline mixing (about
100\ K versus 220\ K, respectively; cf. Figs.\ \ref{fig:mp125ms0750dm0050rdt_hrd}
and \ref{fig:mp125ms0750dm0050rd_hrd}).

\subsection{Abundance anomalies near the turn-off\label{subsec:results-summary}}

We now attempt to characterize how the surface composition changes
(from the composition of the accreted material) in all of our simulations
collectively. We refer to these changes as abundance anomalies. In
models where atomic diffusion dominates, the abundance anomalies are
usually largest around the main sequence turn-off, which is thus a
convenient point of reference for comparing different systems \citep{2016A&A...592A..29M}.
In models without diffusion, or when diffusion is inhibited, the abundances
of most elements instead change monotonically throughout the main
sequence and beyond, as the accreted material gets more and more diluted
(Fig.\ \ref{fig:mp125ms0750dm0050r_X-vs-logl}; this is not true
for elements like nitrogen that can undergo further nuclear processing).
Since a similar point of reference in these models thus cannot be
identified, we adopt the same point, the main sequence turn-off, for
convenience.

\begin{figure}
\includegraphics[width=1\columnwidth]{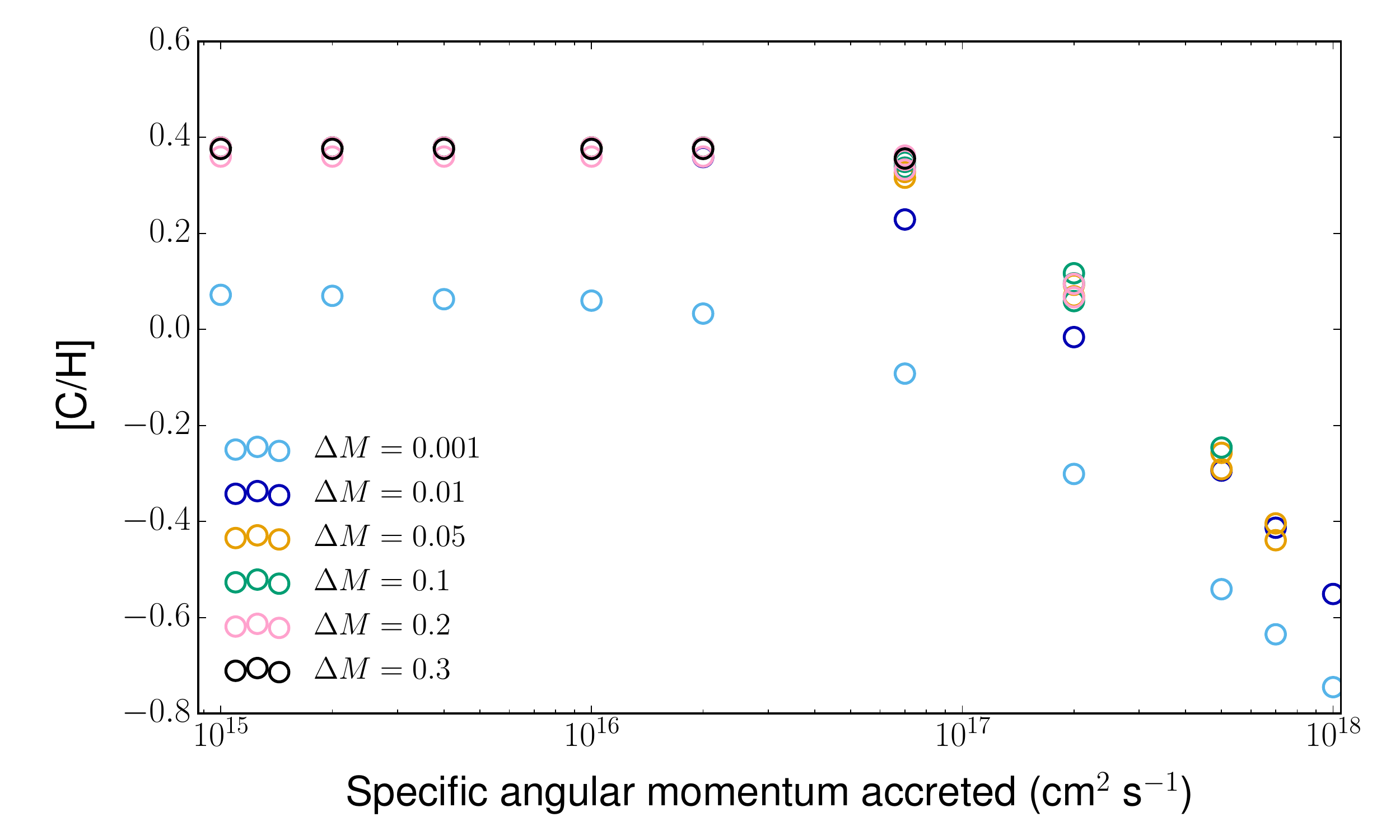}

\caption{Carbon abundance at main sequence turn-off as a function of specific
angular momentum of accreted material in models with rotational mixing
only. To avoid excessive crowding, only models with $M_{1}=1.25\ M_{\odot}$
are plotted, in which $\text{[C/H]}\simeq0.38$ after mass transfer
in all cases. The general result of little to no dilution for $j_{\text{a}}\lesssim2\times10^{16}\ \text{cm}^{2}\thinspace\text{s}^{-1}$
holds for systems with other primary masses.\label{fig:mp125r_ja-vs-CH_TO}}
\end{figure}

In the system discussed above, accretion of material with specific
angular momentum $j_{\text{a}}\lesssim2\times10^{16}\ \text{cm}^{2}\thinspace\text{s}^{-1}$
has little influence on the evolution following mass transfer, if
atomic diffusion is ignored. Figure\ \ref{fig:mp125r_ja-vs-CH_TO}
shows that this is the case in other systems (with different values
of $M_{2,\text{i}}$, $\Delta M$, and $M_{2,\text{f}}$) as well.
It is the specific angular momentum (instead of, e.g., the total angular
momentum accreted) that is decisive in determining whether material
will mix, because material with higher specific angular momentum establishes
a greater gradient in the angular velocity, which aids the shear instability.

\begin{figure}
\includegraphics[width=1\columnwidth]{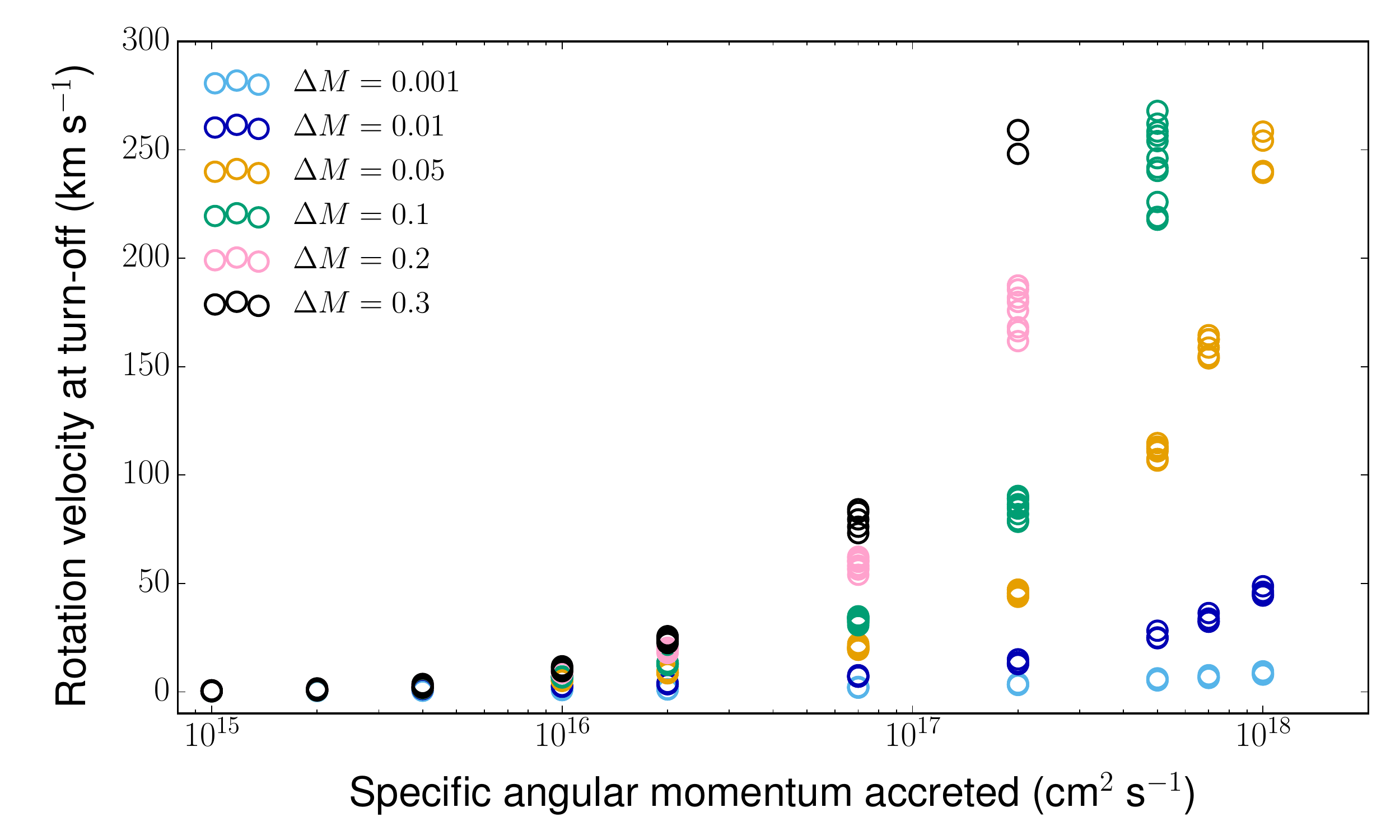}

\caption{Rotation velocities at main sequence turn-off (`4') resulting from
accretion of material of different specific angular momentum in all
simulations with rotational mixing only. The critical velocity at
turn-off is between about 280 and $300\ \text{km}\,\text{s}^{-1}$
in the vast majority of cases.\label{fig:ja-vs-vrotTO_r}}
\end{figure}

In the same system, accretion of material with specific angular momentum
$j_{\text{a}}=2\times10^{16}\ \text{cm}^{2}\thinspace\text{s}^{-1}$
results in a turn-off velocity of about $10\ \text{km}\thinspace\text{s}^{-1}$.
In systems with other combinations of $M_{2,\text{f}}$, $\Delta M$
and $M_{1}$ the turn-off velocity can be anywhere between $v_{\text{rot}}\simeq1\text{--}30\ \text{km}\thinspace\text{s}^{-1}$
(increasing with $\Delta M$), with the range of possible turn-off
velocities increasing with $j_{\text{a}}$ (Fig.\ \ref{fig:ja-vs-vrotTO_r}).
That is because the rotation velocity of the star following mass accretion
reflects the total angular momentum accreted. It does not constrain
the amount of mass accreted, because the same amount of angular momentum
can be obtained by accreting a small amount of material with high
specific angular momentum, or a large amount of material with low
specific angular momentum.

\begin{figure}
\includegraphics[width=1\columnwidth]{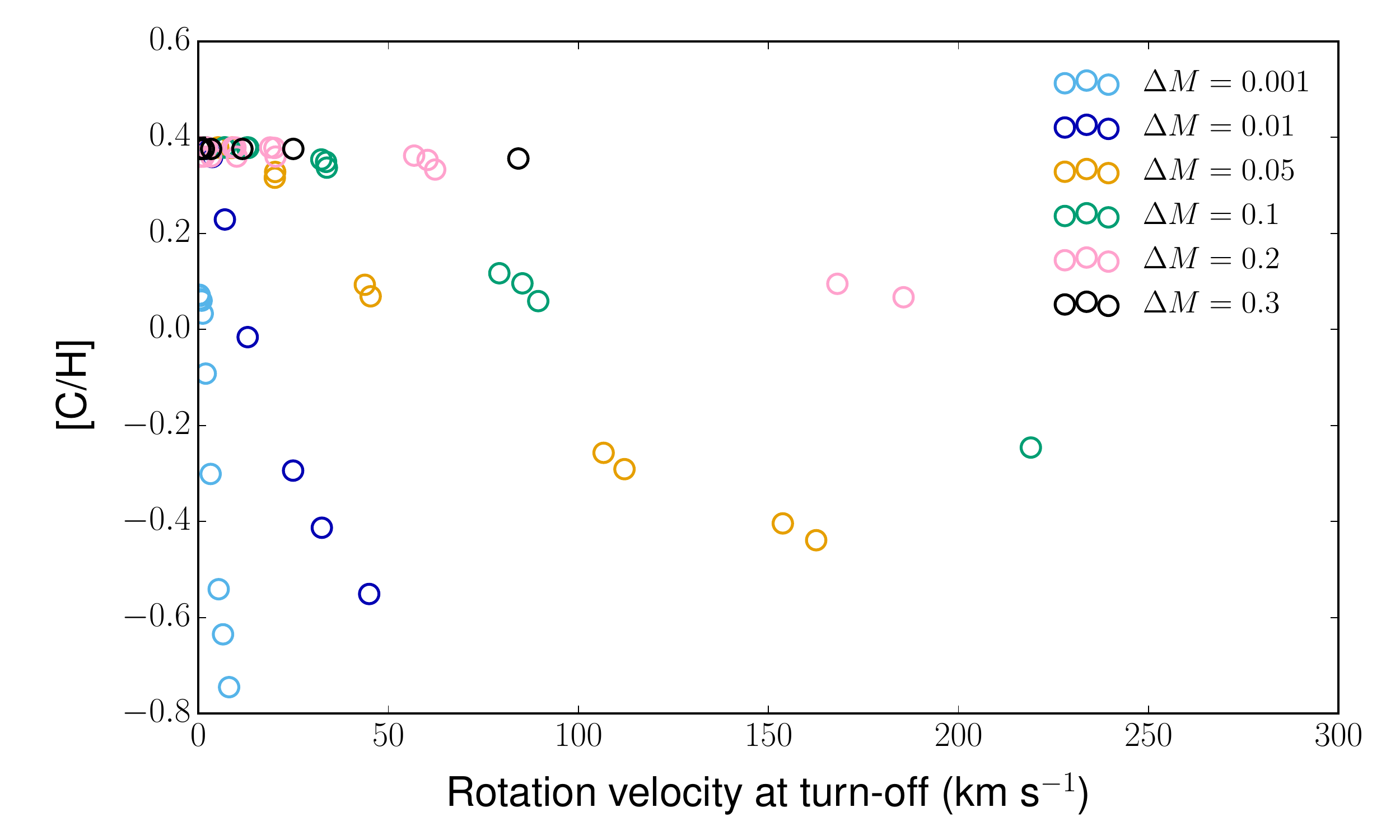}

\caption{Carbon abundance at turn-off (`4') as a function of rotation velocity
in models with $M_{1}=1.25\ M_{\odot}$ and rotational mixing only.
After mass transfer $\text{[C/H]}\simeq0.38$ in all cases.\label{fig:mp125r_vrot-vs-CH_TO}}
\end{figure}

\begin{figure}
\includegraphics[width=1\columnwidth]{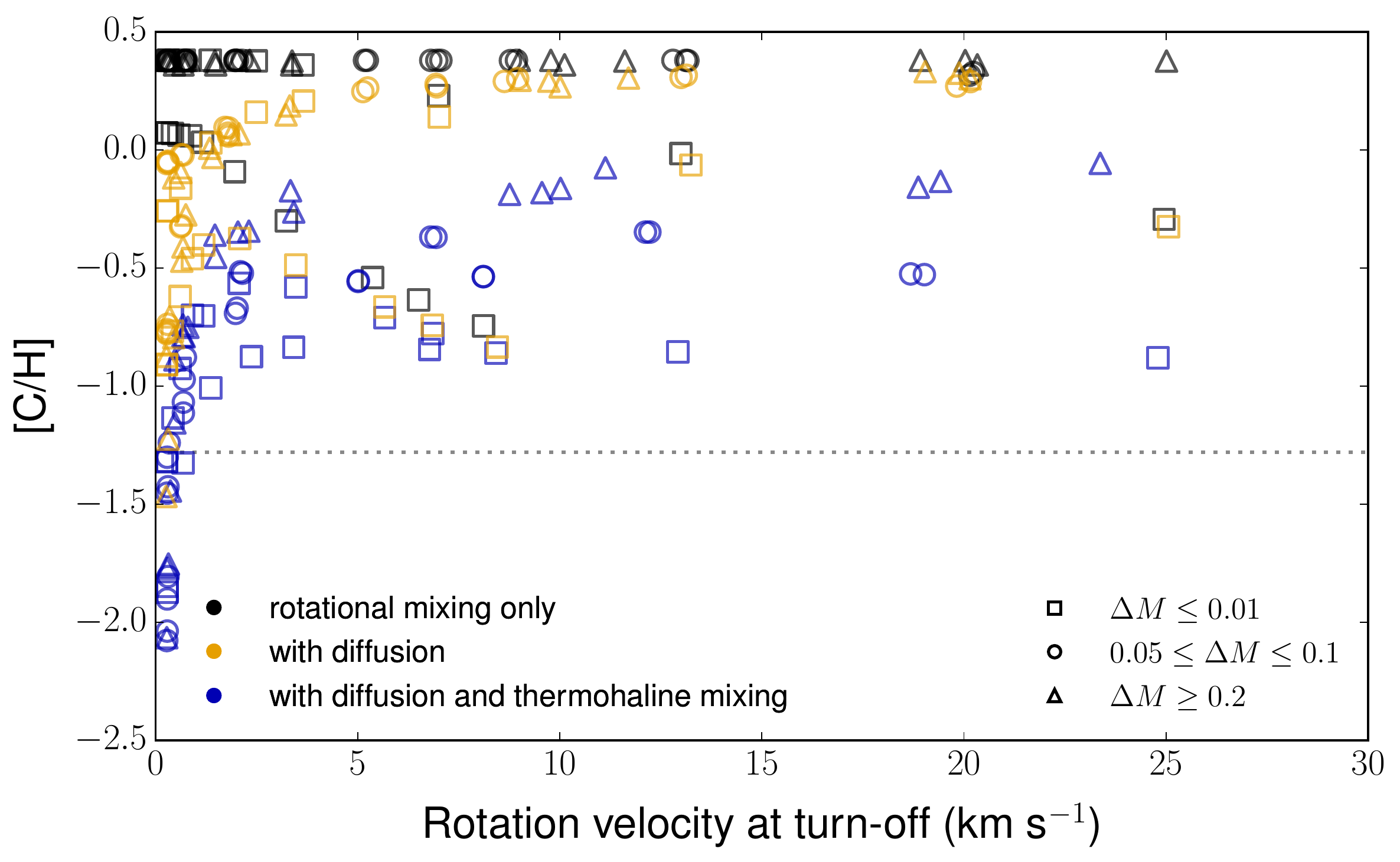}

\caption{Carbon abundance at turn-off (`4') as a function of rotation velocity
in all models with $M_{1}=1.25\ M_{\odot}$. After mass transfer $\text{[C/H]}\simeq0.38$
in all cases. Models above the dotted line are carbon-enhanced \citep[$\text{[C/Fe]}>0.9$;][]{2010A&A...509A..93M}.\label{fig:mp125rdt_vrot-vs-CH_TO}}
\end{figure}

The rotation velocity is therefore not very informative of the amount
of rotational mixing expected, which is unfortunate given that the
rotation velocity is an observable quantity. A rotation rate of, e.g.
$50\ \text{km}\thinspace\text{s}^{-1}$ could correspond to a carbon
dilution of more than a dex if $\Delta M$ is small ($\lesssim0.01\ M_{\odot}$)
or negligible dilution if $\Delta M$ is large ($\gtrsim0.2\ M_{\odot}$;
Fig.\ \ref{fig:mp125r_vrot-vs-CH_TO}). The rotation velocity does,
however, serve as a good indicator of whether atomic diffusion should
be important (Fig.\ \ref{fig:mp125rdt_vrot-vs-CH_TO}). If one compares
models with rotation alone (black symbols) to those with rotation
and diffusion (orange symbols), one sees that only below $v_{\text{rot}}\simeq1\ \text{km}\thinspace\text{s}^{-1}$
does atomic diffusion lead to abundance anomalies of a dex or more.
The amount of dilution of carbon drops below a factor of two above
rotation velocities of a mere $2\text{--}3\ \text{km}\thinspace\text{s}^{-1}$
and basically disappears above $5\ \text{km}\thinspace\text{s}^{-1}$.
The most severe change in abundances in models with such rotation
velocities results from thermohaline mixing (blue symbols), which
typically reduces {[}C/H{]} by a factor of three or more, depending
on the mass accreted and the molecular weight of the accreted material
\citep{2007A&A...464L..57S,2016A&A...592A..29M}.

For the largest values of specific angular momentum ($j_{\text{a}}\gtrsim7\times10^{17}\ \text{cm}^{2}\thinspace\text{s}^{-1}$)
abundance differences between models with and without thermohaline
mixing also disappear. This is because, when rotational mixing is
rapid enough, it can dilute the material to a similar extent as thermohaline
mixing, and the abundances near the turn-off end up being similar
(cf. Figs.\ \ref{fig:mp125ms0750dm0050r_5-vs-30} and \ref{fig:mp125ms0750dm0050rdt_5-vs-30}).
Nevertheless, thermohaline mixing is still by far the more rapid of
the two mixing processes and responsible for most of the dilution
when $\Delta M\gtrsim0.01\ M_{\odot}$ (as in the high-$j_{\text{a}}$
cases in Fig.\ \ref{fig:mp125ms0750dm0050rdt}).

Overall then, in models with rotational mixing only, it is the specific
angular momentum of the accreted material, and not the rotation velocity
or the total angular momentum accreted, that best predicts whether
rotational instabilities will directly lead to chemical mixing. For
a given progenitor system (combination of $M_{1}$, $M_{2,\text{i}}$),
the dilution on the main sequence is determined almost entirely by
the specific angular momentum of the accreted material if more than
a couple of hundredths of a solar mass are accreted. The rotation
velocity, while not a reliable indicator of the importance of rotational
mixing, does reflect the importance of atomic diffusion – large abundance
anomalies above rotation velocities of a few $\text{km}\thinspace\text{s}^{-1}$
are not expected. Thermohaline mixing is responsible for most of the
abundance changes occurring on the main sequence after mass transfer,
unless very little mass is accreted (of the order of $10^{-3}\ M_{\odot}$).\footnote{A quantitative summary of the models presented in this paper (mainly the stellar properties and surface abundances at key points of the evolution) will be made available in electronic form at the CDS.}

\section{Discussion\label{sec:discussion}}

Following up on our previous work, we have modelled the evolution
of a large number of CEMP-\emph{s} stars originating from a range
of putative progenitor systems, for the first time considering in
detail the accretion and internal transport of angular momentum by
these stars. We now discuss the applicability of the models to real
CEMP-\emph{s} stars (Sect.\ \ref{subsec:real-cemps}) and then turn
to the abundance evolution predicted by the models in context of observations
(Sect.\ \ref{subsec:observations}). We end with an examination of
the importance of the free parameters inherent in our adopted prescription
for angular momentum evolution (Sect.\ \ref{subsec:fcfmu}).

\subsection{Angular momentum content of real CEMP-\emph{s} stars\label{subsec:real-cemps}}

We know of no CEMP-\emph{s} stars rotating at a substantial fraction
of their critical velocity. Indeed, typical velocities of CEMP dwarfs
seem to be in the $5\text{--}15\ \text{km}\thinspace\text{s}^{-1}$
range \citep{2012ApJ...751...14M}. These velocities are probably
higher than the typical velocities of old Halo dwarfs \citep[by a factor of about two;][]{2003A&A...406..691L,2009ApJ...704..750C},
which supports the idea of angular momentum accretion by these stars.
Even so, these velocities are quite low (no more than a few percent
of the critical velocity), which indicates that either the stars lose
angular momentum after mass transfer, or they accrete little of it
to begin with. There are issues with both possibilities.

If a substantial amount of mass is to be accreted, accreting a small
amount of angular momentum requires that the specific angular momentum
of the accreted material is low, notably much below the Keplerian
value of $j_{\text{K}}=\sqrt{GMR}\simeq2\times10^{18}\ \text{cm}^{2}\thinspace\text{s}^{-1}$
(Fig.\ \ref{fig:ja-vs-vrotTO_r}). But multi-dimensional hydrodynamic
simulations of representative progenitor systems routinely predict
accretion disk formation \citep[e.g.][]{1996MNRAS.280.1264T,2013MNRAS.433..295H,2017MNRAS.468.4465C}
or otherwise find the specific angular momentum of the material flowing
around the accretor to be close to $j_{\text{K}}$ \citep{Liu2017subm}.
Although these simulations neglect physical processes that likely
play an important role in regulating the angular momentum accreted
by the star \citep[particularly magnetic fields;][]{1996MNRAS.280..458A,2005ApJ...632L.135M,2005MNRAS.356..167M},
the angular momentum would have to effectively be wrong by two orders
of magnitude to reconcile the simulations with the observations. This
issue is discussed in more detail in \citet{2017arXiv170708224M}.

Alternatively, the stars may have rotated rapidly shortly after mass
transfer, but lost much of the accreted angular momentum subsequently.
Given that their interior structure (a radiative interior with a surface
convective envelope) is qualitatively similar to solar-like stars,
one might expect that CEMP stars too lose angular momentum by magnetized
winds \citep{1967ApJ...148..217W,1987MNRAS.226...57M}. At solar metallicity
this magnetic braking is found to cause the gradual spin-down of stars
up to masses of $M\simeq1.4\ M_{\odot}$ on timescales of about $0.1\text{--}1\ \text{Gyr}$
\citep[e.g.][]{1972ApJ...171..565S,1987PASP...99.1322K,2011ApJ...733L...9M,2015Natur.517..589M,2014prpl.conf..433B}.
If magnetic braking operates similarly in CEMP stars, they should
have had a fair amount of time to spin down, as even the youngest
CEMP stars are probably at least a gigayear old.\footnote{The youngest Halo stars are about 10~Gyr old, while the lifetime
of the lowest-mass star that undergoes third dredge-up during the
AGB stage at $Z=10^{-4}$ is about a gigayear less \citep{2010MNRAS.403.1413K}.}

The magnetized winds are believed to be powered by a dynamo sustained
by the interaction between differential rotation near the base of
the envelope and convection \citep{2005PhR...417....1B,2010LRSP....7....3C}.
The successful operation of the dynamo thus depends on the properties
of the surface convection zone. As a prerequisite, one has to exist,
and the progressively longer spin-down timescales of stars in the
$1.1\lesssim M/M_{\odot}\lesssim1.4$ range is attributed to the gradual
thinning of the surface convection zone \citep{1967ApJ...150..551K,1987PASP...99.1322K},
after the disappearance of which no braking occurs. We find that by
various measures that could be relevant for the efficiency of the
dynamo \citep[e.g.][]{1984ApJ...279..763N,1993A&A...269..446S} –
the mass contained ($10^{-5}\lesssim M_{\text{env}}/M_{\odot}\lesssim10^{-3}$),
the convective turn-over timescale ($1\lesssim\tau_{\text{conv}}(\text{d})\lesssim10$),
the fractional radius ($0.05\lesssim R_{\text{env}}/R\lesssim0.2$)
and volume ($0.15\lesssim V_{\text{env}}/V\lesssim0.5$) – the convective
envelopes of $Z=10^{-4}$ models with $M\simeq0.75\text{--}0.85\ M_{\odot}$
resemble those of $Z=0.02$ models with $M\simeq1.15\text{--}1.4\ M_{\odot}$
throughout much of the main sequence. The envelopes of CEMP stars
are still more sizeable because of the increased metallicity ($Z$).
It thus seems plausible that CEMP stars too could sustain a dynamo
and undergo magnetic braking, although possibly on fairly long (gigayear)
timescales.

Assuming CEMP stars do undergo magnetic braking, what are the consequences
for their evolution? Obviously, one consequence is that their surface
rotational velocity decreases over time, but what about the internal
transport of angular momentum and chemical elements? To gain some
insight into this question, we return to the illustrative case of
Sect.\ \ref{subsec:illust-mod-seq-r} with $j_{\text{a}}=5\times10^{17}\ \text{cm}^{2}\thinspace\text{s}^{-1}$.
We restart the model sequence from the end of mass transfer (labelled
`2' in Fig.\ \ref{fig:mp125ms0750dm0050r}), this time including
angular momentum loss following \citet{1988ApJ...333..236K}:

\begin{equation}
\frac{\mathrm{d}{J}}{\mathrm{d}{t}}=-K\left(\frac{R/R_{\odot}}{M/M_{\odot}}\right)^{\frac{1}{2}}\Omega\min(\Omega,\Omega_{\text{sat}})^{2}.\label{eq:mb}
\end{equation}
$\Omega_{\text{sat}}$ is the surface angular velocity, above which
angular momentum loss is found to saturate in rapidly rotating solar-like
stars \citep{1995ApJ...441..865C}, here taken to be $10\Omega_{\odot}$
\citep{2016A&A...587A.105A}, and $K=2.5\times10^{47}$ ($\text{g}$~$\text{cm}^{2}$~s)
is a calibrating constant, chosen to reproduce the solar rotation
rate at the solar age.

\begin{figure*}
\subfloat[Evolution of surface rotation velocity]{\includegraphics[width=1\columnwidth]{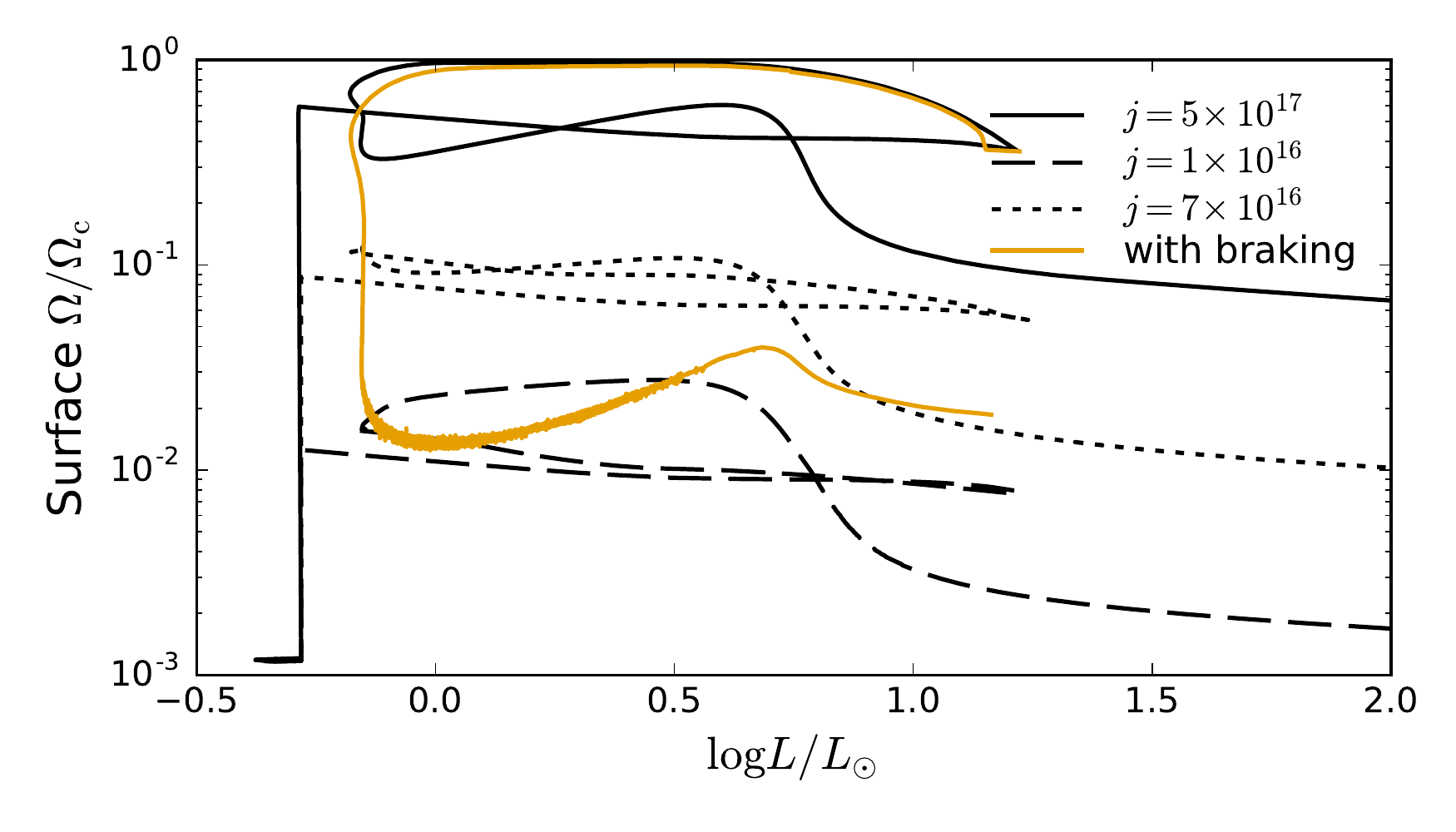}

\label{fig:mp125ms0750dm0050r_5-vs-89_mb}}\hspace{\columnsep}\subfloat[Evolution of carbon]{\includegraphics[width=1\columnwidth]{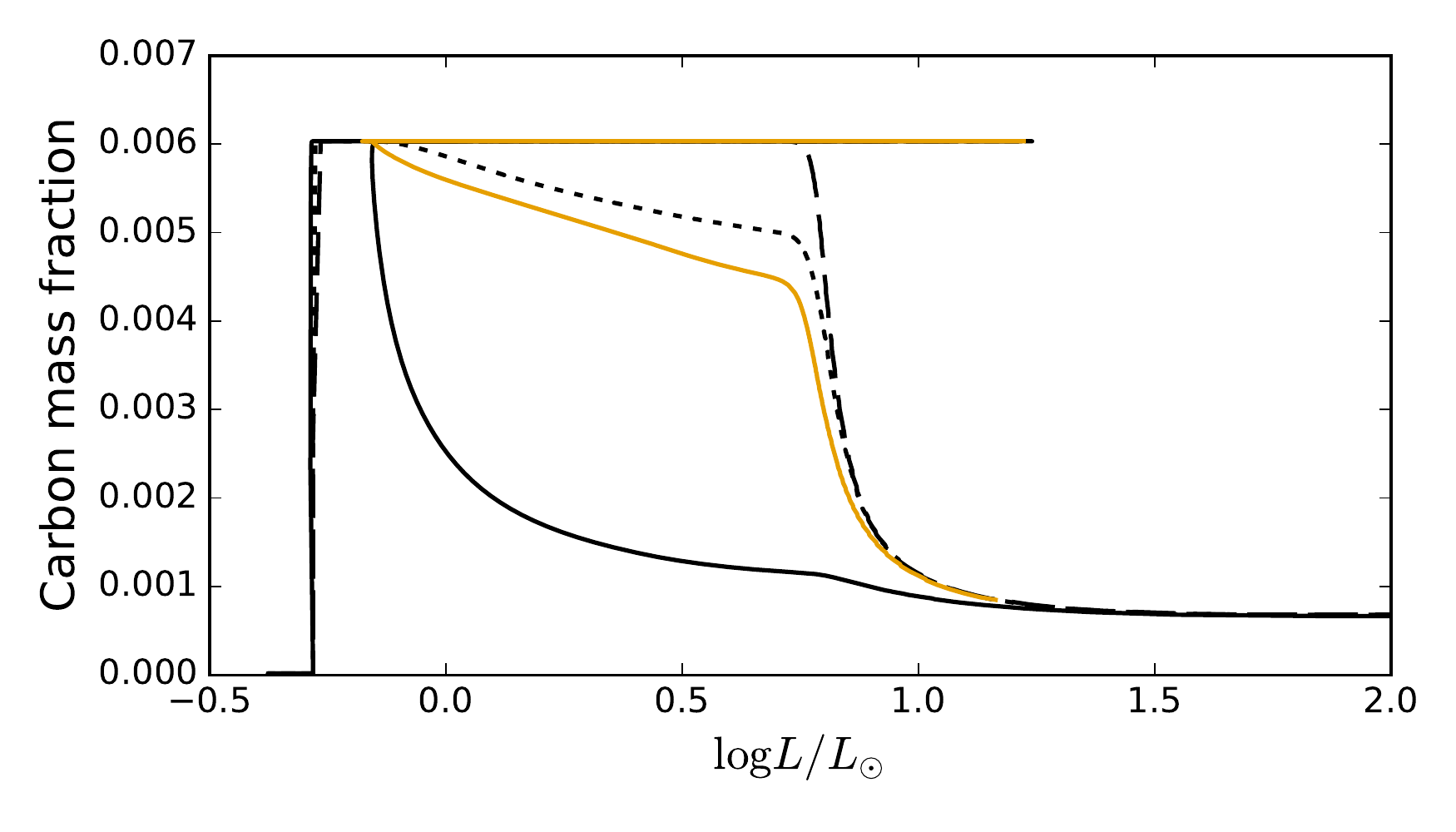}

\label{fig:mp125ms0750dm0050r_5-vs-30_mb}}

\caption{Influence of magnetic braking on the $M_{1}=1.25\ M_{\odot}$, $M_{2,\text{i}}=0.75\ M_{\odot}$,
$\Delta M=0.05\ M_{\odot}$ model with $j_{\text{a}}=5\times10^{17}\ \text{cm}^{2}\thinspace\text{s}^{-1}$
and rotational mixing only. Out of the models without braking, on
the main sequence the $j_{\text{a}}=1\times10^{16}\ \text{cm}^{2}\thinspace\text{s}^{-1}$
model is the one most similar in terms of surface rotation velocity,
and the $j_{\text{a}}=7\times10^{16}\ \text{cm}^{2}\thinspace\text{s}^{-1}$
model is the one most similar in terms of surface abundances.\label{fig:mp125ms0750dm0050r_mb}}
\end{figure*}

The default $M_{2,\text{f}}=0.8\ M_{\odot}$ model rotates at a velocity
of $v_{\text{rot}}\simeq100\ \text{km}\thinspace\text{s}^{-1}$ throughout
the post-mass-transfer main sequence (Fig.\ \ref{fig:mp125ms0750dm0050r_5-vs-84}).
Such rapid rotation implies a large torque according to Eq.\ \eqref{eq:mb}.
Its application to the relatively thin envelope ($0.004\ M_{\odot}$
shortly after mass transfer) of the star results in a very rapid spin-down:
50~Myr after the end of mass transfer the surface rotation velocity
has fallen to $v_{\text{rot}}<10\ \text{km}\thinspace\text{s}^{-1}$
and after about a gigayear it levels off to $v_{\text{rot}}\simeq4\ \text{km}\thinspace\text{s}^{-1}$
(Fig.\ \ref{fig:mp125ms0750dm0050r_5-vs-89_mb}). After that the
angular momentum of the envelope no longer changes appreciably, as
the angular momentum loss is balanced by the outward transport from
the radiative core. In other words, the angular momentum is then extracted
from the core \citep{1991ApJ...367..239P,2014ApJ...780..159E}.

\begin{figure*}
\subfloat[Evolution of internal rotation profiles]{\includegraphics[width=1\columnwidth]{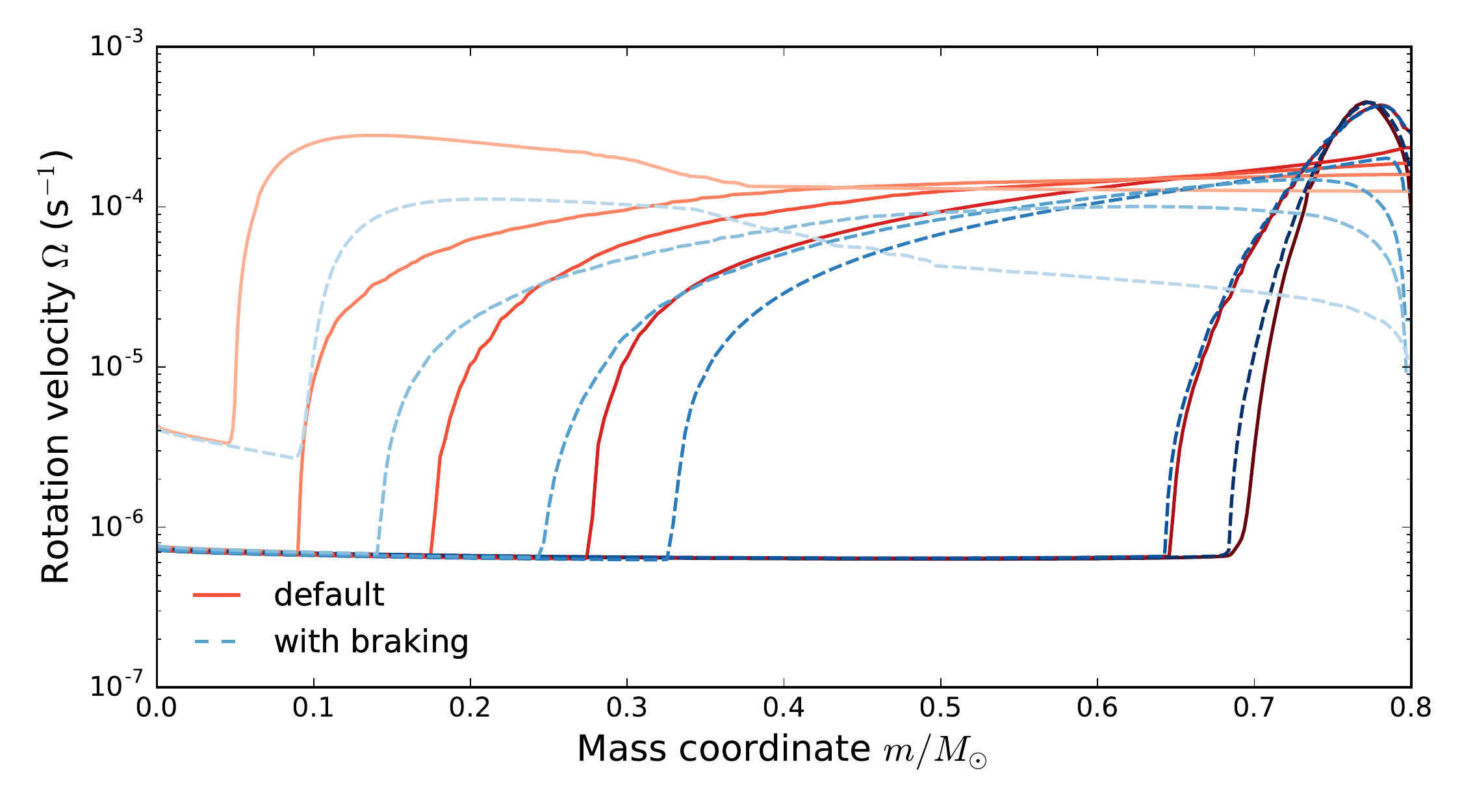}

\label{fig:py_z1e-4mp125ms0750dm0050j5e17r_omega-vs-m_evo_with-mb}}\hspace{\columnsep}\subfloat[Evolution of internal carbon profile]{\includegraphics[width=1\columnwidth]{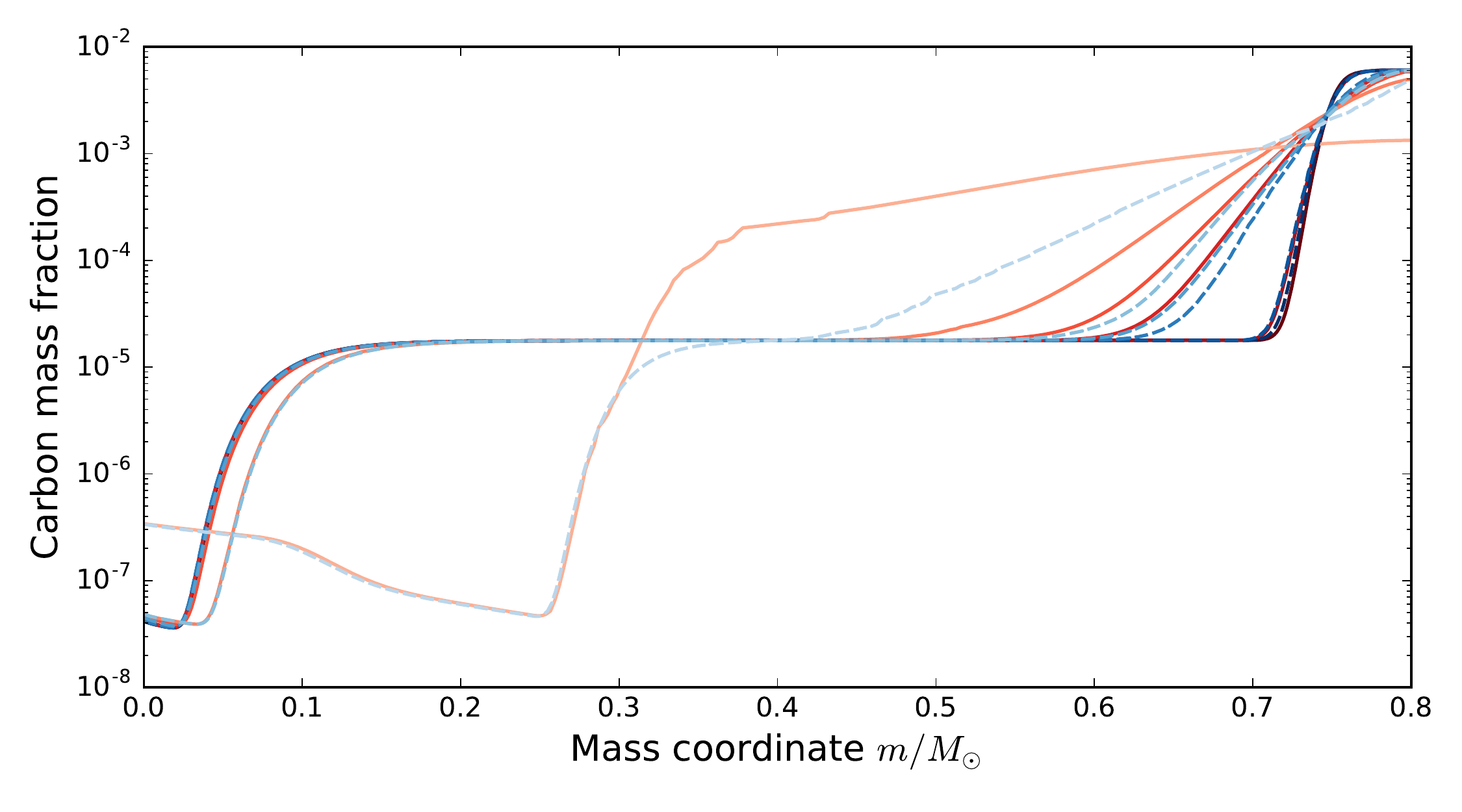}

\label{fig:py_z1e-4mp125ms0750dm0050j5e17r_XC-vs-m_evo_with-mb}}

\caption{Internal evolution of the model from Fig.\ \ref{fig:mp125ms0750dm0050r_mb}
with and without magnetic braking. The sets of profiles correspond
to very similar ages in both cases, and later times are plotted in
progressively lighter shades.\label{fig:mp125ms0750dm0050r_evo_mb}}
\end{figure*}

The surface rotational velocity of the model with magnetic braking
during the main sequence is most similar to that with $j_{\text{a}}=10^{16}\ \text{cm}^{2}\thinspace\text{s}^{-1}$
and no braking. But the internal angular momentum evolution is not
very different from the default model with $j_{\text{a}}=5\times10^{17}\ \text{cm}^{2}\thinspace\text{s}^{-1}$
because the accreted layer is much larger than the convective envelope.
The instabilities at the base of this accreted layer thus still occur,
although the continuous removal of angular momentum does reduce the
depth to which angular momentum has been transported at a given time
(Fig.\ \ref{fig:mp125ms0750dm0050r_evo_mb}). Still, the angular
momentum content in the model with braking remains larger than in
the $j_{\text{a}}=10^{16}\ \text{cm}^{2}\thinspace\text{s}^{-1}$
model without braking, and this is reflected by its more rapid rotation
during post-main-sequence evolution.

Since the transport of chemical species occurs over a longer timescale,
it is more affected by the angular momentum loss at the surface. The
depth to which elements are mixed is much smaller in the case with
braking ($m\simeq0.45\ M_{\odot}$ instead of $m<0.3\ M_{\odot}$
at turn-off; Fig.\ \ref{fig:py_z1e-4mp125ms0750dm0050j5e17r_XC-vs-m_evo_with-mb}),
and the surface abundance evolution on the main sequence is closer
to the $j_{\text{a}}=7\times10^{16}\ \text{cm}^{2}\thinspace\text{s}^{-1}$
case, which has a surface rotational velocity of only $v_{\text{rot}}\simeq20\ \text{km}\thinspace\text{s}^{-1}$
(Fig.\ \ref{fig:mp125ms0750dm0050r_mb}). Internally, however, the
abundance profiles are very smeared out in the model with braking.
This results in a prolonged first dredge-up, compared to the $j_{\text{a}}=7\times10^{16}\ \text{cm}^{2}\thinspace\text{s}^{-1}$
case.

For this system then, including magnetic braking in the case with
$j_{\text{a}}=5\times10^{17}\ \text{cm}^{2}\thinspace\text{s}^{-1}$
gives a surface rotational velocity similar to the $j_{\text{a}}=10^{16}\ \text{cm}^{2}\thinspace\text{s}^{-1}$
case, and surface abundance evolution similar to the $j_{\text{a}}=7\times10^{16}\ \text{cm}^{2}\thinspace\text{s}^{-1}$
case. We expect that the former would remain true, if we had applied
magnetic braking to a model with a different value of $j_{\text{a}}>10^{16}\ \text{cm}^{2}\thinspace\text{s}^{-1}$,
or to some extent even a different system (i.e. combination of $M_{1}$,
$M_{2,\text{i}}$, $\Delta M$) altogether. The reason is that the
surface velocity after the envelope has spun down largely depends
on the constant $K$ in Eq.\ \eqref{eq:mb}. But the non-braking
model with the most similar evolution of surface abundances would
however change (to a case somewhere between that which gives the most
similar rotation velocity, and that from which we start). But generally,
we expect magnetic braking to reduce the surface abundance anomalies
stemming from rotational mixing. For example, all the points in Fig.\ \ref{fig:mp125r_vrot-vs-CH_TO}
would shift towards lower velocities and to higher {[}C/H{]}, the
shift in velocity being more important for models with $\Delta M\gtrsim0.05\ M_{\odot}$,
and the shift in {[}C/H{]} dominating for models with $\Delta M\lesssim0.01\ M_{\odot}$.

\subsection{Comparison to observations\label{subsec:observations}}

In \citet{2016A&A...592A..29M} we showed that atomic diffusion should
lead to very large abundance anomalies (e.g. $\text{[C/Fe]}<-1$)
near the main sequence turn-off, a result clearly at odds with observational
data. Here we again juxtapose some of our model sequences to the measured
carbon abundances ({[}C/H{]}) of CEMP stars from the Sloan Digital
Sky Survey \citep[SDSS;][]{2013AJ....146..132L}, the largest homogeneous
data set of carbon abundances in metal-poor stars. We restrict the
comparison to models with $v_{\text{rot}}\lesssim100\ \text{km}\thinspace\text{s}^{-1}$
at turn-off. While this limit considerably exceeds the highest observed
velocities of CEMP stars, these rapidly rotating models serve to illustrate
the effect of rotational mixing in systems other than the one discussed
so far, and can be taken to mimic the surface chemical evolution of
initially still more rapidly rotating models with magnetic braking.

\begin{figure*}
\includegraphics[width=1\textwidth]{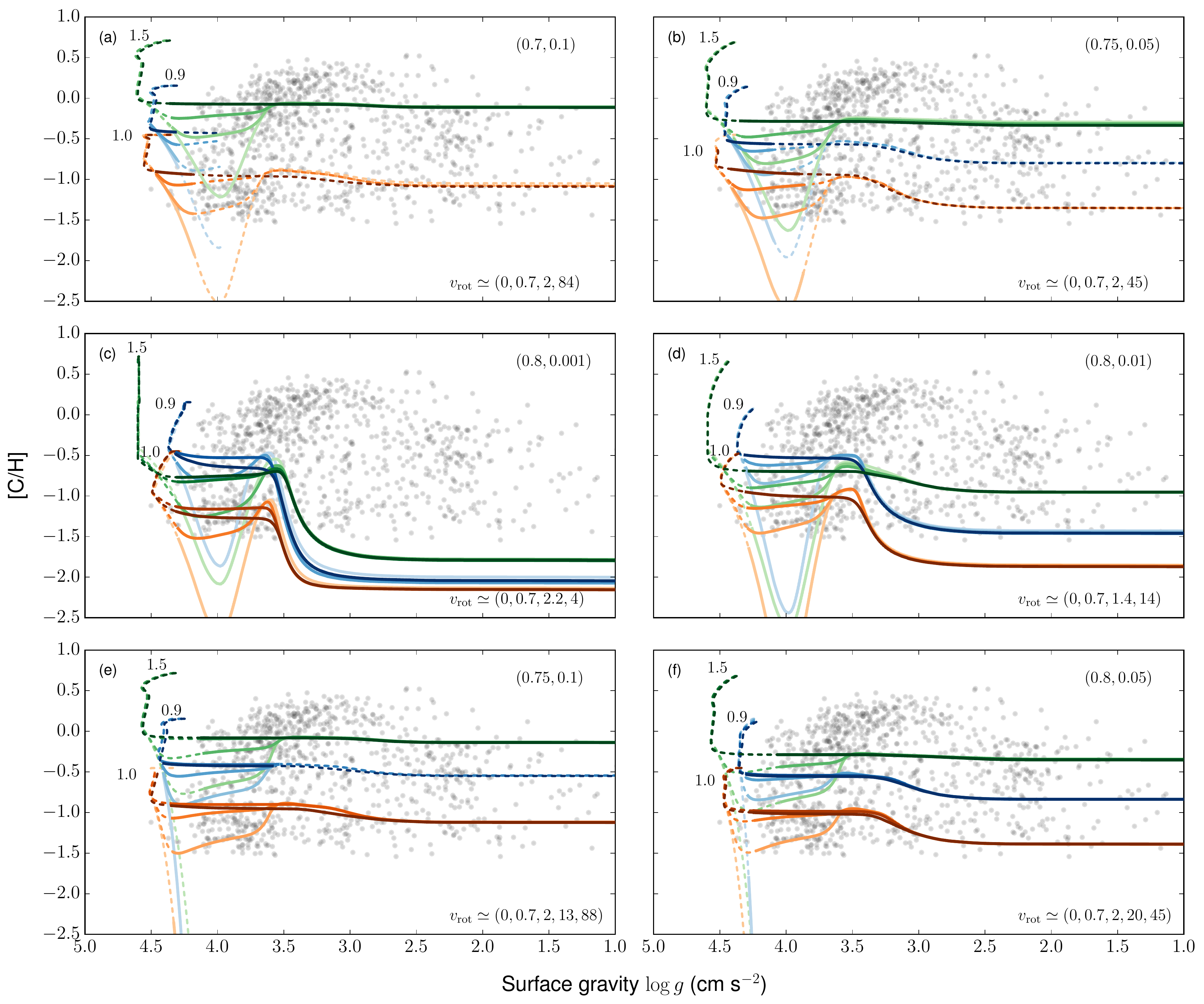}

\caption{Evolution of {[}C/H{]} in CEMP-\emph{s} star models of different initial
and accreted masses (given in solar masses in the top right corner
of each panel) with rotational mixing, gravitational settling, and
thermohaline mixing. The three sets of lines correspond to different
AGB donor masses (marked on the left). The colour intensity indicates
the specific angular momentum of accreted material. From brightest
to darkest, the lines correspond to $j_{\text{a}}=0$ (no rotation),
$0.02,0.04,0.2,0.7,2.0$ ($\times10^{17}\ \text{cm}^{2}\thinspace\text{s}^{-1}$).
The corresponding rotation velocities are listed in the bottom right
corner of each panel (to reduce crowding not all $j_{\text{a}}$ cases
are shown in every panel). The solid part of each line delimits ages
between 10 and 13.8\ Gyr (for panel a computations were stopped at
$t=16$\ Gyr). The points are CEMP stars \citep[$\text{[C/Fe]}\ge 0.9$;][]{2010A&A...509A..93M}
from SDSS with $-2.5\leq\text{[Fe/H]}\leq-2.0$ \citep{2013AJ....146..132L}.\label{fig:logg-vs-CHrdt_with-SDSS}}
\end{figure*}

We also restrict the comparison to models with atomic diffusion. We
can easily see when diffusion, or more accurately gravitational settling
and thermal diffusion, becomes important by considering the {[}C/H{]}
abundance evolution in models with different rotation velocities.
In all cases in Fig.\ \ref{fig:logg-vs-CHrdt_with-SDSS} we see that
the hallmark pattern of atomic diffusion – continuous decrease of
heavy element abundances until the turn-off (\textbf{$\log g\simeq4$}
in these stars) followed by a reversal as the convective envelope
moves inwards – is severely disrupted, compared to the non-rotating
case, already at $v_{\text{rot}}\simeq0.7\ \text{km}\thinspace\text{s}^{-1}$.
For $v_{\text{rot}}\simeq2\ \text{km}\thinspace\text{s}^{-1}$ the
variation of {[}C/H{]} over the main sequence is below 0.3\ dex.
This is what we already concluded in Sect.\ \ref{subsec:results-summary}
(Fig.\ \ref{fig:mp125rdt_vrot-vs-CH_TO}). Although in this study
we have ignored radiative levitation, which can drastically alter
the relative abundances of metals \citep[e.g. the \text{[C/Fe]} ratio;][]{2016A&A...592A..29M},
levitation will only be important together with the other microscopic
diffusion processes, that is, in the slowly rotating models ($v_{\text{rot}}\lesssim2\ \text{km}\thinspace\text{s}^{-1}$).

If rotational mixing is indeed responsible for inhibiting atomic diffusion
in metal-poor stars, the lack of stars with $\text{[C/H]}<-2.5$ between
$-2.5<\text{[Fe/H]}<-2$ seems to require that all stars rotate, even
if very slowly ($v_{\text{rot}}\gtrsim0.5\ \text{km}\thinspace\text{s}^{-1}$).
Otherwise we should observe some carbon-depleted stars around the
main sequence turn-off. Whether there are stars that rotate still
slower seems unclear. Spectroscopically such low velocities are difficult
to disentangle from other line broadening mechanisms \citep{2005ApJS..159..141V},
and photometric missions are currently restricted to rotation rates
above about the same limit \citep[$P_{\text{rot}}\lesssim100~\text{d}$;][]{2012MNRAS.424...11A,2014ApJS..211...24M}.

Whether thermohaline mixing should activate in rotating stars is a
matter of debate. Following \citet{2010A&A...521A...9C} and \citet{2010A&A...522A..10C},
we have treated the thermohaline and rotational instabilities independently,
neglecting their possible interaction (although, they still influence
each other by changing the structure of the stellar models). Thus
modelled, thermohaline mixing results in a relatively immediate and
substantial ($\Delta\text{[C/H]}>0.3$) reduction of {[}C/H{]} following
mass transfer. Therefore, large quantities ($\Delta M>0.2\ M_{\odot}$)
of high molecular weight material are required to reproduce the largest
observed carbon enhancements. Also, mixing proceeds to slightly greater
depths in more rapidly rotating models. In models with $\Delta M\gtrsim0.05\ M_{\odot}$
the depth of thermohaline mixing generally exceeds the maximum depth
reached by the convective envelope at the end of first dredge-up.
In these models there is thus little to no change in {[}C/H{]} during
FDU \citep[but \text{[N/H]} can increase substantially; see][]{2007A&A...464L..57S}.

\begin{figure*}
\includegraphics[width=1\textwidth]{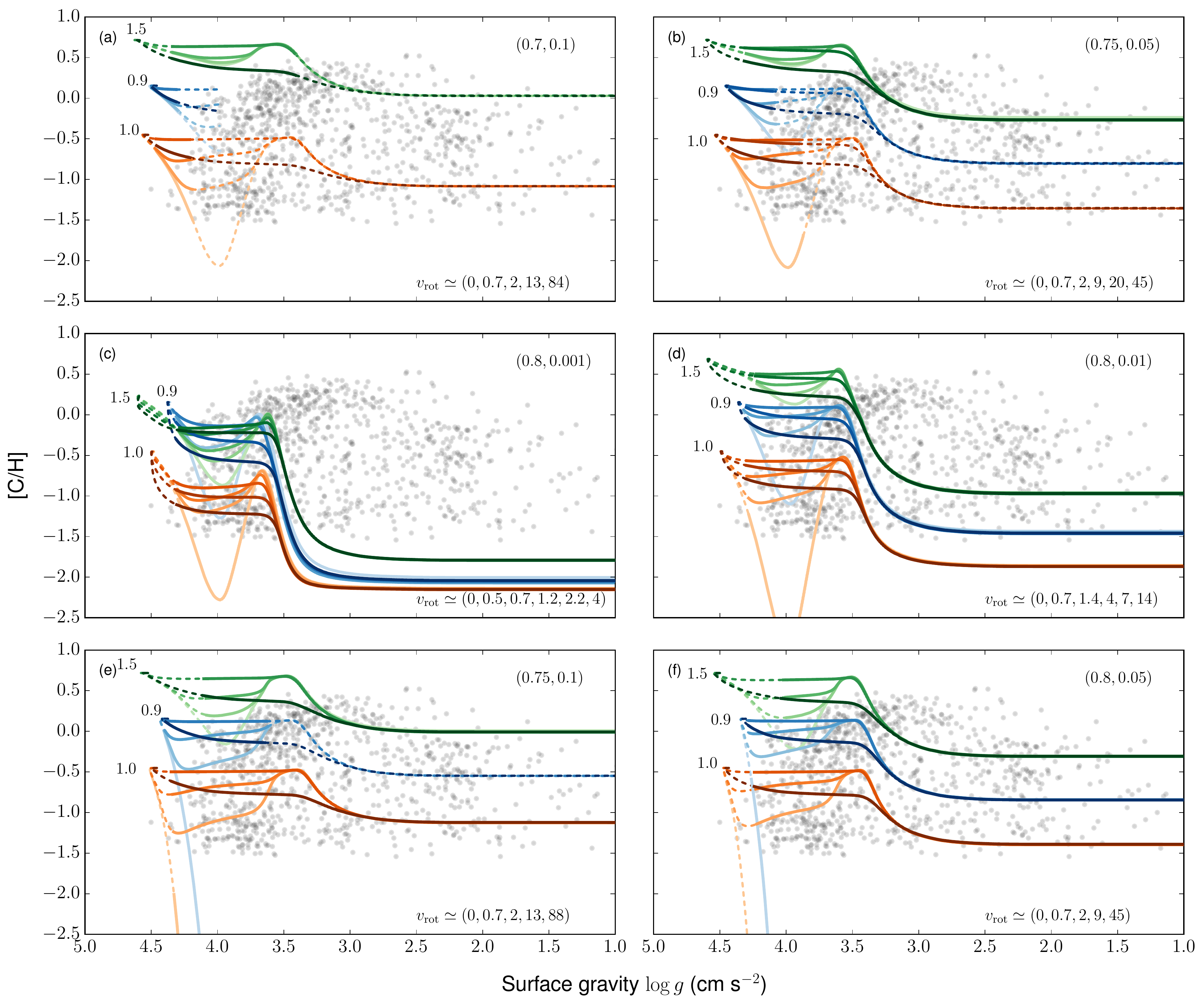}

\caption{As Fig.\ \ref{fig:logg-vs-CHrdt_with-SDSS} but without thermohaline
mixing.\label{fig:logg-vs-CHrd_with-SDSS}}
\end{figure*}

However, many authors \citep{2008ApJ...684..626D,2012ApJ...753...49V,2013A&A...553A...1M}
have argued that the strong horizontal turbulence expected in rotating
stars \citep{1992A&A...265..115Z} should at least curtail thermohaline
mixing, if not outright suppress it \citep[also see][]{2014ApJ...792L..30M}.
If so, it might be more appropriate to exclude thermohaline mixing.
But it is worth noting that the $\mu$-inversion on the red giant
branch, on which much of the cited discussion is focused, is much
smaller than established by accretion of material ($\Delta\mu/\mu\propto10^{-4}$
versus $\Delta\mu/\mu\gtrsim0.01$) and develops gradually as a result
of $^{3}\text{He}$-burning instead of instantaneously, so complete
suppression of thermohaline mixing seems less likely in CEMP-\emph{s}
stars. Given this theoretical uncertainty, in Fig.\ \ref{fig:logg-vs-CHrd_with-SDSS}
we also show models without thermohaline mixing.

At rotation velocities typical of CEMP stars ($v_{\text{rot}}\simeq5\text{--}10\ \text{km}\thinspace\text{s}^{-1}$)
these models look very similar to previously published models with
only convective mixing \citep{2007A&A...464L..57S,2008MNRAS.389.1828S}.
That is, no significant changes in surface abundances occur in these
models before FDU dilutes the accreted material. Before that point,
rotational mixing prevents atomic diffusion but does not cause any
significant dilution of the accreted material by itself (unless very
little mass is accreted as in Figs.\ \ref{fig:logg-vs-CHrd_with-SDSS}c
and \ref{fig:logg-vs-CHrd_with-SDSS}d). Based on the SDSS data, it
is difficult to ascertain whether models with or without thermohaline
mixing ought to be preferred. Some dilution of the accreted material
around FDU may be required \citep[e.g. few stars have $\text{[C/H]}\gtrsim 0$ beyond $\log g\simeq 3$; see also][]{2008ApJ...679.1541D},
but without any thermohaline mixing the dilution is too large unless
$\Delta M$ commonly exceeds $0.2\ M_{\odot}$.

There is a conspicuous lack of unevolved ($\log g\gtrsim4.2$) CEMP
stars in the SDSS data, particularly ones with $\text{[C/Fe]}\gtrsim1.5$.
As discussed in detail in \citet{2016A&A...592A..29M}, this dearth
cannot be explained as an effect of (inhibited) atomic diffusion.
Unsurprisingly then, the new models with rotational mixing also predict
the existence of stars with $\log g\gtrsim4.2$ and $\text{[C/Fe]}>1$.
In particular, lower-mass CEMP-\emph{s} stars ($M_{2,\text{f}}\lesssim0.8\ M_{\odot}$)
and most CEMP-\emph{s} stars with low-mass AGB companions ($M_{1}\lesssim1\ M_{\odot}$)
should populate this region, assuming the ages of these stars are
between 10 and 13.8~Gyr (solid sections of the lines in Figs.\ \ref{fig:logg-vs-CHrdt_with-SDSS}
and \ref{fig:logg-vs-CHrd_with-SDSS}). Since a similar scarcity of
CEMP dwarfs is not evident from high-resolution studies \citep[e.g. as compiled  in the SAGA database;][]{2008PASJ...60.1159S,2011MNRAS.412..843S,2017arXiv170310009S},
we do not believe the issue rests with the models.

\subsection{Influence of rotational mixing parameters\label{subsec:fcfmu}}

The diffusion coefficients in Eq.\ \eqref{eq:Drot} have been derived
making use of order-of-magnitude estimates of some of the length-
and timescales involved, and are thus rather uncertain. The rotational
mixing parameters, $f_{c}$ and $f_{\mu}$, have been introduced to
somewhat correct for this \citep{1989ApJ...338..424P,2000ApJ...528..368H}.
These efficiency parameters are typically calibrated against observed
surface chemical enrichment of rapidly rotating massive stars \citep[e.g.][]{2006A&A...460..199Y,2011A&A...530A.115B}
or the destruction of fragile elements in the Sun and other stars
\citep{1989ApJ...338..424P,1995ApJ...441..876C,1996A&A...308L..13F,2002ApJ...565..571V},
and our default values \citep{2000ApJ...528..368H} fall in their
typical ranges: $0.01\lesssim f_{c}\lesssim0.1$ and $0\lesssim f_{\mu}\lesssim0.2$.

In effect, $f_{c}$ simply sets the timescale for the chemical transport
compared to angular momentum transport. Therefore, increasing $f_{c}$
increases the rate and extent of chemical mixing. But the influence
of $f_{\mu}$, which sets the sensitivity of the rotational instabilities
to molecular weight gradients, is more subtle. Since the various instabilities
depend on $\mu$-gradients in different ways, changing $f_{\mu}$
alters their relative importance in different regions of the star
and over time. Normally stars build up positive molecular weight gradients
in their interiors as they evolve, i.e. $\nabla_{\mu}>0$. Accretion
of AGB ejecta instead forms a negative $\nabla_{\mu}$ in the transition
region between the original and accreted material. In that region
shear instabilities are more likely to set in as a result, independently
of $f_{\mu}$.\footnote{Strictly, increasing $f_{\mu}$ when $\nabla_{\mu}<0$ does make the
dynamical shear instability more likely, whereas the secular shear
instability is favoured for any $\nabla_{\mu}<0$ (regardless of $f_{\mu}$).
However, we find the secular instability to set in far more often.} But in the central regions, where $\nabla_{\mu}>0$, increasing $f_{\mu}$
stabilizes the medium against shear. For the ES circulation and the
GSF instability $\mu$-gradients are always considered inhibiting,
so only the absolute value of $\nabla_{\mu}$ matters. Reducing $f_{\mu}$
thus always helps these transport processes, while increasing $f_{\mu}$
suppresses them. Overall then, increasing $f_{\mu}$ will reduce the
extent of both angular momentum and chemical transport.

To glean the importance of the rotational mixing parameters in our
CEMP star models, we have experimented with changing one of them at
a time. For $f_{\mu}=0.05$ we have run additional models with $f_{c}=0.01$
or $f_{c}=0.1$, and for $f_{c}=1/30$ models with $f_{\mu}=0$ or
$f_{\mu}=1$.%

\begin{figure*}
\subfloat[Evolution of internal rotation profile at different $f_{c}$]{\includegraphics[width=1\columnwidth]{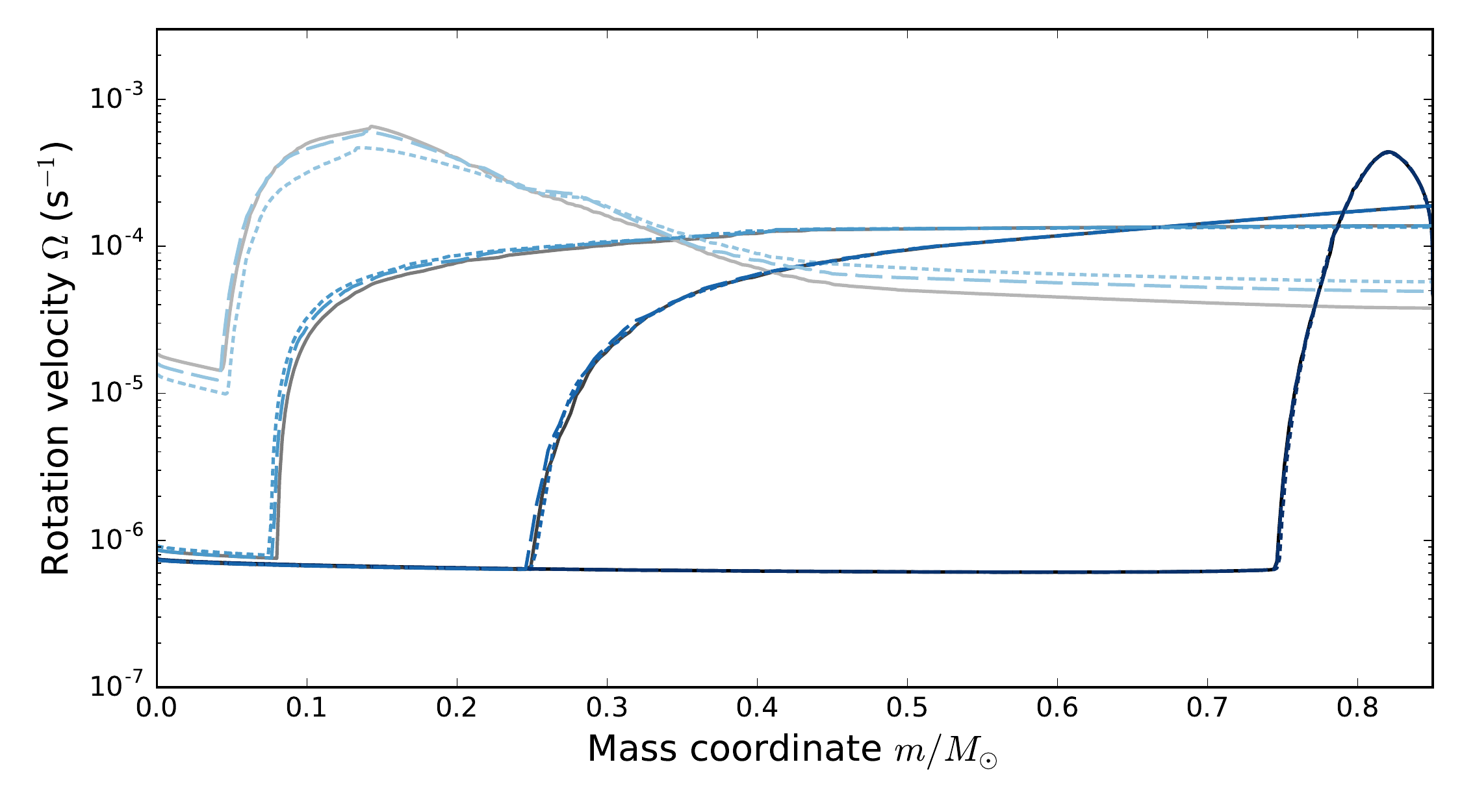}

\label{fig:mp125ms0800dm0050j5e17r_evo-omega_fc-test}}\hspace{\columnsep}\subfloat[Evolution of internal carbon profile at different $f_{c}$]{\includegraphics[width=1\columnwidth]{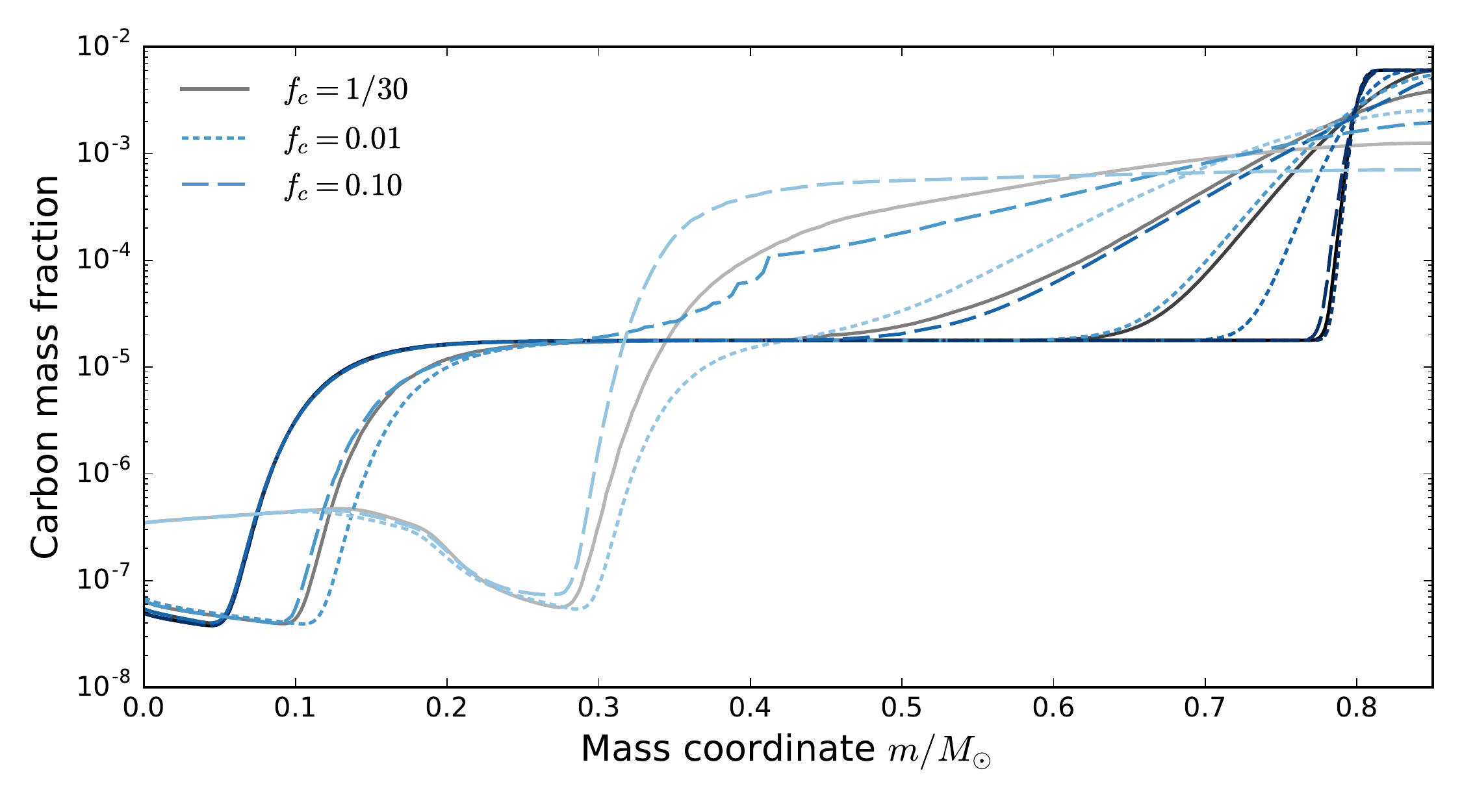}

\label{fig:mp125ms0800dm0050j5e17r_evo-XC_fc-test}}

\subfloat[Evolution of internal rotation profile at different $f_{\mu}$]{\includegraphics[width=1\columnwidth]{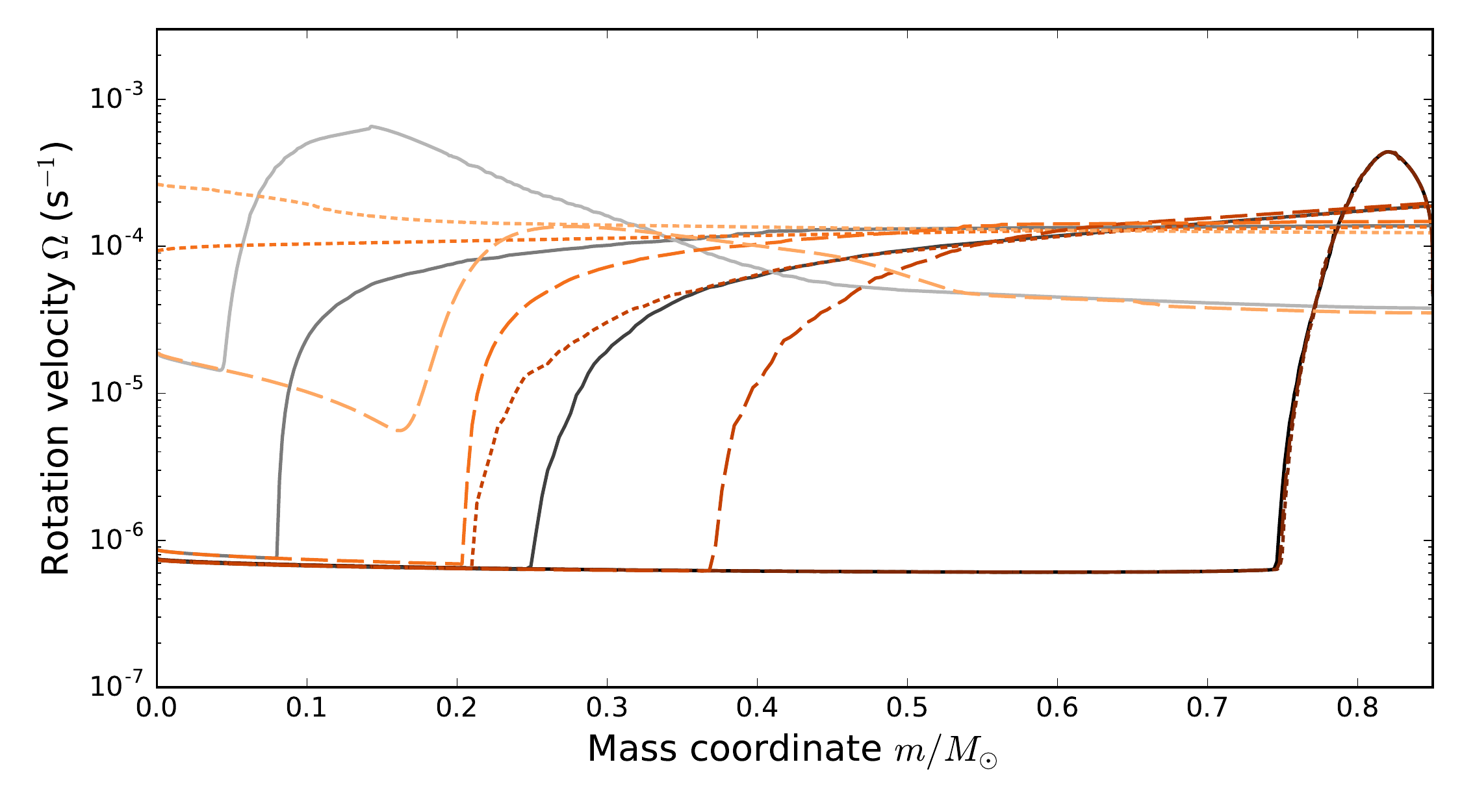}

\label{fig:mp125ms0800dm0050j5e17r_evo-omega_fmu-test}}\hspace{\columnsep}\subfloat[Evolution of internal carbon profile at different $f_{\mu}$]{\includegraphics[width=1\columnwidth]{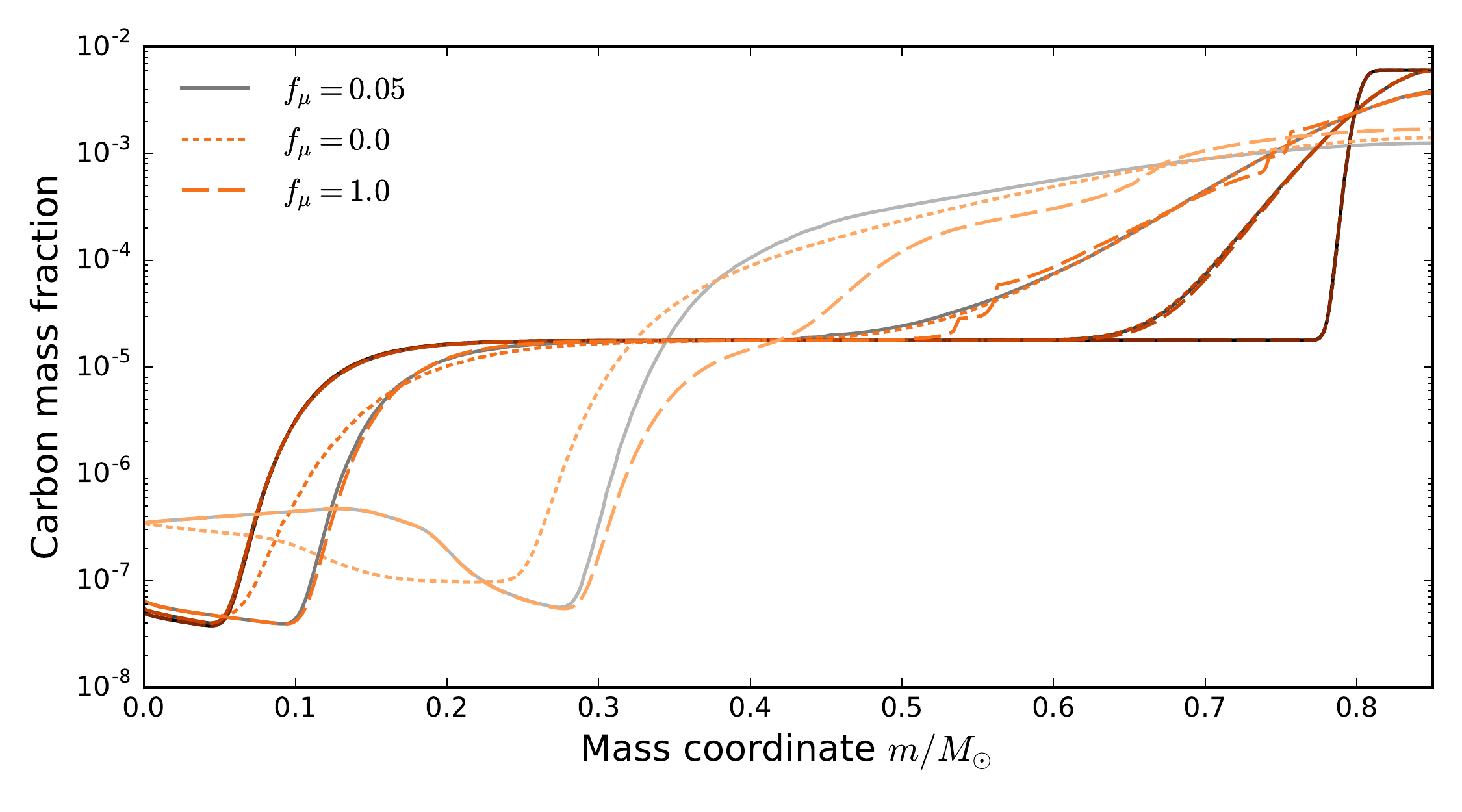}

\label{fig:mp125ms0800dm0050j5e17r_evo-XC_fmu-test}}

\caption{Internal profiles for different rotational mixing parameters $f_{c}$
(panels a, b) and $f_{\mu}$ (c, d) in the system with $M_{1}=1.25\ M_{\odot}$,
$M_{2,\text{i}}=0.8\ M_{\odot}$, $\Delta M=0.05\ M_{\odot}$, $j_{\text{a}}=5\times10^{17}\ \text{cm}^{2}\thinspace\text{s}^{-1}$.
The sets of profiles correspond to ages of about 3.06005 (darkest),
3.1, 4.5, and 11~Gyr (brightest) in all cases. Models with rotational
mixing only.\label{fig:mp125ms0800dm0050j5e17r_evo_fcfmu-test}}
\end{figure*}

Figure\ \ref{fig:mp125ms0800dm0050j5e17r_evo_fcfmu-test} shows the
effect $f_{c}$ and $f_{\mu}$ have on the internal evolution of a
CEMP star model characterized by $M_{1}=1.25\ M_{\odot}$, $M_{2,\text{i}}=0.8\ M_{\odot}$,
$\Delta M=0.05\ M_{\odot}$, and $j_{\text{a}}=5\times10^{17}\ \text{cm}^{2}\thinspace\text{s}^{-1}$.
As expected, while $f_{c}$ has almost no effect on the angular momentum
evolution within the star (Fig.\ \ref{fig:mp125ms0800dm0050j5e17r_evo-omega_fc-test}),
the extent and rate of chemical mixing correlates with $f_{c}$ (Fig.\ \ref{fig:mp125ms0800dm0050j5e17r_evo-XC_fc-test}).
Meanwhile, variation of $f_{\mu}$ affects the transport of both angular
momentum and material. At $f_{\mu}=1$ the transport of both is reduced,
and by the end of the main sequence less of the star is mixed than
in the default case.

\begin{figure*}
\subfloat[Evolution of the central hydrogen abundance at different $f_{\mu}$]{\includegraphics[width=1\columnwidth]{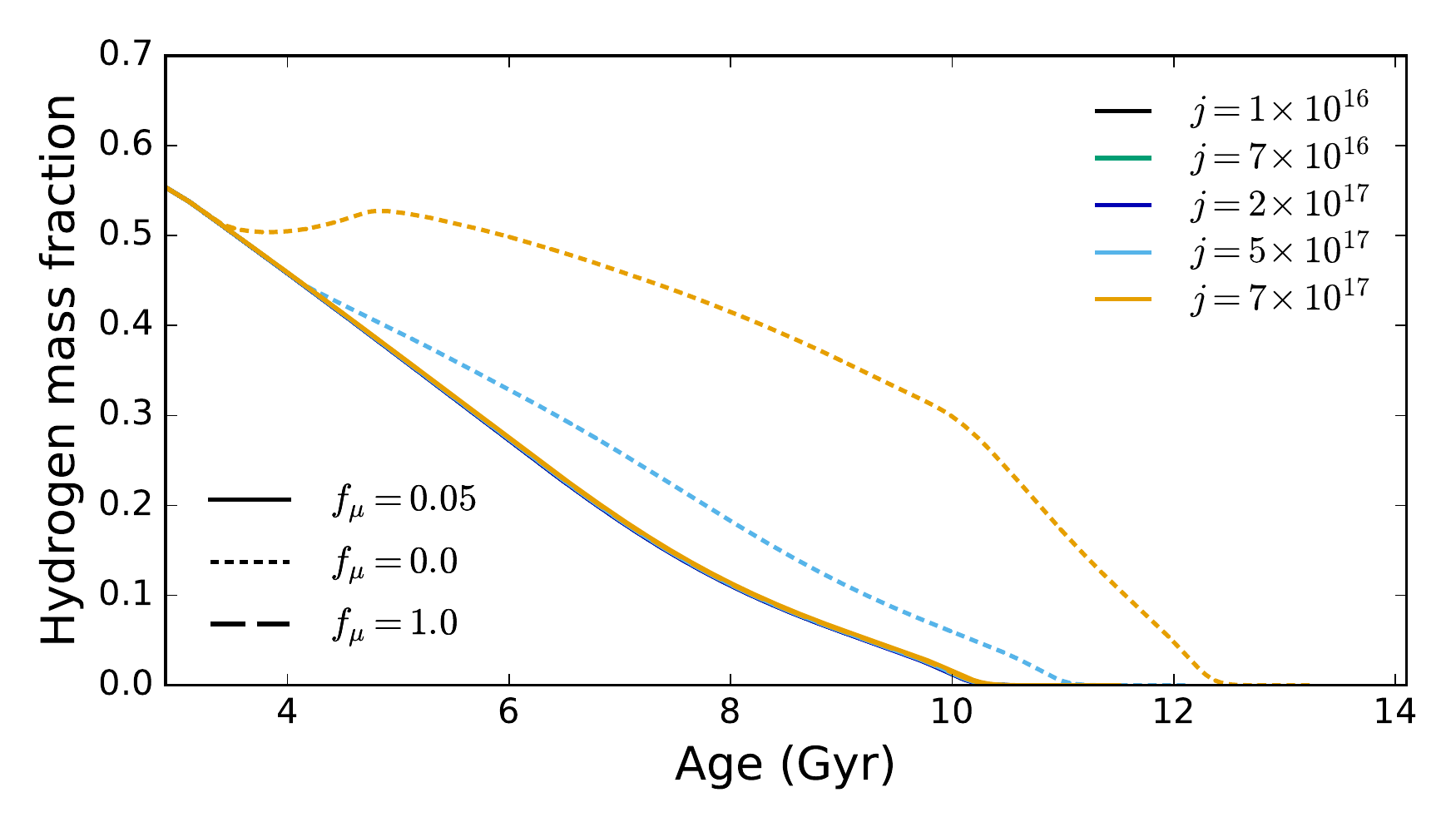}

\label{fig:mp125ms0800dm0050r_2-vs-100_fmu-test}}\hspace{\columnsep}\subfloat[Hertzsprung-Russell diagram at different $f_{\mu}$]{\includegraphics[width=1\columnwidth]{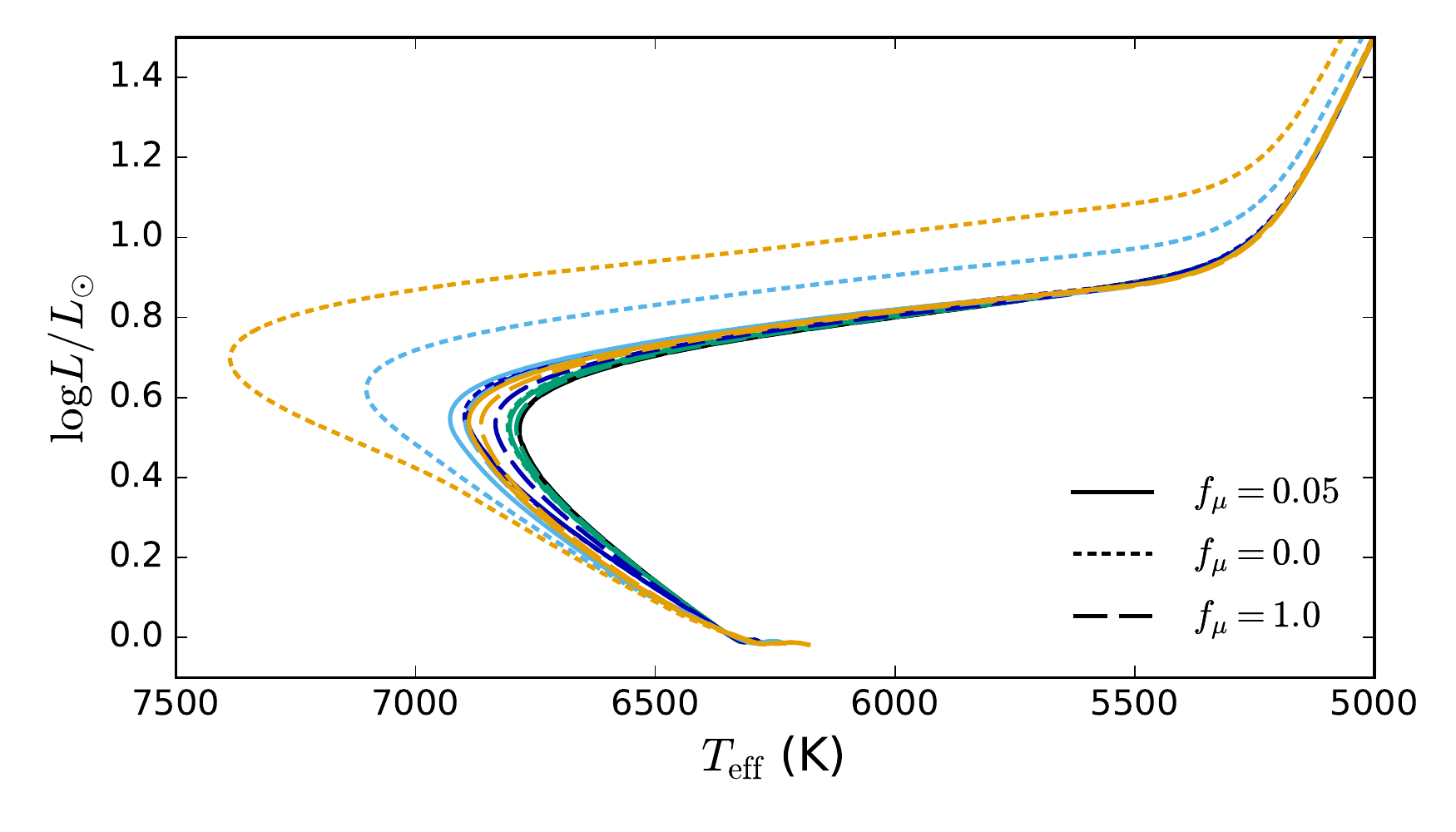}

\label{fig:mp125ms0800dm0050r_hrd_fmu-test}}

\subfloat[Carbon abundances at different $f_{\mu}$]{\includegraphics[width=1\columnwidth]{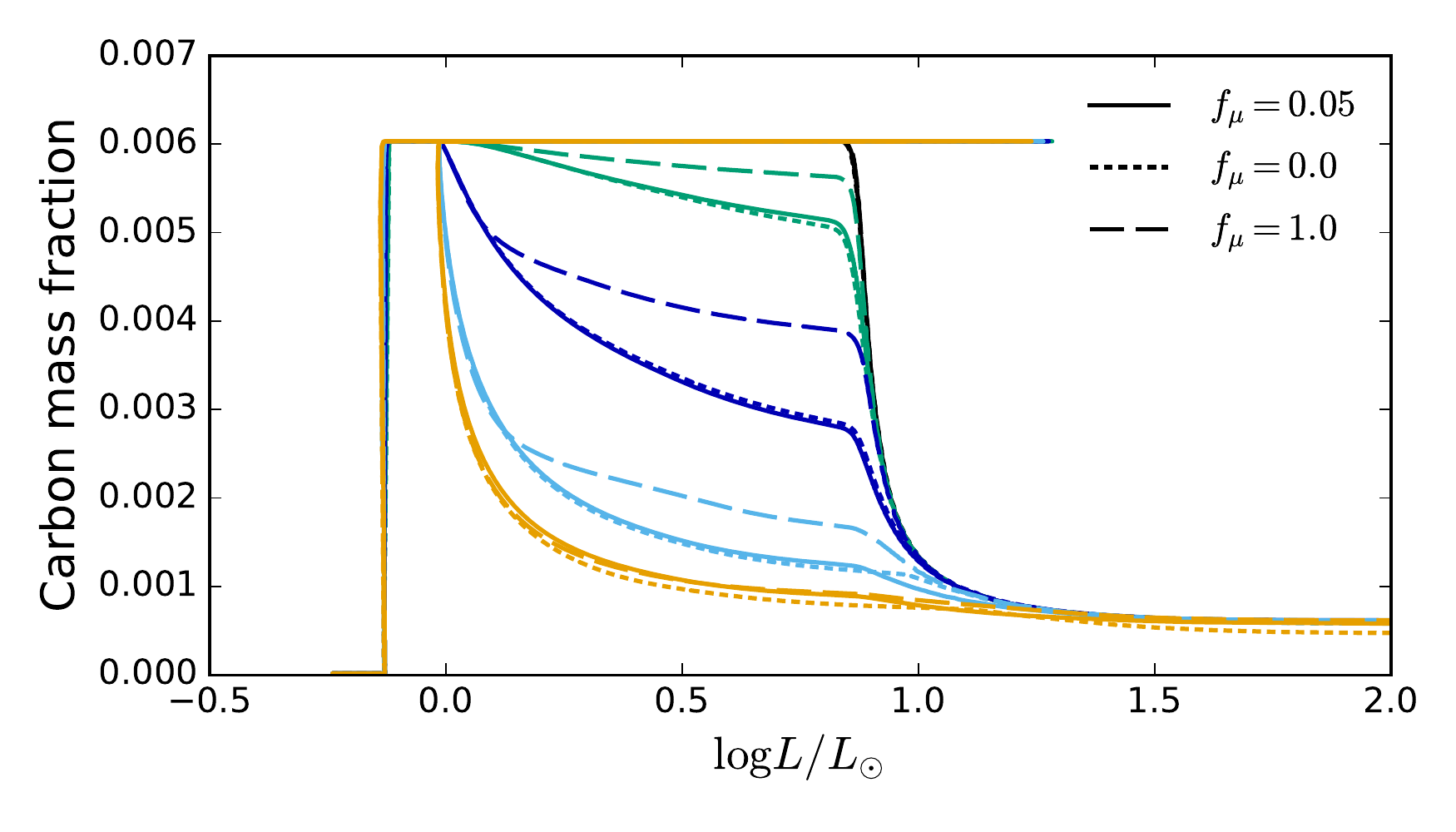}

\label{fig:mp125ms0800dm0050r_lgL-vs-XC_fmu-test}}\hspace{\columnsep}\subfloat[Carbon abundances at different $f_{c}$]{\includegraphics[width=1\columnwidth]{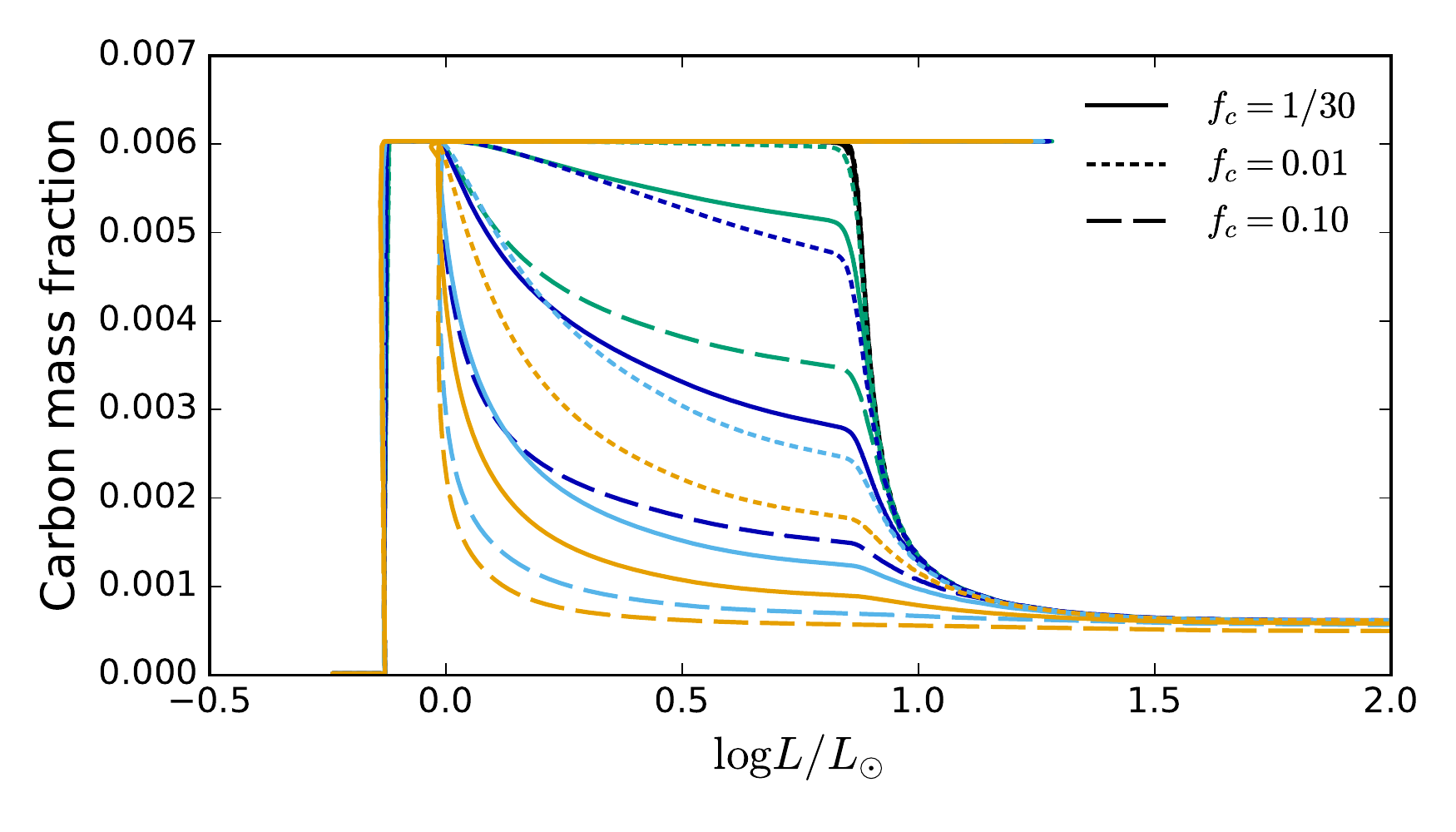}

\label{fig:mp125ms0800dm0050r_lgL-vs-XC_fc-test}}

\caption{Influence of changing the rotational mixing parameters $f_{\mu}$
(panels a-c) and $f_{c}$ (d) in the system with $M_{1}=1.25\ M_{\odot}$,
$M_{2,\text{i}}=0.8\ M_{\odot}$, $\Delta M=0.05\ M_{\odot}$. Models
with rotational mixing only.\label{fig:mp125ms0800dm0050r_fcfmu-test}}
\end{figure*}

The case with $f_{\mu}=0$ is theoretically the most interesting one.
Since the $\mu$-gradients do not interfere with the transport, the
extent of mixing depends primarily on its timescale. When it is sufficiently
short compared to the evolutionary timescale, as is the case here,
the angular momentum can be transported all the way to the centre
of the star. In this particular case, starting from about a gigayear
after mass transfer, the central regions rotate about a factor of
a hundred faster than in the default model. This allows mixing of
additional hydrogen into the burning regions and extends the main
sequence lifetime by about $0.7\ $Gyr. Not only are such models longer
lived (Fig.\ \ref{fig:mp125ms0800dm0050r_2-vs-100_fmu-test}), they
also evolve considerably hotter because of their larger helium content,
reaching much higher surface temperatures at turn-off ($T_{\text{eff}}>7000\ \text{K}$;
Fig.\ \ref{fig:mp125ms0800dm0050r_hrd_fmu-test}). Such temperatures
are not measured in CEMP stars, so rotational mixing in these stars
cannot be efficient enough to cause substantial chemical mixing of
the central regions. Observations would thus seem to rule out models
with $f_{\mu}=0$, but perhaps CEMP stars never acquire enough angular
momentum, or lose it too rapidly, to allow for extensive rotational
mixing in the first place (Sect.\ \ref{subsec:real-cemps}).

\begin{figure*}
\subfloat[$f_{\mu}=1.0$]{\includegraphics[width=1\columnwidth]{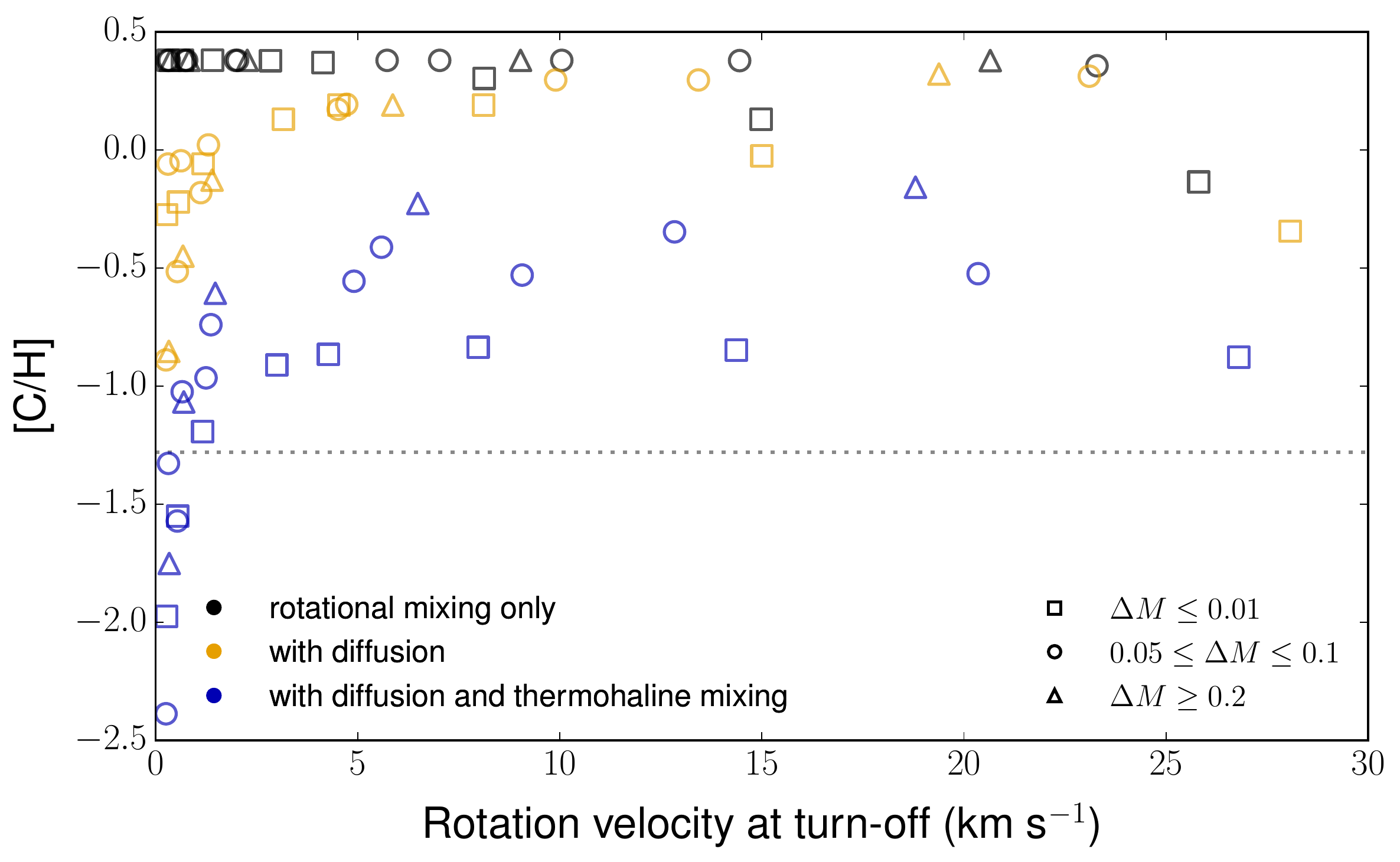}

\label{fig:mp125rdt_vrot-vs-CH_TO_fmu1.0}}\hspace{\columnsep}\subfloat[$f_{c}=0.01$]{\includegraphics[width=1\columnwidth]{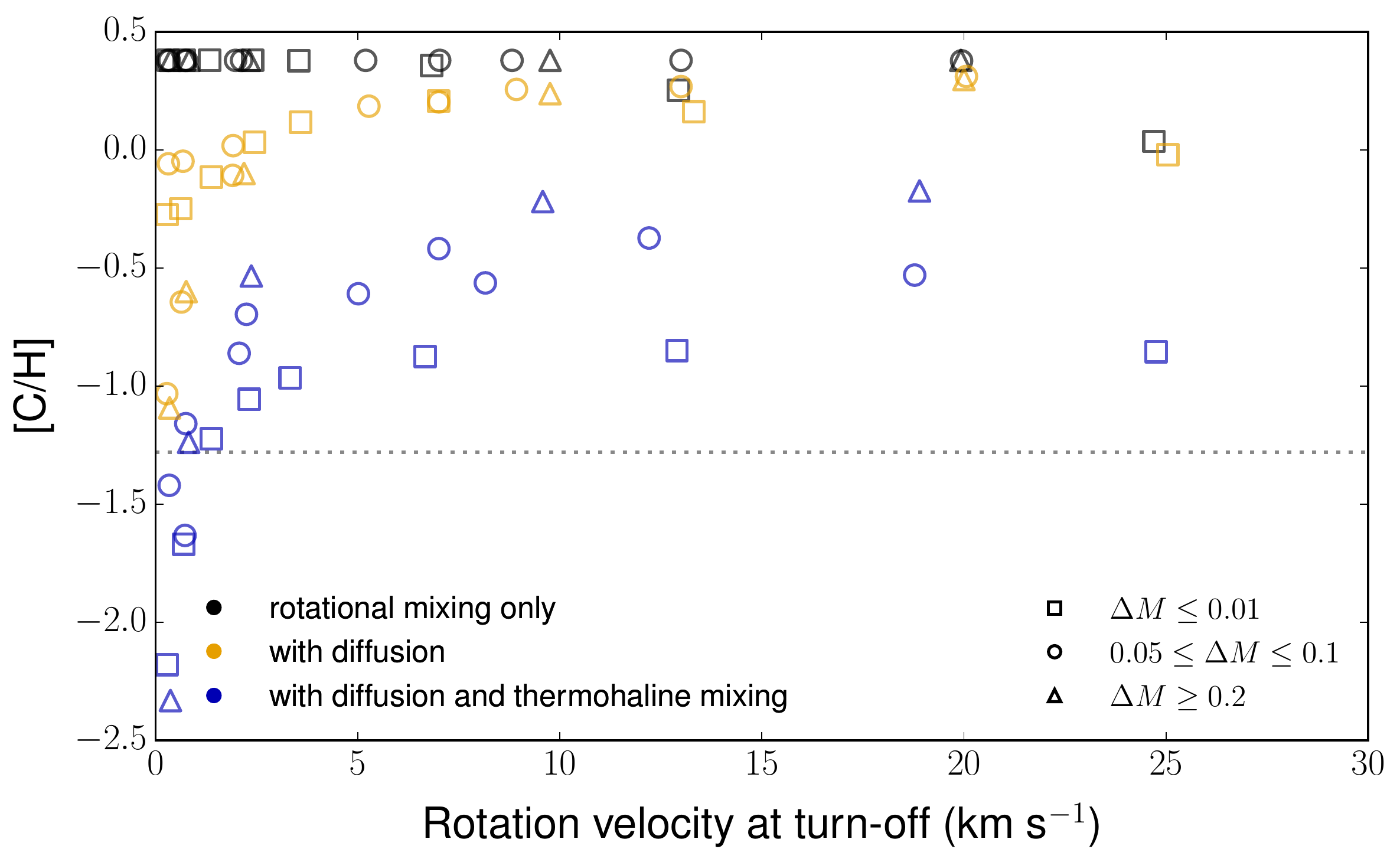}

\label{fig:mp125rdt_vrot-vs-CH_TO_fc0.01}}

\caption{As Fig.\ \ref{fig:mp125rdt_vrot-vs-CH_TO} but with different values
of $f_{c}$ and $f_{\mu}$.\label{fig:mp125rdt_vrot-vs-CH_TO_fcfmu-test}}
\end{figure*}

The internal evolution is naturally reflected by the surface abundances,
as Figs.\ \ref{fig:mp125ms0800dm0050r_lgL-vs-XC_fmu-test},\subref*{fig:mp125ms0800dm0050r_lgL-vs-XC_fc-test}
demonstrate. In terms of surface abundances there is thus somewhat
of a degeneracy between the specific angular momentum assigned to
the accreted material and $f_{c}$ (and $f_{\mu}$ to a lesser extent).
For example, models with $f_{c}=0.1$ resemble models with $f_{c}=1/30$
and higher $j_{\text{a}}$. While this ambiguity is difficult to disentangle
on a case-by-case basis, it does not influence some of the broader
conclusions reached in previous sections. For example, in models with
diffusion substantial abundance anomalies are still expected only
at rotational velocities $v_{\text{rot}}\lesssim2\ \text{km}\thinspace\text{s}^{-1}$
even in the two unfavourable cases with $f_{c}=0.01$ and $f_{\mu}=1.0$
(cf. Figs.\ \ref{fig:mp125rdt_vrot-vs-CH_TO} and \ref{fig:mp125rdt_vrot-vs-CH_TO_fcfmu-test}).

Thus there are no strong constraints on $f_{c}$ in the range {[}0.01,
0.1{]}. Outside of this range we expect that at least from the lower
end $f_{c}$ could be constrained (assuming that rotational mixing
is indeed responsible for suppressing atomic diffusion). Eventually,
for $f_{c}\ll0.01$, the chemical transport due to rotational instabilities
must become so slow that atomic diffusion would be expected to dominate
the surface abundance evolution of CEMP stars, at odds with observations.
Overall, because of the many steps involved in creating CEMP-\emph{s}
stars (nucleosynthesis in the AGB donor, mass and angular momentum
accretion, and subsequent mixing of the accreted material), they could
offer only loose constraints on $f_{c}$ (and $f_{\mu}$), which would
in any case be consistent with more stringent constraints from other
types of stars.

\section{\label{sec:Conclusions}Summary and conclusions}

We present a large number of models of \emph{s}-process-rich carbon-enhanced
metal-poor (CEMP-\emph{s}) stars under the standard paradigm of mass
accretion from an asymptotic giant branch donor. As a follow-up to
\citet{2016A&A...592A..29M}, for the first time we investigate what
effect angular momentum accretion has on the chemical evolution of
CEMP-\emph{s} stars. The angular momentum deposited in the outer layers
of the stars triggers rotational instabilities that induce mixing
of angular momentum and the stellar material. We model the combined
action of this rotational mixing with atomic diffusion (gravitational
settling), and thermohaline mixing.

We can broadly summarize the relevance each of these processes has
to different CEMP-\emph{s} stars as follows. In the slowest rotating
($v_{\text{rot}}\lesssim1\ \text{km}\thinspace\text{s}^{-1}$) massive
($M_{2,\text{f}}\gtrsim0.85\ M_{\odot}$) stars the greatest abundance
changes are caused by atomic diffusion near the main sequence turn-off
($\log L\simeq0.5$; $\log g\simeq4$). Depending on the amount of
mass accreted (and also the mean molecular weight of the accreted
material), either first dredge-up (occurring during $3.5\gtrsim\log g\gtrsim3$)
or thermohaline mixing ($\log g\simeq4.5$) is more important in stars
with moderate rotation velocities ($v_{\text{rot}}\lesssim20\text{\ km}\,\text{s}^{-1}$),
and also less massive CEMP-\emph{s} stars ($M_{2,\text{f}}\lesssim0.8\ M_{\odot}$).
Rotational mixing could be important for (internally) more rapidly
rotating stars, but only if thermohaline mixing is somehow rendered
ineffective, or the accreted mass is only of the order of $10^{-3}\ M_{\odot}$.
Then rotational mixing could lead to similar abundances as thermohaline
mixing but gradually, over timescales comparable to the main sequence
lifetime.

We find that in models with rotation velocities characteristic of
CEMP-\emph{s} stars ($2\lesssim v_{\text{rot}}(\text{km}\,\text{s}^{-1})\lesssim15$),
rotational mixing suppresses the significant abundance anomalies (e.g.
$\text{[C/Fe]}<-1$ from a post-mass-transfer abundance of $\text{[C/Fe]}>2$)
that in absence of rotation are expected to develop near the main
sequence turn-off from uninhibited atomic diffusion \citep{2016A&A...592A..29M}.
The models thus remain carbon-enhanced ($\text{[C/Fe]}\gtrsim1$)
throughout the evolution, as long as the rotation rates are high enough
($v_{\text{rot}}\gtrsim1\ \text{km}\thinspace\text{s}^{-1}$; Figs.\ \ref{fig:mp125rdt_vrot-vs-CH_TO},
\ref{fig:logg-vs-CHrdt_with-SDSS}, \ref{fig:logg-vs-CHrd_with-SDSS}).
It is not known whether any CEMP-\emph{s} (or in general metal-poor)
stars rotate at still lower rates. But, if rotational mixing is indeed
normally responsible for countering atomic diffusion in low-mass stars,
such slowly rotating stars should have large abundance anomalies.
These conclusions are rather insensitive to the parameters characterizing
the efficiency of rotational mixing in our models (Fig.\ \ref{fig:mp125rdt_vrot-vs-CH_TO_fcfmu-test}). 

There is plenty of room for improvement in the treatment of angular
momentum accretion and evolution. In particular, we have treated the
ratio of angular momentum to mass accreted, i.e. the specific angular
momentum of the accreted material, as a free parameter spanning three
orders of magnitude. In real systems this range could be much more
restricted, and dedicated multi-dimensional simulations are required
to constrain it. Particularly, if the specific angular momentum is
high, a mechanism for angular momentum loss during accretion must
be identified. Otherwise, it is impossible to explain the most carbon-enhanced
objects. Also, the nature and outcome of the mutual interaction between
rotational and other instabilities, such as thermohaline convection,
should be settled. Finally, we have only briefly considered the possibility
of angular momentum loss following the mass transfer. But, since in
terms of surface abundances such stars evolve similarly to more rapidly
rotating stars without angular momentum losses, we do not expect this
to invalidate our main conclusions.

\begin{acknowledgements}
We thank Adrian Potter for sharing his rotating version of the \textsc{stars}
code (RoSE) and Young Sun Lee and Timothy Beers for sharing the SDSS
CEMP data. We also thank Carlo Abate for constructive comments on
this manuscript. RJS is the recipient of a Sofja Kovalevskaja Award
from the Alexander von Humboldt Foundation.
\end{acknowledgements}

\bibliographystyle{aa}

\bibliography{AA-2017-31272.bib}
\end{document}